\documentclass[sigconf]{acmart}

\usepackage{makecell}
\usepackage{multirow}

\definecolor{PartnerOrange}{HTML}{FF651E} 
\definecolor{DirectBlue}{HTML}{0059FF}    
\definecolor{BasicGray}{HTML}{666666}     

\newcommand{\BP}{\textcolor{BasicGray}{\textbf{BP}}}
\newcommand{\DS}{\textcolor{DirectBlue}{\textbf{DS}}}
\newcommand{\PS}{\textcolor{PartnerOrange}{\textbf{PS}}}

\newcommand{\BPbox}{\colorbox{BasicGray}{\textbf{\textcolor{white}{BP}}}}
\newcommand{\DSbox}{\colorbox{DirectBlue}{\textbf{\textcolor{white}{DS}}}}
\newcommand{\PSbox}{\colorbox{PartnerOrange}{\textbf{\textcolor{white}{PS}}}}

\AtBeginDocument{%
  }

\copyrightyear{2026}
\acmYear{2026}
\setcopyright{cc}
\setcctype{by}
\acmConference[CHI '26]{Proceedings of the 2026 CHI Conference on Human Factors in Computing Systems}{April 13--17, 2026}{Barcelona, Spain}
\acmBooktitle{Proceedings of the 2026 CHI Conference on Human Factors in Computing Systems (CHI '26), April 13--17, 2026, Barcelona, Spain}
\acmPrice{}
\acmDOI{10.1145/3772318.3791370}
\acmISBN{979-8-4007-2278-3/2026/04}

\acmISBN{978-1-4503-XXXX-X/2018/06}

\begin{document}

\title{Scaffolded Vulnerability: Chatbot-Mediated Reciprocal Self-Disclosure and Need-Supportive Interaction in Couples}

\author{Zhuoqun Jiang}
\affiliation{%
  \institution{Singapore University of Technology and Design}
  \country{Singapore}
}
\email{zhuoqun_jiang@mymail.sutd.edu.sg}
\orcid{0000-0002-4296-0996}
\authornote{Corresponding author.}

\author{ShunYi Yeo}
\affiliation{%
  \institution{Singapore University of Technology and Design}
  \country{Singapore}}
\email{yeoshunyi.sutd@gmail.com}
\orcid{0000-0002-8067-3402}

\author{Dorien Herremans}
\affiliation{%
  \institution{Singapore University of Technology and Design}
  \country{Singapore}}
\email{dorien.herremans@gmail.com}
\orcid{0000-0001-8607-1640}

\author{Simon Tangi Perrault}
\affiliation{%
  \institution{Télécom Paris, Institut Polytechnique}
  \country{Paris}}
\email{simon.perrault@telecom-paris.fr}
\orcid{0000-0002-3105-9350}

\renewcommand{\shortauthors}{Zhuoqun Jiang, ShunYi Yeo, Dorien Herremans, and Simon Perrault}

\begin{abstract}
  While reciprocal self-disclosure drives intimacy, digital tools seldom scaffold autonomy, competence, and relatedness---the motivational underpinnings defined by Self-Determination Theory (SDT) that enable deep exchange. We introduce a chatbot employing dual-layer scaffolding to satisfy these needs: first providing enabling affordances (instrumental support) for vulnerability, then mediating affordances (relational support) for responsiveness. In a randomized study ($N=72$; 36 couples) comparing Partner Support (\PS{}: both layers), Direct Support (\DS{}: enabling only), and Basic Prompt (\BP{}: questions only), results reveal a critical distinction. While enabling affordances (\PS{}, \DS{}) were sufficient to deepen disclosure, only mediating affordances (\PS{}) reliably elicited partner-provided need support and increased perceived closeness. Furthermore, controlled motivation decreased across conditions, and scaffolding buffered vitality, which remained stagnant in \BP{}. We contribute empirical evidence that SDT-guided mediation fosters connection, offering a practical framework for designing AI-mediated conversations that support, rather than replace, human intimacy.
\end{abstract}



\begin{teaserfigure}
  \centering
  \includegraphics[width=\textwidth]{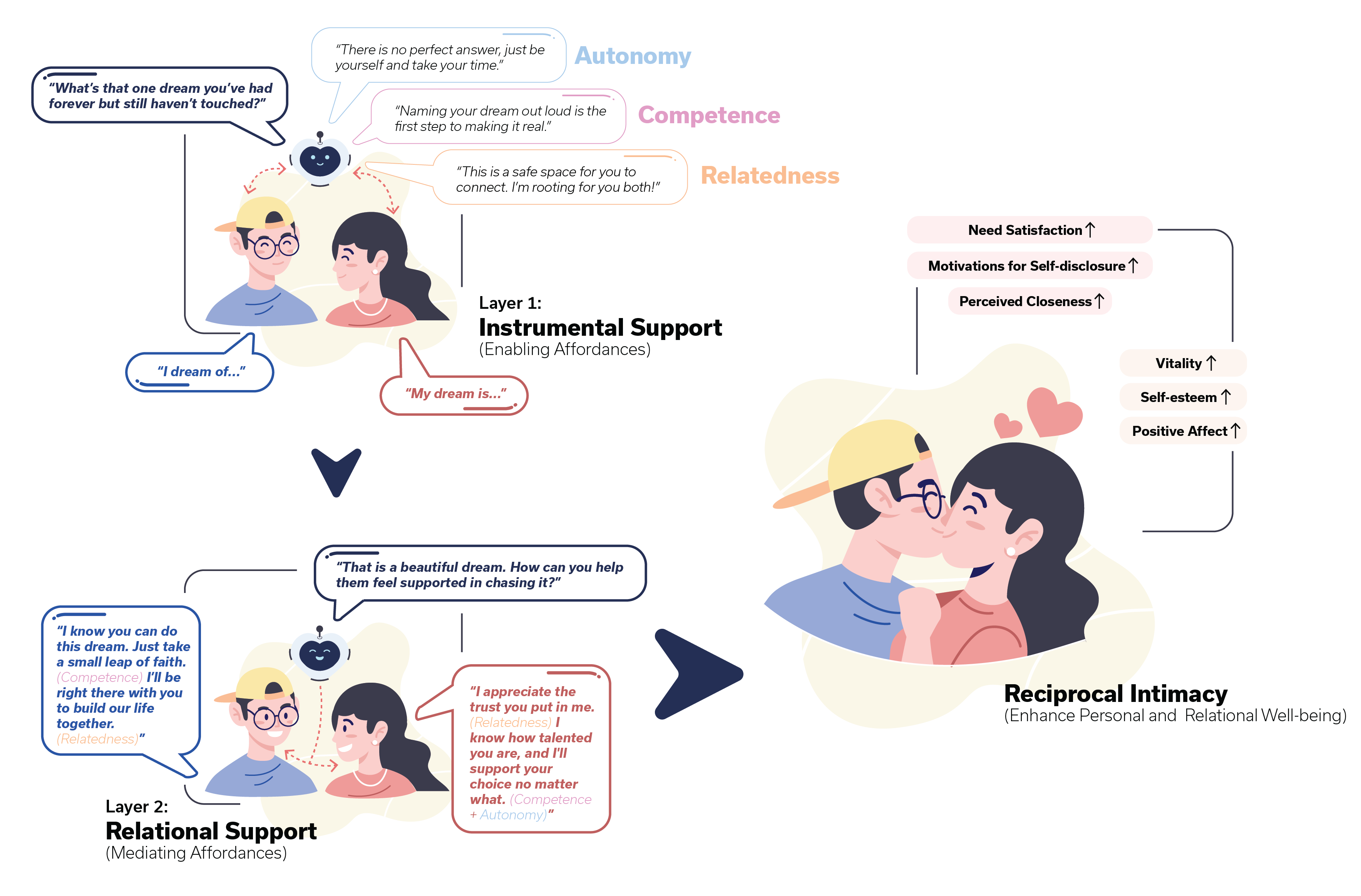}
  \caption{Conceptual vision of our dual-layer scaffolding framework. Layer 1 provides Enabling Affordances (Instrumental Support) to the individual, establishing the safety and structure required for disclosure. Layer 2 deploys Mediating Affordances (Relational Support) to scaffold the partner's reciprocal validation. Together, this sequence transforms individual disclosure into reciprocal intimacy, fostering relational and personal well-being.}
  \label{fig:visual_summary}
  \Description{A conceptual flow diagram in three parts. On the left, a chatbot interacts with a couple by supporting three basic psychological needs: autonomy, shown with options, pacing, and rationales; competence, shown with guidance and affirmations; and relatedness, shown with warmth and reciprocity. In addition, the chatbot delivers prompts in speech bubbles, such as ``What's that one dream you've had forever but still haven't touched?'' and ``Is there something small you could do to help your partner feel more supported in working toward it?'' These illustrate how the system guides self-disclosure and encourages supportive partner responses. On the right, a couple is depicted affectionately, representing enhanced outcomes. Labels indicate the expectation of increased relational well-being (need satisfaction, motivation for self-disclosure, and closeness) and personal well-being (vitality, self-esteem, and positive affect).}
\end{teaserfigure}

\maketitle

\section{Introduction}

Communication is central to maintaining close relationships~\cite{stafford2019communication, matthew2022within}. Specifically, self-disclosure serves as a critical mechanism for this maintenance, allowing partners to deepen their bond through the sharing of personal thoughts and feelings~\cite{laurenceau1998intimacy, reis1988intimacy}. While self-disclosure fosters intimacy and trust, its impact depends heavily on the motivational quality of the exchange~\cite{gaine2009theunique, pusch2022motivational}. Self-Determination Theory (SDT) posits that conversations build genuine closeness only when they support three basic needs: autonomy (volition), competence (efficacy), and relatedness (connection)~\cite{ryan2000selfdetermination, laguardia2008selfdetermination, kluwer2020autonomy}. Yet, everyday constraints often thwart these needs as daily talk is frequently dominated by routine logistics, and partners often possess mismatched levels of comfort with vulnerability~\cite{duck2006someevident, ryjova2024aday, ignatius2007factors, gaine2009theunique}. Consequently, opportunities for deep disclosure are missed, and even well-intentioned attempts to deepen connection can backfire if partners feel pressured (low autonomy), unsure of how to respond (low competence), or unheard (low relatedness)~\cite{kil2022autonomy, gaine2009theunique, ignatius2007factors}. The challenge, therefore, is to provide gentle scaffolding that renders disclosure voluntary, confident, and responsive.

To bridge these barriers, couples increasingly rely on Computer-Mediated Communication (CMC) to maintain contact~\cite{kwok2020cyberintimacy, foster2022rules}. While standard messaging platforms support logistics and lightweight conversation~\cite{daantje2008therole}, HCI research has developed dedicated relatedness technologies designed specifically to amplify intimacy~\cite{wenhart2025relatedness}. Yet, when we analyze these systems through an SDT lens, a clear trade-off emerges between intimacy and instrumental support. Technologies focused on affective expressivity (e.g., biosignals, lights) succeed in surfacing emotions but often struggle to sustain reciprocity; by emphasizing passive sensing over active co-production, they leave the interaction loop fragile~\cite{brueckner2018empathy, liu2021significant, kelly2017demanding}. Conversely, systems facilitating genuine conversation (e.g., telepresence) foster togetherness but often lack the scaffolding for competence, rarely providing the guidance required to navigate emotionally rich topics~\cite{vaanen2011carpe, heshmat2020family}. Finally, text-enrichment tools aim to enforce depth but often compromise autonomy; by imposing constraints like character counts, they risk framing care as a procedural burden rather than a volitional act~\cite{kelly2018letter, kim2019love}.

We argue that conversational agents are uniquely positioned to resolve this trade-off. Unlike passive signals or static question banks, chatbots can dynamically regulate the conversation flow, offering a unique opportunity to scaffold reciprocity while preserving interpretability~\cite{zheng2021pocketbot, jiang2025remini}. However, prior work largely treats such facilitation as valuable in its own right, without explicitly grounding it in underlying psychological processes. We argue that to fully leverage a chatbot's potential for intimacy, design must follow a dual-layer scaffolding approach. We posit that the limitations of prior systems often stem from attempting to elicit relational support (intimacy between partners) without first providing the necessary instrumental support (the psychological conditions for engagement). Specifically, if a user feels pressured by the system, unsure of how to respond, or emotionally unsafe in the environment, their capacity to support their partner diminishes. Therefore, our chatbot functions sequentially: \textbf{instrumental support} serves as the foundation, providing \emph{enabling affordances} to support the user's needs for volition (Autonomy), clarity (Competence), and psychological safety (Relatedness). Only once the user feels capable and unpressured does the system deploy \emph{mediating affordances} to scaffold \textbf{relational support}, guiding partners to enact the mutual and reciprocal need support which is essential for sustaining close relationships.

To operationalize this, we built an SDT-grounded chatbot that acts as a facilitative mediator. The system integrates the 36 Questions paradigm~\cite{jones2015thirtysix} into a phased conversational structure. It implements our dual-layer model by: (i) providing \emph{enabling affordances} via direct need support (e.g., offering choice, rationales, and warmth) to establish a need-supportive environment; and (ii) deploying \emph{mediating affordances} via structured follow-up prompts that explicitly guide partners to provide autonomy, competence, and relatedness support to one another. We evaluated this model with 36 couples ($N = 72$) in a randomized study isolating these layers across three conditions: Partner Support (\PS{}: enabling + mediating), Direct Support (\DS{}: enabling only), and Basic Prompt (\BP{}: questions only).

Our results show that engagement is the highest in \PS{}, yielding the longest, richest conversations. \PS{} and \DS{} also deepened self-disclosure versus \BP{}, but only \PS{} reliably elicited partner-provided need-supportive behaviors and increased perceived closeness. Furthermore, while subjective vitality remained statistically stagnant in the baseline, it improved in the scaffolded conditions. Couples described the chatbot as creating rare opportunities for deeper talk, modeling supportive phrasing, and scaffolding reflections that helped them validate and encourage one another. We also discuss how mediating affordances were crucial for eliciting need-supportive behaviors, how autonomy can be preserved even with structured guidance, and how chatbots may best function as facilitators that help couples build lasting supportive practices.

Our research makes three contributions: \begin{enumerate} \item \textbf{Empirical Evidence:} We demonstrate that while chatbot-provided instrumental support is sufficient to deepen disclosure, mediating affordances are necessary to transform that disclosure into reciprocal care and relational closeness. \item \textbf{Design Insights:} We contribute concrete interaction strategies that resolve the tension between structural guidance and user agency, showing how enabling and mediating affordances can foster competence and relatedness without compromising autonomy. \item \textbf{Design Framework:} We extend the design space of relatedness technologies by introducing Dual-Layer Scaffolding, an SDT-grounded design framework that articulates how system-provided support capacitates partner-provided care to sustain meaningful engagement. \end{enumerate}

\section{Background and Related work}

\subsection{Innate Motivational Underpinnings in Self-Disclosure within Close Relationships}

\subsubsection{Self-disclosure in Close Relationships}

Self-disclosure, the act of sharing personal information with another person, is central to how close relationships develop~\cite{derlega1993selfdisclosure, sprecher2004selfdisclosure}. It has been consistently linked to intimacy~\cite{finkenauer2015toknowyou, laurenceau1998intimacy}, relationship satisfaction~\cite{meeks1998communication}, and long-term stability~\cite{sprecher1987theeffects}. Drawing on \textbf{Social Penetration Theory} (SPT), self-disclosure can be described in terms of both breadth and depth: early interactions often involve broad but superficial talk about everyday topics, whereas closer relationships increasingly involve sharing beliefs, values, and personally meaningful experiences~\cite{altman1973social, ayres1979uncertainty}. Disclosures that express feelings, rather than just facts, are especially predictive of intimacy and satisfaction~\cite{reis1988intimacy}. Self-disclosure is also inherently reciprocal. When one partner's openness is met with attentive listening and a comparable level of sharing, trust is strengthened and the exchange is mutual and rewarding~\cite{sprecher2013takingturns, collins1994selfdisclosure}. In contrast, mismatched or one-sided disclosure can feel uncomfortable, making people more hesitant to open up in future. This work focuses on conversations that invite both partners to move beyond surface-level talk toward more balanced, emotionally rich self-disclosure.

\subsubsection{Psychological Needs and Motivation in Relationships}

Beyond reciprocity, responses quality matters: disclosures met with acceptance, validation, and care promote relationship growth and maintenance~\cite{laurenceau2005theinterpersonal}. Self-disclosure is not just an exchange of information but a way for partners to seek recognition and support of their psychological needs. \textbf{Self-Determination Theory} (SDT) provides a useful lens for understanding this process, explaining how the motivational basis of relational behaviors shapes their impact, and why even well-intentioned behaviors can be experienced as more or less beneficial depending on the context~\cite{laguardia2008selfdetermination}.

SDT posits three basic psychological needs that are especially relevant in close relationships: autonomy, competence, and relatedness~\cite{ryan2000selfdetermination}. In the context of disclosure, \textbf{Autonomy} involves respecting the partner's volition and perspective. It manifests as acknowledging the other's internal frame of reference, minimizing pressure, and encouraging them to share at their own pace rather than directing the flow of disclosure. \textbf{Competence} entails scaffolding the partner's sense of efficacy. It is enacted by providing structure, concrete encouragement, and clear feedback that helps partners feel confident in their ability to articulate difficult emotions or achieve their personal aspirations. \textbf{Relatedness} focuses on emotional connection. It is demonstrated through expressions of warmth, validation, and genuine interest, which signal to the partner that they are understood and their feelings are safe~\cite{ryan2000selfdetermination, baumeister1995theneed, ryan2000thedarker, laguardia2000within, ryan1995autonomy}. Crucially, SDT suggests that the satisfaction of these needs is essential for high-quality interaction: their fulfillment is linked to greater emotional awareness, openness in processing emotions, and willingness to share them with a partner, as well as higher self-esteem, vitality, positive affect, and stronger relationship quality and closeness~\cite{patrick2007therole, la2007end}.

SDT also emphasizes that need support shapes why people engage in relational behaviours. When conversations support autonomy and competence, partners are more likely to talk because they personally value these exchanges and see them as part of their relationship, rather than because they feel pressured, guilty, or afraid of disappointing the other~\cite{ryan2000thedarker, deci2000thewhat}. Organismic Integration Theory (OIT), a subtheory of SDT, describes this process of internalization: over time, people can move from more controlled forms of motivation (e.g., ``I have to talk about this so my partner won't be upset'') towards more autonomous forms (e.g., ``we talk about this because it matters to us''), which is associated with more sustainable engagement and better well-being~\cite{deci2013intrinsic, ryan2000selfdetermination}.

People also differ in their tendency to disclose~\cite{ignatius2007factors}, and within romantic relationships, overall motivation to maintain the bond may diverge from motivation for specific activities such as talking about personal worries or long-term goals~\cite{gaine2009theunique}. Partners often differ in their comfort with sharing private information, which can lead to tension~\cite{kil2022autonomy}. Naturalistic studies further show that couples' daily conversations are often routine and focused on logistics, media, or household tasks, so opportunities for deep, affective disclosure may not arise~\cite{duck2006someevident, ryjova2024aday, jess2005mapping}. These findings suggest that disclosure, despite being central to relationship quality, is unevenly distributed and the conditions for need-supportive conversations may not arise.

This gap highlights the need for external scaffolding to break the cycle of routine talk. Motivated by this, we designed our conversational agent to act as a facilitative mediator. As a design lens~\cite{beck2016examining}, SDT draws attention to how autonomy, competence, and relatedness are supported in the moment, guiding us to adopt a warm, non-controlling chatbot tone that avoids pressuring participants to disclose and instead prompts partners to acknowledge, validate, and concretely support one another. As a methodological and analytical lens~\cite{beck2016examining}, SDT informs what we measure and how we interpret it: we assess participants' perceived autonomy, competence, and relatedness support in the conversation, and their autonomous and controlled motivation, allowing us to evaluate whether the system successfully shifts interaction dynamics from routine exchange to deep, volitional connection.

\subsection{Technologies for Intimacy and Relatedness in Close Relationships}

\subsubsection{Technologies Facilitating Self-disclosure and Expressivity}

In romantic relationships, \textbf{computer-mediated communication} (CMC) is used across multiple channels~\cite{kwok2020cyberintimacy, foster2022rules}, including text messages and voice and video calls as a way to maintain relationship~\cite{daantje2008therole, chou2022why, neustaedter2012intimacy}. CMC can help partners stay connected, bridge periods of physical separation, and foster emotional closeness and relationship quality~\cite{gomes2021theimpact, aljasir2022present, holtzman2021longdistance}. Messaging apps are mostly used for logistical coordination, lightweight check-ins, and affectionate banter, while video calls, shared media or games help couples create a sense of being together even when physically apart~\cite{shen2024seeking}.

Building on these general-purpose platforms, HCI research has developed dedicated \emph{relatedness technologies} designed specifically to amplify intimacy. A common design strategy is to foreground affective self-disclosure and expressivity, allowing partners to share feelings across different modalities, e.g. through physiological and biosignal sharing, like heart rate~\cite{buschek2018personal, liu2019animo, liu2021significant}; IoT ecosystems and connected artifacts~\cite{hansen2020thedelay, leong2023social, yang2020autobiographical, ambe2022from}; ambient or metaphorical displays such as colored smart lights~\cite{morris2017augmenting, gaver2023living, gaver2022yoyo}; AR experiences~\cite{zhang2022auggie}; and messaging platforms augmented with emotional layers, for example through expressive message bubbles~\cite{an2024emowear, jarus2020always}. These systems demonstrate how technology can make emotions more visible and create small moments of shared emotional experience.

This body of work reveals recurring challenges. Reciprocity is often fragile and uneven: technical setups may give one partner more opportunities or responsibility: e.g. when telepresence or smart-home tools exist only in one home, leaving invisible labor to the local partner~\cite{yang2020autobiographical}. Some systems prompt constant exchanges, such as daily light signals or ambient data, making missed responses feel like neglect rather than a normal pause~\cite{gaver2023living}. Even biosignal-sharing wearables, though bidirectional, often emphasize passive sensing with little support for balanced, supportive acknowledgement~\cite{min2014biosignal}. Authenticity and interpretability can also be compromised. Biosignals and metaphorical representations may exaggerate feelings, create ambiguity, or raise doubts about reliability, sometimes leading to intrusiveness or anxiety~\cite{liu2021significant, min2014biosignal, brueckner2018empathy}. These tensions suggest limits of opaque or non-textual signals without transparent, user-controlled channels for reciprocity. Building on this, we examine how self-disclosure can be scaffolded as a mutual process within a familiar text-based medium.

In contrast to these often passive or ambiguous signals, research on family dynamics highlights the value of active co-production. Work on co-parenting frames support as a reciprocal labor where partners coordinate tasks and sustain solidarity~\cite{lin2024ultimately, lin2025parental}, while mobile interventions such as \emph{Love Every Day} use brief daily prompts to encourage caring behaviours to improve relationship quality~\cite{lucier2018enhancing}. These studies suggest that couples benefit from being active agents of care rather than passive broadcasters of data. Building on this insight, we examine how self-disclosure can be scaffolded as a reciprocal, co-produced process within a familiar text-based medium, moving toward the clarity of explicit support.

\subsubsection{Technologies Supporting Genuine Conversation}

Some studies treats \emph{genuine conversation} as the central strategy for fostering relatedness, deeper emotional engagement and empathy building~\cite{wenhart2025relatedness}. These systems do so by (a) transmitting nonverbal cues, e.g. facial expression or gaze awareness~\cite{misawa2012mapetite, vaanen2011carpe}; (b) supporting shared activities like witnessing events or exchanging photos~\cite{tang2013homeproxy, baishya2017inyoureyes, lenz2016four}; (c) offering ambient awareness of a partner's context~\cite{dziabiola2022qude, angelini2015multisensory}; (d) bridging environmental barriers such as time-zone differences to sustain everyday talk~\cite{heshmat2020family, raju2021pvpix, heshmat2017connecting}; and (e) easing spontaneous connection by suggesting topics or facilitating call initiation~\cite{wang2023chatterbox, yarosh2013almost}. By making it easier to grasp the context around a partner (activity, look), these systems lower the friction of starting a conversation, help partners feel present in one another's daily lives, and support practices such as sharing small experiences, listening, and expressing needs~\cite{wenhart2025relatedness}. Yet many of these systems struggle to support conversational depth and emotional richness. For example, users of eye-based video conferencing found it difficult to interpret emotions from a single eye~\cite{vaanen2011carpe}; and in FamilyStories, participants often shared short asynchronous snippets while reserving more meaningful topics for real-time conversations~\cite{heshmat2020family}.

Within this broader space, a subset of \emph{genuine connection} systems focuses on improving text itself rather than adding new communication channels~\cite{calvo2014positive}. These designs enrich written messages by making them more expressive and intentional~\cite{kelly2018letter, kim2019love, kelly2018rewriting, podlubny2017synchronous}. For example, Lily suggests semantically aligned song lyrics to refine affectionate expression~\cite{kim2019love}, while Message Builder promotes effortful composition by requiring longer messages as a signal of care~\cite{kelly2018letter}, challenging CMC's efficiency bias~\cite{kelly2017demanding}. Yet effort can also feel burdensome, as seen in Message Builder's escalating character counts~\cite{kelly2018letter}, Lily's non-clickable suggestions~\cite{kim2019love}, and Curtains Messenger's enforced synchronicity~\cite{podlubny2017synchronous}. Defining meaningful ``effort'' is similarly difficult: metrics like character counts can be gamed, and system-generated prompts may undermine authenticity~\cite{kelly2018letter, kim2019love}. Critically, many of these interventions focus on modulating the \emph{received} message to preserve perceived genuineness, rather than scaffolding the \emph{sender's} internal capacity to articulate vulnerable thoughts. Our work takes the latter approach: we posit that by supporting the sender's autonomy and competence to express themselves authentically, the system naturally fosters the receiver's need satisfaction and relational connection, without requiring artificial enhancement of the message artifact.

The stakes of these design gaps become visible when couples face sensitive or high-stakes topics. Van Greevenbroek et al. show that discussing declining or uneven sexual desire requires more than just connection or enforced effort; it demands shared, protected spaces and gentle prompts that ease initiation and support non-judgemental understanding~\cite{van2023like}. Their work illustrates that emotionally charged topics require support for timing, tone, and a sense of safety—forms of competence and autonomy support often missing in purely presence-based or constraint-based tools.

In sum, prior work reveals a trade-off: presence-based systems foster relatedness but often lack the competence support needed for deep engagement, while text-enrichment tools enforce depth but at the cost of autonomy. This highlights a critical design challenge: how to scaffold the relational process without rigidly constraining the textual form of their communication.

\subsubsection{Technologies Leveraging Chatbots for Relational Communication}

With rapid advances in conversational agents, chatbots have emerged as a promising medium-specific intervention in text-based CMC, offering ways to scaffold reciprocity, preserve interpretability, and reframe effort as meaningful. While most studies have focused on \emph{individual} interactions with chatbots, their role in improving expressivity and facilitating self-disclosure is well established~\cite{lee2020ihearyou, park2021wrote, seo2024chacha}. More recent work situates chatbots \emph{within close relationships}, pointing to their potential to mediate or scaffold conversations in ways that strengthen reciprocity loops while maintaining authenticity and interpretability~\cite{yuksel2023conversational, jiang2025remini, zheng2021pocketbot, balaji2024designing}. For example, PocketBot, a rule-based chatbot, offers prompts for humor, conflict repair, and ``deep talk,'' encouraging emotionally richer exchanges through simple, pre-scripted questions~\cite{zheng2021pocketbot}. Its deep talk feature was explicitly designed to help long-distance couples reopen conversations about values, feelings, and aspirations, and in a field deployment couples reported that these prompts helped them surface ``surprising unknowns''. Remini, an AI-powered chatbot, scaffolds mutual reminiscence and reciprocal self-disclosure around shared positive memories using multi-turn, context-aware prompts; by prompting them to elaborate on emotional details and reflect on what memories mean to each other, it increases narrative detail, and participants report higher perceived emotional synchrony and partner responsiveness compared to a minimally guided baseline~\cite{jiang2025remini}. 

Together, these systems suggest a design stance in which the chatbot acts as a lightweight facilitator of deep, reciprocal self-disclosure, helping partners find words, elicit acknowledgment, and time follow-ups to close reciprocity loops, while keeping communication authentic and interpretable. However, prior work largely treats such facilitation as valuable in its own right, without explicitly grounding it in underlying psychological processes such as partner-provided need support. Our work takes this design stance as a starting point and examines how chatbot-mediated conversation can be anchored in these processes.

\subsubsection{Self-determination Theory as a Design Lens}
\label{sec:lens}

While SDT has become a widely adopted framework in HCI, its application remains fragmentary. As noted in recent reviews in HCI, SDT is frequently utilized as a convenient set of post-hoc evaluation scales rather than as a generative lens that structures design decisions~\cite{wenhart2025relatedness, ballou2022self, tyack2020sdt, tyack2024sdt}. In the context of close relationships, this theoretical gap limits our ability to reason about why specific design features succeed in one context but feel pressuring or ambiguous in another.

When we apply a generative SDT lens to prior work, a clear trade-off emerges between intimacy and instrumental support. Existing systems often stem from attempting to elicit relational support (intimacy between partners) without first providing the necessary instrumental support (the psychological conditions for engagement). Specifically, existing designs tend to neglect one or more conditions required for reciprocal care. Technologies focused on affective expressivity succeed in surfacing emotions but often struggle to sustain reciprocity; by emphasizing passive sensing over active co-production, they leave the interaction loop fragile or ambiguous. Conversely, systems facilitating genuine conversation via presence cues foster togetherness but often lack the scaffolding for competence; while they lower the friction of contact, they rarely provide the structural guidance required to navigate emotionally rich topics. Finally, text-enrichment tools aim to enforce depth but often compromise autonomy; by imposing constraints like character counts or forced synchronicity, they risk framing care as a procedural burden rather than a volitional act.

We address these gaps by proposing a dual-layer scaffolding approach. We posit that effective design must function sequentially: \textbf{instrumental support} serves as the foundation, where the system provides \emph{enabling affordances} to support the user's needs for volition (Autonomy), clarity (Competence), and safety (Relatedness). Only once the user feels emotionally safe, capable, and unpressured can the system successfully deploy \emph{mediating affordances} to scaffold \textbf{relational support}, guiding partners to enact the reciprocal care that is the ultimate goal.

We situate this approach within mutual self-disclosure because disclosure is the primary vehicle for intimacy yet remains fragile without reciprocal care. A chatbot is uniquely suited for this scenario because, unlike static prompts or passive signals, it can dynamically regulate the flow, providing the structure of a game with the sensitivity of a moderator. Despite extensive work on relatedness technologies, few systems explicitly operationalize SDT to cultivate this full loop of reciprocal support. This gap motivates our study: examining how chatbot-mediated conversations, when explicitly grounded in SDT, shape both the provision of need support between partners and the psychological outcomes of their interactions. Specifically, we investigate how the chatbot's instrumental scaffolding translates into relational reciprocity through the following research questions:

\textbf{RQ1:} How do chatbot-mediated conversations during positive relational self-disclosure influence partner-provided need support and participants' outcomes in need satisfaction, well-being, and motivation?

\textbf{RQ2:} How do different forms of chatbot mediation shape the dynamics of partner self-disclosure and participants' perceptions of reflective conversation?

\section{Method}

We designed a chatbot that acts as a facilitative mediator, guiding positive relational self-disclosure and need-supportive conversation between romantic partners. The system was evaluated in a comparative study with three experimental conditions.

\subsection{Design Rationale}

In our design, we operationalize Self-Determination Theory (SDT) using a dual-layer scaffolding framework to facilitate the internalization of self-disclosure. We utilize prompts adapted from the ``36 Questions'' paradigm~\cite{jones2015thirtysix} as the core activity, supported by two distinct layers:

\begin{itemize} \item \textbf{Layer 1: Enabling Affordances (Instrumental Support).} To lower barriers to vulnerability, the chatbot wraps questions in direct support---providing rationales, choices, and a warm, non-judgmental tone---to create a psychologically safe environment for the act of disclosure. \item \textbf{Layer 2: Mediating Affordances (Relational Support).} To scaffold reciprocity, the chatbot uses follow-up reflection prompts that explicitly guide partners to interpret disclosures and provide autonomy, competence, and relatedness support to one another for the partner's disclosed experiences. \end{itemize}

We operationalized this stance through five specific strategies (detailed below). Appendix~\ref{app:snippet} presents example chat snippets drawn from our experiment. We implemented the chatbot using large language models. To ensure experimental consistency, the three core self-disclosure questions and the reflection logic were specified as fixed system prompts. However, the exact wording of the chatbot's direct responses was generated dynamically at runtime to maintain a natural, need-supportive tone (Layer 1). The underlying prompts were iteratively tuned through internal testing and early pilots to ensure they consistently achieved a clear, non-pressuring tone and SDT-consistent behavior. The dialogue examples referenced throughout this paper are anonymized excerpts from actual study interactions, with labels (e.g., G1, G2) corresponding to the group IDs used in our study.

\subsubsection{\textbf{Targeting Disclosure for Specific Needs}}

To move conversations beyond routine logistics, we adapted three questions from the ``36 Questions'' paradigm~\cite{arthur1997the, altman1973social}, selecting one from each depth level to align with specific SDT needs (Table~\ref{tab:questions}). Unlike prior work that uses these questions generically~\cite{zheng2021pocketbot, zhang2023icebreaking}, we selected prompts specifically to elicit disclosures that would prime the listener to provide distinct forms of support:

\begin{itemize} \item \textbf{Autonomy (Volition):} The first question (``Perfect Day'') invites participants to describe a scenario of pure choice, unconstrained by obligation. This targets \textbf{Autonomy} because it reveals the user's intrinsic values and ideal self, creating an opportunity for the partner to validate their volition rather than just their schedule. \item \textbf{Competence (Efficacy):} The second question (``Untouched Dream'') focuses on aspirations and the barriers hindering them. This targets \textbf{Competence} by surfacing gaps between current ability and desired goals, inviting the partner to provide encouragement, structure, or affirmation of capability. \item \textbf{Relatedness (Connection):} The third question (``What people miss'') targets \textbf{Relatedness} by exposing a vulnerable, unseen aspect of the self. This acts as a direct bid for emotional verification, inviting the partner to offer empathy and deep acceptance. \end{itemize}

\begin{table*}[t]
\centering
\caption{Self-disclosure questions adapted from the 36 Questions, each targeting one psychological need at a different depth level.}
\label{tab:questions}
\footnotesize
\renewcommand{\arraystretch}{1.1}
\setlength{\tabcolsep}{6pt}
\begin{tabular}{p{0.70\linewidth}|c|c}
\toprule
\textbf{Question} & \textbf{Targeted Need} & \textbf{Depth Level (36Q)} \\
\midrule
What's a day you remember as close to perfect? & Autonomy & Level 1 \\
\midrule
What's that one dream you've had forever but still haven't touched? What's holding you back? & Competence & Level 2 \\
\midrule
What's something meaningful about you that people often miss, but you wish they understood? & Relatedness & Level 3 \\
\bottomrule
\end{tabular}
\end{table*}

\begin{figure*}[t]
  \centering
  \includegraphics[width=\linewidth]{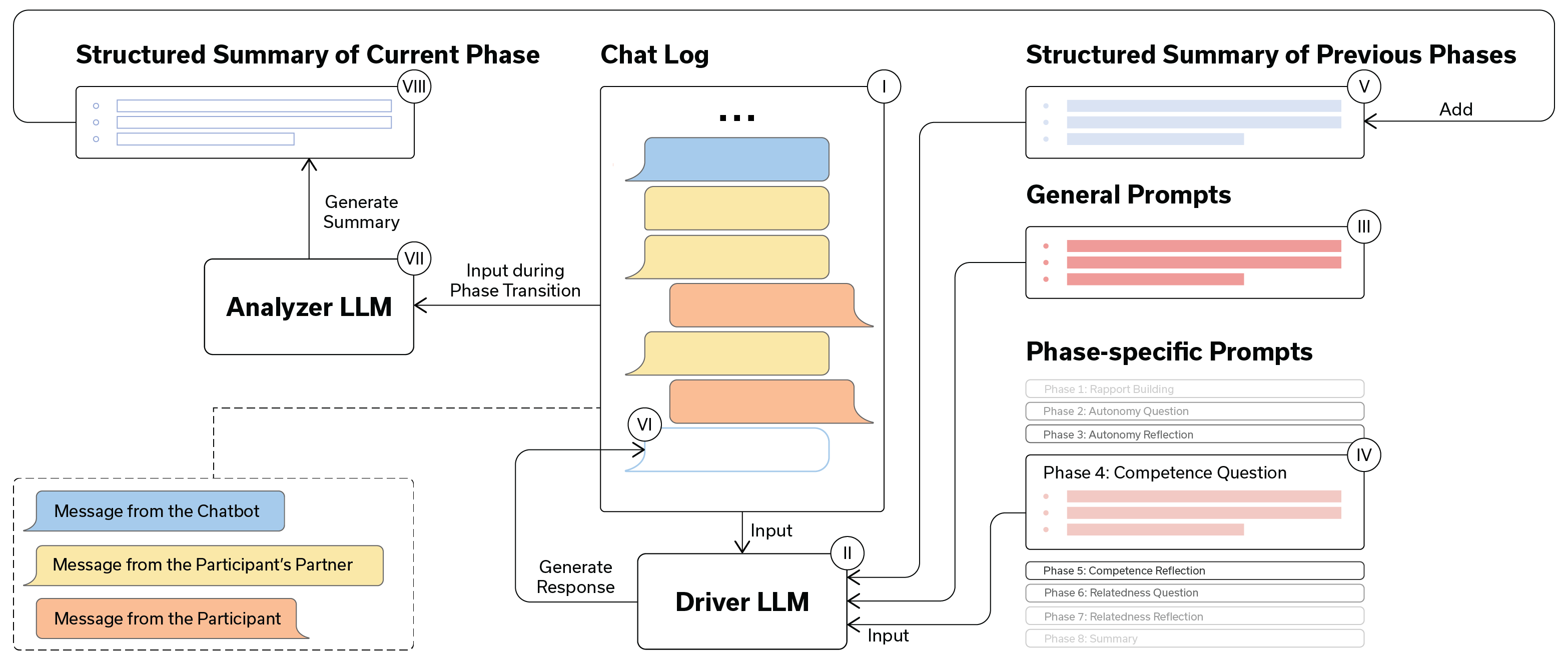}
  \caption{Example workflow of the chatbot's conversation system in Phase 4 (Competence Question). (1) After being called in the Telegram group chat, the Driver LLM (II) receives input consisting of the current chat log (I), general prompts (III), phase-specific prompts for the active phase (IV), and a structured summary of prior interactions (V). (2) Based on these inputs, the Driver LLM generates a response (VI) and delivers it back to the group chat. (3) When the phase transitions, the Analyzer LLM (VII) processes the completed chat log and produces a structured summary (VIII), which is added to the conversation context for subsequent phases.}
  \label{fig:system}
  \Description{A workflow diagram of the chatbot conversation system during Phase 4, labeled Competence Question. In the center, a chat log shows alternating messages: blue from the chatbot, yellow from the participant's partner, and orange from the participant. The Driver LLM receives four inputs: the current chat log, general prompts defining tone and role, prompts specific to the active phase, and a structured summary of previous phases. It generates a response, which appears in the chat log. At phase transitions, the Analyzer LLM processes the completed chat log to create a structured summary of the current phase. This summary is then added to the collection of prior summaries, ensuring continuity and context across phases.}
\end{figure*}

\subsubsection{\textbf{Providing Enabling Affordances.}}

To create the necessary safety for disclosure, the chatbot provided direct need support through its own responses (Layer 1). By acknowledging feelings, offering rationales, and validating disclosures, it modeled the supportive communication it aimed to elicit, establishing a climate where users felt safe and heard. Prior work confirms that chatbots can effectively support autonomy, competence, and relatedness in individual interactions~\cite{cai2023listen, zhu2025benefits}. Our design extends this capacity to the dyadic level, cultivating a holistic need-supportive environment for couples.

\paragraph{Autonomy (Volition).} Specifically, the chatbot validated participants’ perspectives, offered options, and framed effort as voluntary, following examples of autonomy-supportive behaviors~\cite{slemp2024interpersonal}. Within each activity, it provided meaningful choices. For example, asking if users \emph{want a silly chatbot joke as a breather, or should we keep rolling? totally your call!''} (G1). And it reminded them there were no right or wrong answers. It also offered rationales for its prompts to foster internalization, helping participants see disclosure as purposeful rather than arbitrary~\cite{ryan2000selfdetermination}. For instance, it explained that \emph{naming [dreams] out loud is how you start building the ladder up''} (G1). Finally, to mitigate the risk of intrusiveness in polyadic settings (where a chatbot participates in a conversation with multiple human partners rather than a single user)~\cite{toxtli2018understanding, zheng2022ux}, the chatbot reinforced autonomy by advancing only when explicitly called upon, giving partners full control over pacing.

\paragraph{Competence (Clarity).}

Drawing on principles of competence support~\cite{slemp2024interpersonal}, the chatbot provided enabling affordances to foster a sense of efficacy and reduce ambiguity. The system clarified goals of each self-disclosure question and provided structure through step-by-step instructions, ensuring that deep disclosure felt attainable rather than overwhelming. Furthermore, the chatbot delivered informational feedback—warm, non-judgmental affirmations that acknowledged effort and reflected specific details to signal understanding. For instance, it validated a user's articulation of complex feelings: \emph{``wow... honestly, these are so vivid... I totally get missing performing arts... the environment just kinda put a mental block there''} (G2). This feedback reinforced competence by affirming the participant's capacity to articulate meaningful experiences, thus lowering the barrier to engagement.

\paragraph{Relatedness (Safety).}

To establish psychological safety, the chatbot prioritized rapport which is the sense of harmony and synchrony in conversation~\cite{gratch2007creating}. Aligning with prior work showing that informal dialogue and social praise enhance perceived warmth and disclosure~\cite{papneja2025self, xu2022enhancing, lee2017enhancing}, the chatbot greeted participants playfully (e.g., \emph{``classic `just woke up' vibes''} [G1]) to reduce social distance. Furthermore, it provided an enabling affordance by modeling reciprocal openness through light self-disclosures, such as joking about \emph{``binge-reading weird internet forums.''} These behaviors positioned the chatbot as a friendly guide rather than a clinical observer, reducing the social risk of vulnerability and encouraging participants to share more openly~\cite{lee2020ihearyou}. Throughout, the chatbot conveyed interest, care, and unconditional positive regard, consistent with relatedness-supportive behaviors in SDT~\cite{slemp2024interpersonal}.

\subsubsection{\textbf{Deploying Mediating Affordances}}

While enabling affordances create the necessary safety for disclosure (Layer 1), the ultimate goal of the system is to foster relational support between partners (Layer 2). To achieve this, the chatbot deploys mediating affordances in the form of structured follow-up reflection prompts. These prompts act as a responsiveness scaffold, explicitly cuing the listening partner to interpret the disclosure through the lens of specific psychological needs rather than simply acknowledging it.

After each main disclosure, the chatbot provided a fixed reflection prompt tailored to the targeted need: \begin{itemize} \item \textbf{Scaffolding Autonomy Support:} To validate volition, the prompt invited the listener to recognize their partner's intrinsic values: \emph{When you hear about their perfect day, what do you think it really says about what lights them up? And hey... is there something you could do to help them live that vibe more often?''} \item \textbf{Scaffolding Competence Support:} To affirm efficacy, it encouraged the partner to identify barriers and offer concrete encouragement: \emph{What do you think this dream says about what really matters to your partner? Is there something small you could do---or say---to help them feel more confident or supported in working toward it?''} \item \textbf{Scaffolding Relatedness Support:} To foster emotional attunement, it asked partners to articulate their sense of connection: \emph{``Did anything your partner shared make you feel closer, or help you get them on a new level?''} \end{itemize}

This design operationalizes the shift in the locus of support. By explicitly guiding partners to enact validation, care, and understanding, the system closes the reciprocity loop, ensuring that intimacy develops through mutual responsiveness rather than just individual disclosure~\cite{reis1988intimacy, laurenceau1998intimacy}.

\subsubsection{\textbf{Human-Centered Ethical Design.}}

Responding to calls for urgent need to embed ethical frameworks and Human-Centered AI~\cite{wenhart2025relatedness, Schmager16022025} , we designed the chatbot as a `cognitive orthosis' that augments rather than replaces human capabilities. It supports couples in the demanding social tasks of reflective self-disclosure and need-supportive responding.

This stance also responds to research on Human–Chatbot Relationships, which shows that people may form meaningful emotional bonds with chatbots, from friendships to familial ties~\cite{pentina2023exploring}. While such relationships can provide comfort, they also risk fostering dysfunctional attachment and undermining offline relationships~\cite{laestadius2024toohuman}. Against this backdrop, we position the chatbot not as a surrogate partner, but as a facilitator of human–human conversation. Concretely, the chatbot never initiated contact on its own, did not offer prescriptive relationship advice (for example, telling participants what decisions to make), and repeatedly redirected questions about the relationship back to the partners. In line with SDT and this ethical orientation, all prompts were phrased as invitations rather than demands (for example, ``if you would like to share''), participants could always skip, rephrase, or stop a question, and the model was instructed to avoid pressuring users to disclose more than they wanted. We also constrained the content scope to positive relational memories and everyday hopes rather than conflict or trauma, in order to reduce the risk of triggering or pathologizing responses. Within these safeguards, the chatbot adopted a non-judgmental, empathetic tone and remained impartial when partners described differences, focusing on helping each partner articulate and acknowledge the other's perspective rather than taking sides.

\subsubsection{\textbf{Scaffold conversation through phased sequencing.}}
\label{sec:designrationale}

Finally, leveraging prior work on staged LLM tasks~\cite{kim2024mindfuldiary, seo2024chacha, jiang2025remini}, we designed the conversation as an eight-phase state machine~\cite{winograd1986alanguage}. The session progresses from rapport building through alternating cycles of self-disclosure and partner reflection targeting autonomy, competence, and relatedness, concluding with joint gratitude. This sequencing gradually guides couples toward deeper mutual reflection while preserving participant control over pacing, creating a safe conversational arc for vulnerable exchanges.

\begin{figure*}[t]
  \centering
  \includegraphics[width=\textwidth]{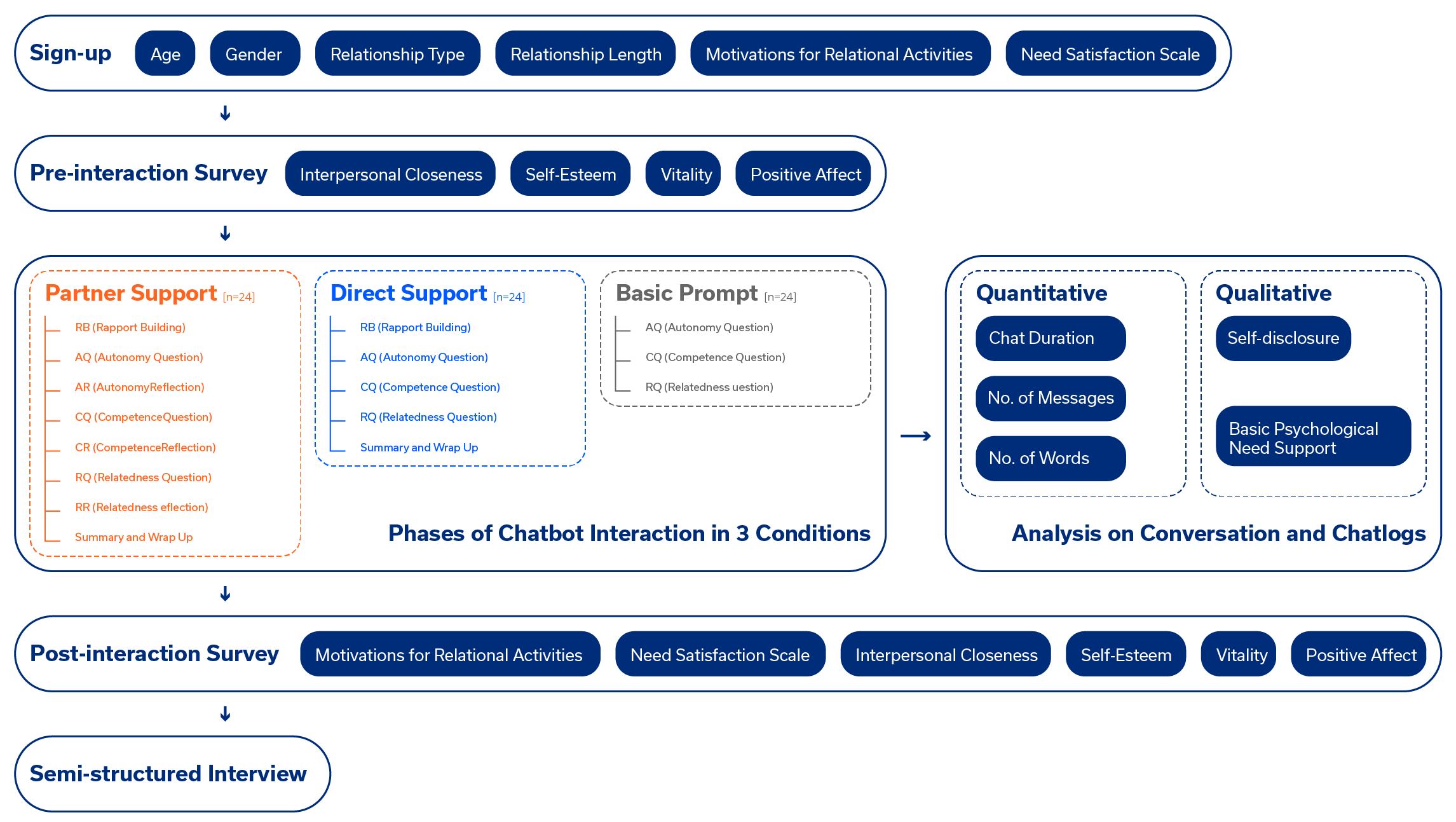}
  \caption{Experiment flow. Participants first completed a sign-up form and a pre-interaction survey assessing demographic information, interpersonal closeness, self-esteem, vitality, positive affect, motivations for relational activities, and need satisfaction. Dyads were then randomly assigned to one of three experimental conditions, \PSbox{}, \DSbox{}, or \BPbox{}, and engaged in a chatbot-mediated conversation following the condition-specific structure. Conversation data were subsequently analyzed using both quantitative measures (chat duration, number of messages, number of words) and qualitative coding (self-disclosure, basic psychological need support). Finally, participants completed a post-interaction survey and a semi-structured interview.}
  \label{fig:expflow}
  \Description{A flow diagram showing the experiment procedure from sign-up to interview. At the top, participants provide demographic details and complete a pre-interaction survey measuring interpersonal closeness, self-esteem, vitality, positive affect, motivations for relational activities, and need satisfaction. Next, dyads are randomly assigned to one of three chatbot conditions. In the Partner Support condition, participants experience all eight phases including rapport building, autonomy, competence, and relatedness questions with reflection prompts. In the Direct Support condition, participants follow the same phases but without partner-reflection prompts. In the Basic Prompt condition, only autonomy, competence, and relatedness questions are asked without scaffolding. Conversation data are then analyzed quantitatively, using chat duration, number of messages, and number of words, and qualitatively, using coding of self-disclosure and basic psychological need support. Finally, participants complete a post-interaction survey with the same measures as before and a semi-structured interview.}
\end{figure*}

\subsection{Implementation}

\subsubsection{System Deployment}

To maximize accessibility and support flexible, user-driven communication patterns~\cite{wenhart2025relatedness, dziabiola2022qude, Li2023systematic}, we deployed the chatbot on \emph{Telegram}. Embedding the system within a widely used messaging platform allowed it to blend naturally into participants' everyday routines, minimizing the friction often associated with standalone prototypes~\cite{podlubny2017synchronous, kim2019love, yarosh2013almost}. The chatbot was built in \emph{Python} using Telegram's Bot API. We utilized \emph{GPT-4.1} (\texttt{gpt-4.1-2025-04-14}) via the OpenAI API, selected for its extended context window and reduced latency which are critical factors for maintaining the seamless flow required.

\subsubsection{Architecture of the Chatbot}

We implemented the chatbot using two coordinated modules: the \emph{Driver LLM} and the \emph{Analyzer LLM} (Figure~\ref{fig:system}). To ensure reliability and prevent hallucinations, we utilized a phase-specific prompting strategy rather than a monolithic instruction~\cite{perez2021truefewshot, brown2020language}, dividing the interaction into eight structured phases to improve output quality~\cite{reynolds2021prompt}. The chatbot was activated via command, with participants identified by anonymized IDs.

\paragraph{Driver LLM.} The Driver manages real-time interaction using four structured inputs: (1) \emph{General prompts} defining the role and tone; (2) \emph{Phase-specific prompts} tailored to the active task; (3) \emph{Current conversation history}; and (4) a \emph{Conversation summary} from the Analyzer. Based on these, it generates responses or instructions. To preserve autonomy, the system re-prompts only when necessary and follows participants' decisions to skip questions. Transitions (``moving on'') are triggered once both partners complete the phase. Full prompt details are in Appendix~\ref{app:prompt}.

\paragraph{Analyzer LLM.} The Analyzer maintains continuity by generating a summary at the end of each phase. Drawing on the conversation history and phase objectives, it produces a concise account of key disclosures and responses. These summaries are fed back to the Driver, creating a feedback loop that ensures subsequent prompts remain context-aware and responsive to partner dynamics.

\subsection{Chatbot Phases across Conditions}
\label{sec:conditions}

To isolate the effects of the dual-layer scaffolding, we designed three experimental conditions that progressively layer the support mechanisms:

\textbf{Partner Support (\PSbox): Full Dual-Layer Scaffolding.} Participants in this condition received the complete intervention, comprising both Enabling Affordances (Layer 1) and Mediating Affordances (Layer 2). The interaction followed an eight-phase structure: rapport building, three self-disclosure questions (autonomy question-AQ, competence question-CQ, relatedness question-RQ), three partner-reflection follow-ups (autonomy reflection-AR, competence reflection-CR, relatedness reflection-RR), and a summary phase. This design tests the combined effect of the chatbot's instrumental support and the partner's reciprocal support.

\textbf{Direct Support (\DSbox): Enabling Affordances Only.} Participants received the same rapport building, three disclosure questions, and supportive chatbot responses as in \PS{}, but without the partner-reflection follow-ups (five phases). This condition isolates Layer 1: it provides the chatbot's instrumental support (safety, structure, warmth) but removes the scaffold for reciprocal partner support.

\textbf{Basic Prompt (\BPbox): Baseline (No Scaffolding).} The chatbot acted as a minimal facilitator, delivering only the three core disclosure questions when prompted (three phases). It provided neither rationales nor supportive feedback (No Layer 1), nor did it prompt partner reflection (No Layer 2). This condition serves as a baseline to control for the content of the "36 Questions" activity itself.

The key difference between \PS{} and \DS{} is the locus of support. While \DS{} relies on the agent to provide validation (Agent-to-Human), the follow-up prompts in \PS{} are designed to deploy mediating affordances, shifting the locus of support from the chatbot to the partner (Human-to-Human). Theoretically, this mechanism serves as a responsiveness scaffold, explicitly cueing the partner to enact the autonomy, competence, and relatedness support that the chatbot modeled in the previous turn.

\subsection{Design Validation and Pilot Testing.} To validate the necessity of our dual-layer scaffolding, we conducted a pilot with 6 dyads (3 \PS{}, 3 \DS{}). Observations revealed that while the enabling affordances (disclosure questions) successfully elicited sharing, they rarely triggered spontaneous relational support; without the explicit mediating affordances of follow-up prompts, partners frequently missed opportunities for validation. This confirmed that the `Reflection' phases in \PS{} are not merely additive but are the critical mechanism required to shift the locus of support from the agent to the partner.

\section{User Study}

To evaluate the three chatbot variants, we conducted a user study. In the following, we describe the setup, procedure, and measures used to assess participants' experiences. All procedures were approved by our institutional ethics review board (approval details omitted for anonymous review).

\subsection{Participants}
We recruited couples (dating, married, or friends with romantic interest) via campus channels, targeting relationships with high reciprocal potential~\cite{laguardia2008selfdetermination}.

Eligibility criteria required that individuals: (a) be part of a couple (dating, friends with romantic interest, engaged, or married); (b) be willing to share personal thoughts with their partner; (c) speak and write in English; (d) own a smartphone with Telegram; and (e) be at least 18 years old.

72 participants met these criteria and enrolled (see demographics in Table~\ref{tab:demographics}, more details in Appendix~\ref{app:demographic}). Dyads were subsequently randomized to one of three conditions. All participants received compensation according to local IRB guidelines.

{\small
\begin{table}[h]
\centering
\caption{Demographic characteristics of participants by condition. FRI = Friends with Romantic Interest.}
\label{tab:demographics}
\begin{tabular}{lcccc}
\toprule
\textbf{Characteristic} & \makecell{Partner \\ Support} & \makecell{Direct \\ Support} & \makecell{Basic \\ Prompt} & \textbf{Total} \\
\midrule
$n$ & 24 & 24 & 24 & 72 \\
Age (M ± SD) & 23.8 ± 2.8 & 24.8 ± 3.6 & 26.3 ± 5.5 & 24.9 ± 4.3 \\
Gender (F/M) & 13/11 & 13/11 & 12/12 & 38/34 \\
Length (yrs), Md (IQR) & 2.5(2.5) & 2(2.5) & 2.25(2) & 2(2.75) \\
Type & & & & \\
\quad Dating & 22 & 18 & 20 & 60 \\
\quad Engaged & 0 & 2 & 0 & 2\\
\quad Married & 0 & 2 & 4 & 6\\
\quad FRI & 2 & 2 & 0 & 4\\
\bottomrule
\end{tabular}
\end{table}
}

\subsection{Procedure}

Participants joined the study remotely from separate locations to prevent  off-channel communication. After receiving explanations of procedures, rights (e.g., withdrawal), and anonymization, they provided informed consent. Following a baseline survey, dyads entered a Telegram group with the chatbot. To minimize demand effects, the experimenter left before the conversation began, instructing participants that they could skip questions or advance prompts at will.

\begin{figure}
  \centering
  \includegraphics[width=\linewidth]{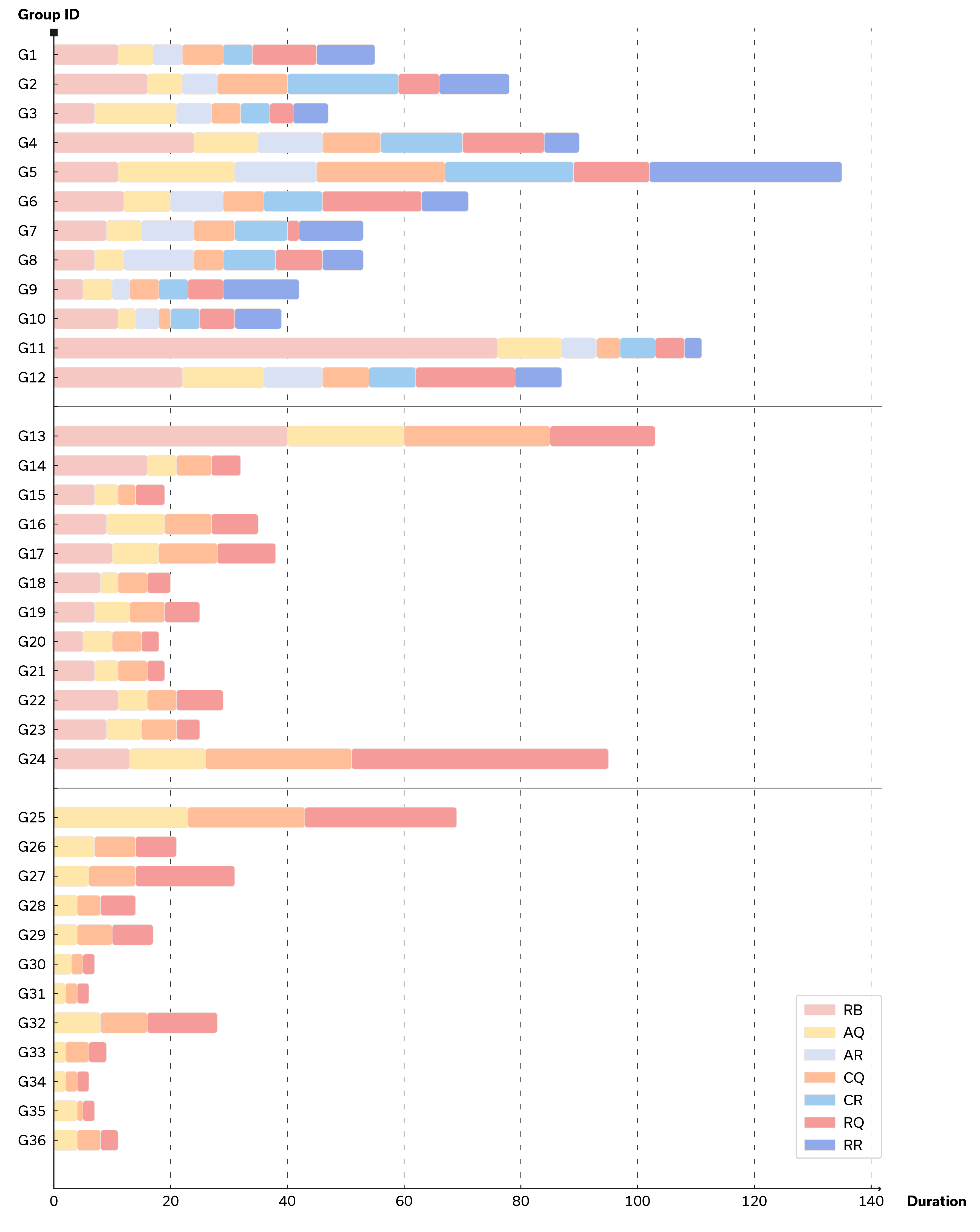}
  \caption{Per-group phase timelines across the three conditions (minutes). Each bar (G1–G36) shows the duration of conversational phases. Partner Support (\PSbox, G1–G12) includes all seven phases; Direct Support (\DSbox, G13–G24) includes rapport and selected questions; Baseline (\BPbox, G25–G36) includes only the question phases.}
  \label{fig:timeline}
  \Description{A timeline chart showing conversation phases for 36 dyads, labeled G1 through G36 on the vertical axis and duration in minutes on the horizontal axis. Each horizontal bar is divided into colored segments that mark the time spent in different chatbot-guided phases. Groups G1 to G12 represent the Partner Support condition, which includes all seven phases: rapport building, autonomy question, autonomy reflection, competence question, competence reflection, relatedness question, and relatedness reflection. Groups G13 to G24 represent the Direct Support condition, which includes rapport building plus autonomy, competence, and relatedness questions but without reflection phases. Groups G25 to G36 represent the Basic Prompt condition, which includes only the autonomy, competence, and relatedness question phases. Segment length indicates the actual duration of each phase, showing variability both across conditions and between groups.}
\end{figure}

The study concluded with post-interaction surveys, followed by a short open-ended interview with each dyad. The post-intervention surveys allow for a comparison with the pre-intervention ones. Session durations varied widely across conditions and were right-skewed, with some dyads engaging much longer than scheduled. The \PS{} condition yielded the longest sessions ($M=107.7$ mins), followed by \DS{} ($M=73.2$ mins)and \BP{} ($M=42.3$ mins) as seen in Table~\ref{tab:durations} with details in Figure~\ref{fig:timeline}. For clarity, we report both median and mean durations, along with interquartile ranges and standard deviations. A visual overview of the procedure is provided in Figure~\ref{fig:expflow}.

\begin{table}[b]
\centering
\caption{Session durations by condition (in minutes).}
\label{tab:durations}
\begin{tabular}{lcccc}
\toprule
\textbf{Condition} & \textbf{Median} & \textbf{IQR} & \textbf{Mean} & \textbf{SD} \\
\midrule
Partner Support (\PS) & 103.5 & 36.5 & 107.7 & 27.8 \\
Direct Support (\DS) & 60.0 & 18.5 & 73.2 & 36.1 \\
Basic Prompt (\BP) & 35.0 & 20.0 & 42.3 & 18.7 \\
\bottomrule
\end{tabular}
\end{table}

\subsection{Independent Variables}
\label{sec:ivs}
This study uses a mixed-design with two independent variables: (1) the Chatbot \textbf{Condition} (between subjects, described in Section~\ref{sec:conditions}) with three levels \{Partner Support, Direct Support, Basic Prompt\} and \textbf{Phase} (within subjects) with two levels \{Pre-interaction, Post-interaction\}.

\subsection{Dependent Variables}

We evaluated both conversational traces and self-reported outcomes to capture how chatbot-mediated interactions shaped engagement, disclosure, need support, and well-being.

\subsubsection{Dependent variables extracted from chat logs.}

This subsection reports measures derived directly from the conversations.

\paragraph{Quantitative.}

All conversations were recorded with participant consent, and we extracted the following measures:

\textbf{Chat duration}: total time spent in the conversation.
\textbf{Word counts},  \textbf{message counts}, and  \textbf{words per message}: computed using LIWC22~\cite{boyd2022thedevelopment, yla2010thepsychological}. These are established metrics for evaluating user engagement in chatbot interactions~\cite{hill2015real, Li2023systematic, lee2020ihearyou, kim2024mindfuldiary}. Longer narratives and more frequent exchanges are interpreted as signs of greater engagement, which was a central design goal of our system.

\paragraph{Qualitative.} 
We conducted a qualitative analysis of the chat logs with ATLAS.ti~\cite{AtlasTi2023}. Two independent coders (with three and five years of prior experience) applied the coding rubrics. The analysis focused on two dimensions:

\textbf{Self-disclosure (SD).} To capture depth and type of disclosure, we adapted rubrics from prior online forum research~\cite{barak2007degree}. Disclosures were coded into informational, thoughts, and feelings, with full rubric details in Appendix~\ref{app:sd}. Conversational phases were used as the unit of analysis, except for the final Summary phase, which largely consisted of brief farewells and was excluded. Inter-rater agreement was strong across categories (Cohen's $\kappa \geq 0.8$), exceeding the .70 benchmark~\cite{viera2005understanding, mchugh2012interrater}.

\textbf{Enacted Partner Support.} To assess how partners actually expressed support for one another's psychological needs in conversation, we developed a rubric grounded in self-determination theory~\cite{slemp2024interpersonal, ahmadi2023classification}. Existing schemes were either context-specific (e.g., classrooms) or relied on frequency counts that are sensitive to conversation length. We instead adopted a depth-based approach, parallel to the SD coding with full details provided in Appendix~\ref{app:ns}. Within each conversational phase, coders identified autonomy-, competence-, or relatedness-supportive behaviors, or coded the absence of support (level 0). Examples behaviors of each category are listed in Appendix~\ref{app:core_behaviors_full}. This allowed us to capture quality in conversational style while keeping dyads comparable. Inter-rater reliability was high (Cohen's $\kappa \geq 0.8$), meeting recommended thresholds.

\subsubsection{Dependent variables measured by surveys.}

Guided by Self-Determination Theory, which links motivational orientations, need satisfaction, and supportive relational processes to relationship quality and individual well-being, we assessed participants' motivations for relational activities, need satisfaction, relational closeness, self-esteem, vitality, and positive affect~\cite{laguardia2008selfdetermination}. Full details of the scales used in the experiment are provided in Appendix~\ref{app:questionnaire}.

\paragraph{\textbf{Self-Disclosure Motivations in Relationships}}
We adapted and combined the self-disclosure of feelings and self-disclosure of thoughts subscales from the Motivations for Relational Activities Scale (MRA)~\cite{gaine2009theunique}, given their strong conceptual alignment and overlap. The MRA has been shown to predict relationship well-being and daily relational functioning in marriage~\cite{gaine2011implications}. The measure assesses four types of motivation: intrinsic, identified, introjected, and external. Intrinsic and identified items were averaged to form an autonomous motivation score, while introjected and external items were averaged to form a controlled motivation score. Participants rated twelve items (e.g., ``\emph{I shared because it helped me understand myself or see things more clearly}'') on a 7-point Likert scale (1 = not at all true; 7 = very true). Each construct was measured using six items, and final scores were the average of those items. Internal consistency was high for autonomous motivation at baseline ($\alpha$ = .87) and post-intervention ($\alpha$ = .88), and for controlled motivation at baseline ($\alpha$ = .84) and post-intervention ($\alpha$ = .82). Scores ranged from 1 to 7, with higher values reflecting stronger endorsement of the given motivational orientation where higher autonomous motivation and lower controlled motivation are considered more desirable.

\paragraph{\textbf{Perceived Need Satisfaction}}  
Relational need satisfaction was measured using the Need Satisfaction Scale~\cite{laguardia2000within}, which assesses the fulfilment of autonomy, competence, and relatedness. This scale has been widely applied in studies of couple relationships~\cite{baker2020therole, kluwer2020autonomy}. Participants rated nine items (e.g., ``\emph{I felt like a competent person}'') on a 7-point Likert scale (1 = strongly disagree; 7 = strongly agree). Each need was assessed by three items, and the overall score was the average across all nine items. Internal consistency was high at baseline ($\alpha$ = .89) and post-intervention ($\alpha$ = .83). Scores ranged from 1 to 7, with higher values indicating greater need satisfaction.

\paragraph{\textbf{Inclusion of Other in the Self}}  
Closeness was measured using the Inclusion of Other in the Self (IOS) scale~\cite{aron1992inclusion}, a single-item pictorial measure widely used in computer-mediated communication research~\cite{an2024emowear, zhang2023icebreaking}. The IOS presents seven Venn diagram-like images depicting two circles (self and partner) with varying degrees of overlap (1–7). Higher scores indicate greater perceived closeness.

\paragraph{\textbf{Self-Esteem}}  
Self-esteem was assessed using the Rosenberg Self-Esteem Scale~\cite{rosenberg1965rosenberg}, a widely used measure of self-worth and self-respect. The scale has been broadly applied in HCI research to track well-being outcomes associated with technology use and interaction~\cite{zhu2025benefits, pereira2017measuring, narain2020promoting}. To reduce participant burden, we used the validated short form~\cite{monteiro2022theefficient}, in which participants rated five items (e.g., ``\emph{I feel I do have much to be proud of}'') on a 4-point Likert scale (1 = strongly disagree; 4 = strongly agree). Internal consistency was acceptable at baseline ($\alpha$ = .73) and post-intervention ($\alpha$ = .78). Scores ranged from 5 to 20, with higher scores indicating greater self-esteem.

\paragraph{\textbf{Vitality}}  
Subjective vitality was measured using the 6-item version of the Subjective Vitality Scale~\cite{Bostic2000Avalidation}, which captures the state of feeling alive and energized as an aspect of psychological well-being. The scale has also been applied in HCI research to evaluate user experiences in interactive settings~\cite{tyack2021thesmall, tyack2020restorative}. Participants rated six items (e.g., ``\emph{At this moment, I feel alive and vital}'') on a 7-point Likert scale (1 = not at all true; 7 = very true). Internal consistency was high at baseline ($\alpha$ = .88) and post-intervention ($\alpha$ = .91). Scores ranged from 6 to 42, with higher scores indicating greater subjective vitality.

\paragraph{\textbf{Positive Affect}}  
Positive affect was measured using the Positive Affect subscale of the Positive and Negative Affect Schedule (PANAS)~\cite{watson1988development}, a widely used instrument in technology-based well-being interventions~\cite{wagener2022mood, alarith2022time, wagener2023selvreflect, schlagowski2023wish}. To reduce participant burden, we used the validated short form~\cite{Mackinnon1999Ashortform}, in which participants rated five items (e.g., ``\emph{Inspired},'' ``\emph{Excited}'') on a 5-point Likert scale (1 = very slightly or not at all; 5 = extremely). Internal consistency was high at baseline ($\alpha$ = .87) and post-intervention ($\alpha$ = .88). Scores ranged from 5 to 25, with higher scores indicating greater positive affect.

\paragraph{\textbf{Perceived Need Support from the Chatbot}}  
To evaluate the efficacy of the system's enabling affordances, perceived need support from the chatbot was measured using an adapted version of the Need Satisfaction Scale~\cite{laguardia2000within}. The scale was tailored to assess how participants experienced instrumental support directly from the chatbot (Layer 1). To reduce participant fatigue, each need was assessed with two items (six items total) instead of the original three per need. An example item is: ``\emph{I felt a sense of closeness or emotional support from the chatbot's responses}.'' Items were rated on a 7-point Likert scale (1 = strongly disagree; 7 = strongly agree). For each need, scores were calculated by averaging the two items, with higher values indicating stronger perceived support. The scale demonstrated high internal consistency ($\alpha$ = .83).

\subsection{Interview}

After the conversation task, we conducted semi-structured interviews to capture participants' subjective impressions of the chatbot-mediated experience. Interview questions invited reflection on overall impressions, the usefulness of the main prompts, and the impact of specific scaffolding mechanisms such as follow-up questions, guidance on how to respond, and rationales. We also probed perceptions of the chatbot's presence, whether it changed how participants talked to each other, and whether it created additional opportunities for partner support compared to their usual conversations. Finally, participants were asked about the potential for future use of such a chatbot in their relationship, including contexts where it might feel meaningful or relevant. All interviews were audio-recorded, anonymized, transcribed verbatim, and subjected to thematic analysis as described by Braun and Clarke~\cite{virginia2006using}. Interviews tended to be longer in \PS{} ($M = 14.6$ min) and \DS{} ($M = 14.8$ min) than in \BP{} ($M = 8.1$ min).

\subsection{Statistical Analysis}

We analyzed the quantitative data using statistical methods appropriate to the distributional properties of each outcome and the inclusion of covariates. For \textbf{motivation} and \textbf{need satisfaction}, we fitted \textbf{random-intercept linear mixed-effects models (LMM)} using the \texttt{lme4} package in R~\cite{JSSv067i01}. These outcomes had conceptually relevant baseline measures (pre-intervention motivation and need satisfaction scores) and demographic covariates (age, gender, relationship length and type) that we controlled for to isolate the effects of \textbf{Condition} (between-subjects: \PS{}, \DS{}, \BP{}) and \textbf{Phase} (within-subjects: pre vs. post). The random-intercept structure accounted for the repeated-measures nature of the data, with each participant contributing two observations.

For the remaining DVs (IOS, self-esteem, vitality, and positive affect), we did not include baseline measures as covariates because they capture momentary states rather than stable individual differences. 
We instead used applied \textbf{nonparametric Aligned Rank Transform (ART) ANOVA} via the \texttt{ARTool} package in R~\cite{wobbrock2011thealigned} as it allows testing interactions without assuming normality of residuals.

For both LMM and ART ANOVA, we conducted post-hoc pairwise comparisons using the \texttt{emmeans} package (EMMs) in R~\cite{searle1980population}, applying either Holm  for (``Time within Condition'') or Tukey adjustment (``Condition within Time'').

We conducted between-subjects comparisons using \textbf{Kruskal Wallis} rank-sum tests~\cite{kruskal1952william}, implemented with the \texttt{rstatix} package in R. When significant effects emerged, Dunn's post-hoc tests with Holm correction were applied. This procedure was used consistently across analyses, including participants' perceived need support from the chatbot, conversation engagement metrics, and differences in self-disclosure depth and need support levels.

To examine differences in depth of self-disclosure and level of need support between question and reflection phases within the Partner Support condition, we conducted \textbf{Wilcoxon signed-rank} tests.

\subsection{Power Analysis}
Given the sample size of $N=72$ (36 dyads) divided across three conditions, we conducted a post-hoc sensitivity analysis using G*Power 3.1~\cite{faul2009statistical}. With an $\alpha$ of .05, and power of .80, our design was sensitive to effects of size $f = 0.37$ (large effects, equivalent to a $\eta_p^2$ of .12). This suggests that while the study was sufficiently powered to detect the robust behavioral changes observed in the chat logs, the statistical power was limited for detecting smaller interaction effects in self-reported well-being measures. Therefore, we report effect sizes ($\eta_p^2$) alongside significance tests to facilitate cautious interpretation of these findings.

\begin{figure}[b]
  \centering
  \includegraphics[width=\linewidth]{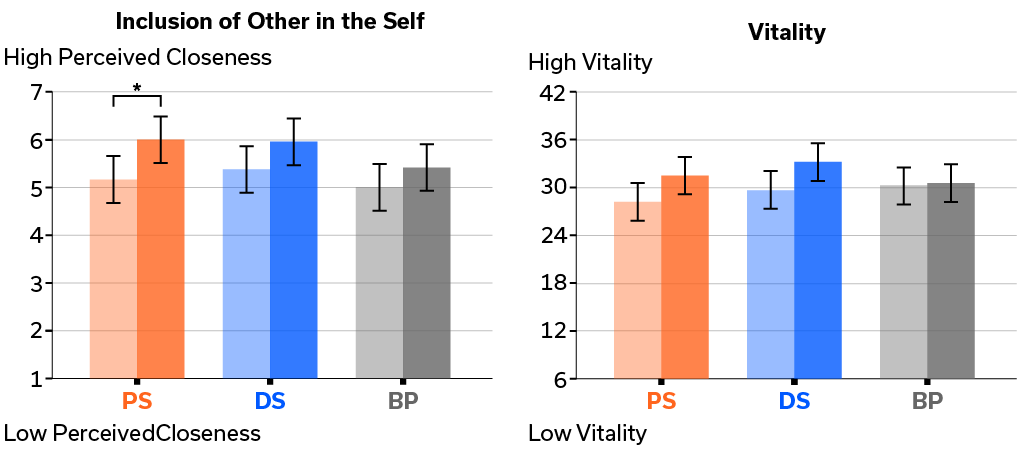}

  \caption{Estimated marginal means of interpersonal closeness (IOS), self-esteem, vitality, and positive affect before and after the interaction. Error bars represent 95\% Confidence Intervals..}
  \label{fig:wellbeing}
  \Description{Two bar charts compare wellbeing outcomes across three chatbot conditions: Partner Support (PS), Direct Support (DS), and Basic Prompt (BP). The vertical axis of both charts ranges from 0 to 100. The left chart shows perceived closeness, measured by the Inclusion of Other in the Self scale. Bars indicate that participants in the Partner Support condition reported higher closeness than those in the Direct Support and Basic Prompt conditions, with a significant difference compared to Baseline. The right chart shows vitality. Bars indicate that vitality scores were lower in the Basic Prompt condition than in the other two conditions, with the difference marked as significant. Error bars on all bars represent standard errors.}
\end{figure}

\section{Results}

To answer RQ1 and RQ2, we analyze conversational traces and survey measures, followed by qualitative analysis of chat logs and interview transcripts.

\subsection{Perceived Need Support from the Chatbot}

We found no significant differences across \emph{Condition} for perceived chatbot autonomy support ($p = .23$). Scores were generally high ($M=5.69$, $SD=1.18$, $Md=6.0$), with more than half of participants (56\%) rating 6 or above on the 7-point scale. In contrast, perceived chatbot competence support differed significantly ($\chi^2(2)=11.80, p<.01$, $\eta^2_H = .14$), with both \DS{} ($p < .01$) and \PS{} ($p < .01$) rated higher than \BP{}. Perceived chatbot relatedness support also differed significantly ($\chi^2(2)=28.43, p<.001$, $\eta^2_H = .38$), with higher ratings in \DS{} ($p < .001$) and \PS{} ($p < .001$) compared to \BP{}. Together, these findings indicate that the chatbot's interaction design effectively supported competence and relatedness needs (Figure~\ref{fig:ns_chatbot}). Full descriptive statistics are provided in Appendix Table~\ref{tab:chatbot_needs_appendix}.

\begin{figure}[b]
  \centering
  \includegraphics[width=\linewidth]{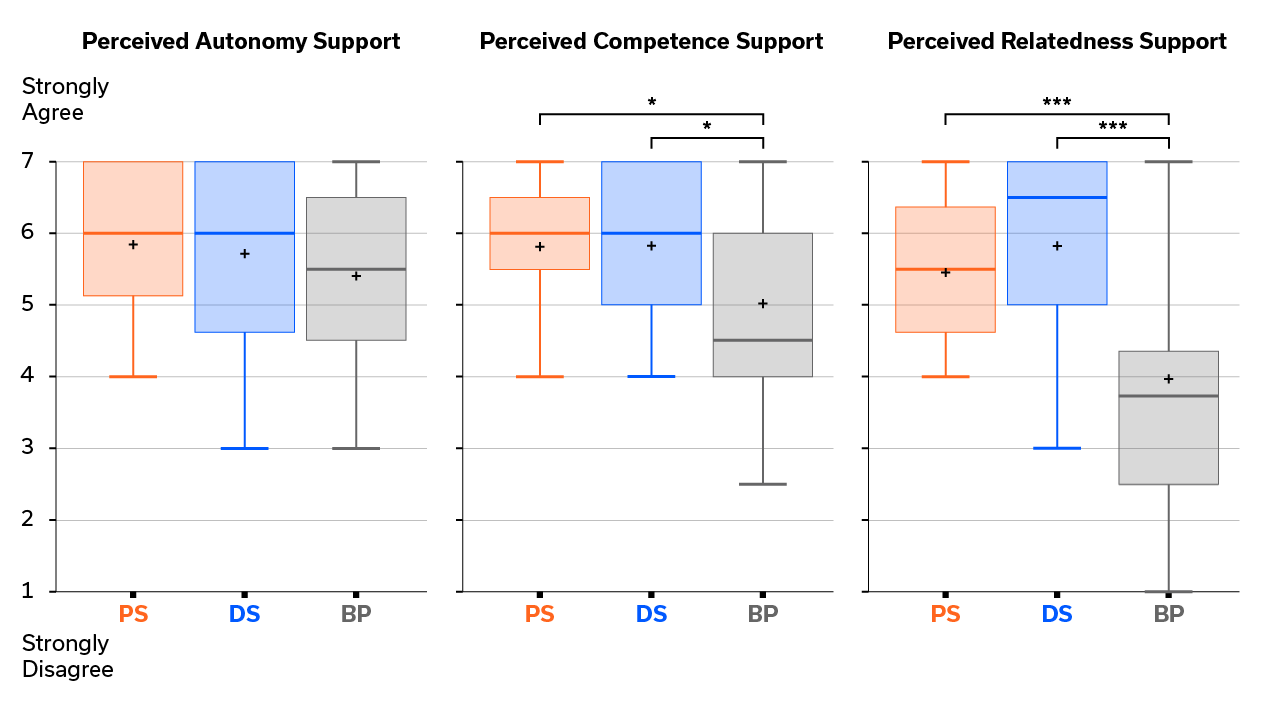}

  \caption{Perceived need support from the chatbot across conditions.}
  \label{fig:ns_chatbot}
  \Description{Three boxplots compare perceived autonomy, competence, and relatedness support from the chatbot across the Partner Support (PS), Direct Support (DS), and Basic Prompt (BP) conditions. The vertical axis ranges from 1, labeled Strongly Disagree, to 7, labeled Strongly Agree. In the autonomy support plot, median ratings are similar across all three conditions with no significant differences. In the competence support plot, PS and DS show higher ratings than BP, with significant differences indicated for both comparisons. In the relatedness support plot, PS and DS again show hight ratings than BP, with highly significant differences marked. Boxes show interquartile ranges, whiskers show full ranges, and plus signs mark means.}
\end{figure}

\subsection{The Impact of Need Support Prompts on Motivation and Need Satisfaction}

\begin{table}[t]
\centering
\small
\setlength{\tabcolsep}{6pt}
\caption{LMM fixed effects (F-tests) for motivation and need satisfaction dependent variable.}
\label{tab:lmm_uniform}
\begin{tabular}{l l c c c}
\hline
\textbf{DV} & \textbf{Effect} & \textbf{F(df)} & \textbf{$p$} & $\eta^{2}_p$\\
\hline
\multicolumn{4}{l}{\textbf{Autonomous Motivation}} \\
\hline
 & Condition         & 0.94 (2)  & .394 & .04 \\
 & Phase              & 0.37 (1)  & .543 & .01 \\
 & Condition $\times$ Phase & 0.91 (2)  & .404 & .01 \\
\hline
\multicolumn{4}{l}{\textbf{Controlled Motivation}} \\
\hline
 & Condition         & 0.85 (2)  & .432 & .02 \\
 & Phase              & 28.94 (1) & $< .001^{***}$ & .35 \\
 & Condition $\times$ Phase & 2.68 (2)  & .073 & .04 \\
\hline
\multicolumn{4}{l}{\textbf{Need Satisfaction}} \\
\hline
 & Condition         & 1.26 (2)  & .288 & .07 \\
 & Phase              & 3.31 (1)  & .071 & .09 \\
 & Condition $\times$ Phase & 0.89 (2)  & .414 & .01 \\
\hline
\end{tabular}
\end{table}

For \textbf{autonomous motivation}, no significant main effects were found. At baseline, participants already reported high autonomous motivation ($M=5.91$, $SD=0.91$, on a 7-point scale; $Md=6.0$), with more than half (57\%) scoring 6 or above.

For \textbf{controlled motivation}, there was a significant main effect of \emph{Phase} ($p<.001$) with controlled motivation decreasing from pre- to post-intervention across conditions.

\textbf{Need satisfaction} showed no significant main effects of Condition or Phase, and no significant Condition~$\times$~Phase interaction. At baseline, need satisfaction was already high ($M=5.76$, $SD=0.95$, on a 7-point scale; $Md=5.78$).
Detailed results can be seen in Table~\ref{tab:lmm_uniform}. These findings directly address RQ1 by clarifying the motivational dynamics of the interaction. Collectively, these results suggest that the intervention did not meaningfully alter autonomous motivation, which remained stable across time and conditions. In contrast, controlled motivation decreased over time across all conditions. Need satisfaction showed no overall condition or time effects. Full descriptive statistics are provided in Appendix Table~\ref{tab:autonomous_motivation_appendix}, ~\ref{tab:controlled_motivation_appendix}, ~\ref{tab:need_satisfaction_appendix}.

\subsection{The Impact of Need Support Prompts on Well-being}

For \textbf{IOS}, we found a significant main effect of \emph{Phase} ($p<.001$) and a \emph{Condition $\times$ Phase} interaction ($p<.05$). Follow-up comparisons indicated that IOS increased significantly from pre- to post-interaction in \PS{}, while no changes were observed in the \DS{} or \BP{}.

For \textbf{Self-esteem}, there was only a main effect of \emph{Phase} ($p<.001$), suggesting overall improvement across conditions.

For Vitality, participants generally reported higher energy levels after the interaction (Main Effect of Phase: $p < .001$), but this improvement was significantly moderated by condition (Interaction: $p<.05$). Specifically, while \DS{} and \PS{} experienced a distinct boost in vitality, the \BP{} group remained statistically stagnant, suggesting that the chatbot's scaffolding was necessary to generate this uplifting effect.

For \textbf{Positive Affect}, there were significant main effects of both \emph{Condition} ($p < .01$) and \emph{Phase} ($p < .001$) but no interaction.

Taken together, these results suggest that three conditions broadly improved participants' well-being over time, with IOS and vitality showing condition-specific effects, while self-esteem and positive affect improved more generally. Main descriptive statistics are shown in Table~\ref{tab:wellbeing}, with corresponding visualizations in Figure~\ref{fig:wellbeing}. Full descriptive statistics are provided in Appendix Table~\ref{tab:wellbeing_appendix}. We note that the observed effect size for the significant interactions in IOS and Vitality are slightly below the sensitivity threshold indicated by our power analysis. While statistically significant in this sample, this suggests these specific interaction effects should be interpreted with a degree of caution pending future replication.

\begin{table}[t]
\centering
\small
\setlength{\tabcolsep}{6pt}
\caption{ART ANOVA results (F-tests) for well-being dependent variables.}
\label{tab:wellbeing}
\begin{tabular}{l l c c c }
\hline
\textbf{DV} & \textbf{Effect} & \textbf{F(df)} & \textbf{$p$} & $\eta^{2}_p$ \\
\hline
\multicolumn{4}{l}{\textbf{IOS}} \\
\hline
 & Condition         & 1.19 (2) & .310 & .03 \\
 & Phase              & 28.23 (1) & $< .001^{***}$ & .29 \\
 & Condition $\times$ Phase & 4.13 (2) & $< .05^{*}$ & .11 \\
\hline
\multicolumn{4}{l}{\textbf{Self-esteem}} \\
\hline
 & Condition         & 1.89 (2) & .159 & .05 \\
 & Phase              & 22.30 (1) & $< .001^{***}$ & .24 \\
 & Condition $\times$ Phase & 0.97 (2) & .384 & .03 \\
\hline
\multicolumn{4}{l}{\textbf{Vitality}} \\
\hline
 & Condition         & 0.46 (2) & .634 & .01 \\
 & Phase              & 18.22 (1) & $< .001^{***}$ & .21 \\
 & Condition $\times$ Phase & 4.22 (2) & $< .05^{*}$ & .11 \\
\hline
\multicolumn{4}{l}{\textbf{Positive Affect}} \\
\hline
 & Condition         & 5.88 (2) & $< .01^{**}$ & .15 \\
 & Phase              & 29.45 (1) & $< .001^{***}$ & .29 \\
 & Condition $\times$ Phase & 1.00 (2) & .373 & .03 \\
\hline
\end{tabular}
\end{table}

\begin{figure}[b]
  \centering
    \includegraphics[width=\linewidth]{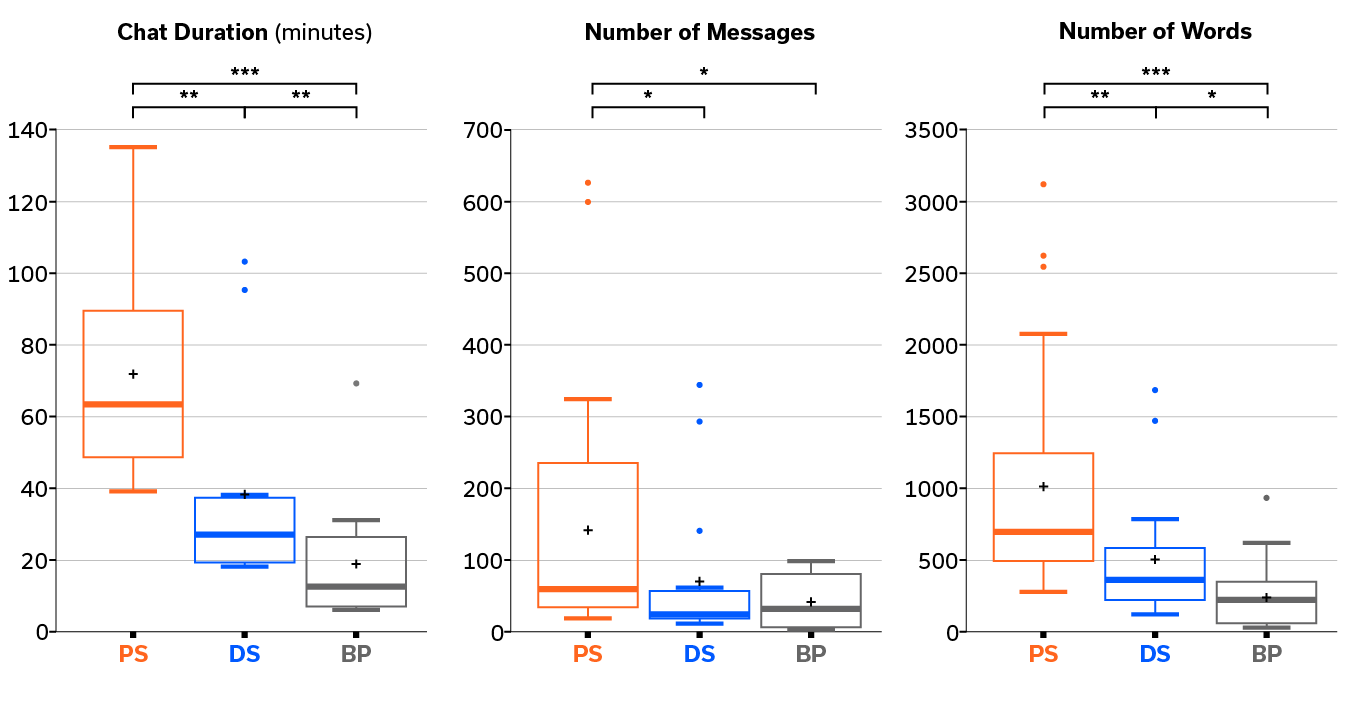}
  \caption{Chat engagement metrics across conditions.}
  \label{fig:con_metrics}
  \Description{Three boxplots compare conversation engagement across Partner Support (PS), Direct Support (DS), and Basic Prompt (BP) conditions. The left plot shows chat duration in minutes. PS conversations are longest, with a median around one hour and some extending over two hours, while DS and BP conversations are notably shorter. The middle plot shows the number of messages. PS has the highest counts, often above 200, whereas DS and BP typically stay below 100. The right plot shows the number of words exchanged. Word counts are highest in PS, frequently exceeding 1,000 words, compared to lower totals in DS and BP. Significant differences are marked above the plots, indicating that PS produced more sustained and content-rich interactions than the other conditions.}
\end{figure}

\subsection{Conversation Engagement}

For \textbf{chat duration}, as shown in Figure~\ref{fig:con_metrics}, chat duration differed significantly across conditions ($H(2)=42.3$, $p<.001$, $\eta^2_H=.58$), with \PS{} sessions lasting significantly longest on average ($Mdn=79.5$ min), followed by \DS{} ($Mdn=32.5$ min) and \BP{} ($Mdn=12.5$ min) (all $p<.01$). We acknowledge that the increased duration in \PS{} is partly attributable to the additional reflection phases. However, visual inspection of the phase timelines (Figure~\ref{fig:timeline}) suggests that participants in \PS{} and \DS{} also spent more time within the common question phases compared to \BP{}, indicating deeper engagement even when structural opportunities were identical.

For \textbf{number of messages}, differences were significant ($H(2)=10.3$, $p < .01$, $\eta^2_H=.12$). \PS{} generated more messages than both \DS{} ($p<.05$) and \BP{} ($p<.01$), while \DS{} did not significantly differ from \BP{}. For \textbf{number of words}, group differences were again significant ($H(2)=30.6$, $p<.001$, $\eta^2_H=.42$). The conditions differ significantly from each other (all $p<.05$) with \PS{} participants producing far more words, followed by \BP{} and \DS{}. Notably, the higher word counts in \PS{} and \DS{} align with the findings on disclosure depth (Section~\ref{sec:SDdepth}), which confirmed that participants in these conditions produced richer, more detailed responses even when analyzing only the shared Question Phases. Corresponding boxplots are presented in Figure~\ref{fig:con_metrics}. Full descriptive statistics are provided in Appendix Table~\ref{tab:engagement_appendix}.

\subsection{Self-disclosure and Enacted Partner Support}
\label{sec:SDdepth}

We first compared self-disclosure depth across the three conditions during the three common question phases. Significant group differences emerged for informational ($p<.001$), thought-based ($p<.001$), and emotional disclosure ($p<.001$). Both \PS{} and \DS{} prompted richer disclosures than \BP{} across all three dimensions (all $p<.05$). No differences were observed between \PS{} and \DS{}, suggesting that the enabling affordances (structured prompts and warm tone) present in both conditions were sufficient to deepen disclosure beyond baseline small-talk levels.

Regarding enacted partner support, no differences were observed for autonomy or competence, but clear contrasts appeared in relatedness support ($p<.001$). Both \PS{} and \DS{} yielded higher enacted relatedness support than \BP{} (both $p<.01$). Main results and pairwise comparisons are provided in Table~\ref{tab:kruskal-matrix} and Figure~\ref{fig:SD}. Full descriptive statistics are provided in Appendix Table~\ref{tab:disclosure_support_appendix}.

\begin{figure}[b]
  \centering
    \includegraphics[width=\linewidth]{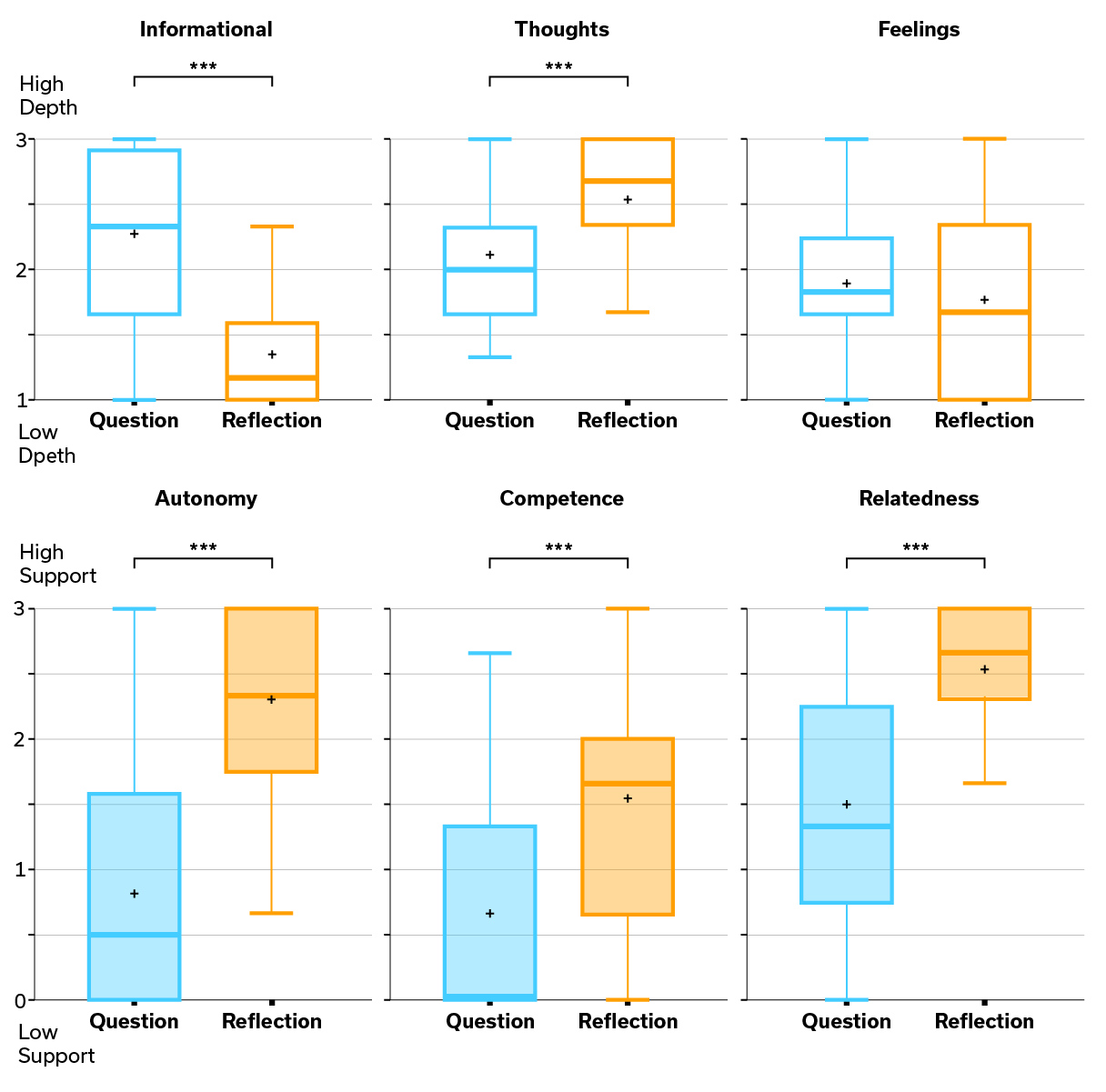}
  \caption{Comparison of self-disclosure depth and need support level between question phases and reflection phases within the Partner Support (\PSbox, G1-12) condition.}
  \label{fig:phase_comparison}
  \Description{Two rows of boxplots compare question phases and reflection phases within the Partner Support condition. The top row shows self-disclosure depth across informational, thoughts, and feelings, rated from 1 for minimal detail to 3 for rich, contextualized sharing. Reflection phases show higher depth than question phases for thoughts and lower depth than questions phases for informational with significant differences. No significant difference was observed for feelings. The bottom row shows partner-provided need support level for autonomy, competence, and relatedness, rated from 0 for low support to 3 for high support. Reflection phases show higher support than question phases across all three needs, with significant differences in each case. Together, the plots indicate that reflection phases fostered stronger need-supportive behaviors than question phases.}
\end{figure}

Within the \PS{} condition, functional distinctions emerged between the disclosure (questions) and scaffolding (reflection) phases. Question phases elicited greater disclosure depth for thought content ($W=188, p<.001, r=.86$), confirming that the enabling prompts effectively encouraged detailed reflection. In contrast, reflection phases were the primary site of enacted partner support: autonomy ($W=250, p<.001, r=.85$), competence ($W=230, p<.001, r=.86$), and relatedness ($W=209, p<.001, r=.87$) were all significantly higher during reflections than during questions. This confirms the intended dual-layer dynamic: participants successfully shifted from narrating personal experiences (self-focus) to providing validation and encouragement (partner-focus) when prompted. Corresponding boxplots are presented in Figure~\ref{fig:phase_comparison}.

\begin{table*}[t]
\centering
\small
\setlength{\tabcolsep}{5pt}
\renewcommand{\arraystretch}{1.2}
\caption{Main effects for all self-disclosure depth and need support DVs. $r$ shows effect size. * $p<.05$, ** $p.<.01$, *** $p<.001$, etc...}
\label{tab:kruskal-matrix}
\begin{tabular}{lcccccc}
\toprule
\multirow{2}{*}{Measure} & \multicolumn{3}{c}{Omnibus (Kruskal--Wallis)} & \multicolumn{3}{c}{Post-hoc (adjusted $p$; $r$)} \\
\cmidrule(lr){2-4} \cmidrule(lr){5-7}
 & $H(2)$ & $p$ & $\eta^{2}_H$ & \BP{} vs. \DS{} & \BP{} vs. \PS{} & \DS{} vs. \PS{} \\
\midrule
Information & 21.2 & $<.001$ & .28 & ** ($r=.45$) & **** ($r=.65$) & n.s. \\
Thoughts    & 19.9 & $<.001$ & .26 & *** ($r=.52$) & *** ($r=.59$) & n.s. \\
Feelings    & 14.8 & $<.001$ & .19 & * ($r=.35$)  & *** ($r=.55$) & n.s. \\
Autonomy    &  --  & n.s.    &  -- & n.s.         & n.s.          & n.s. \\
Competence  &  --  & n.s.    &  -- & n.s.         & n.s.          & n.s. \\
Relatedness & 15.8 & $<.001$ & .20 & ** ($r=.44$) & *** ($r=.54$) & n.s. \\
\bottomrule
\end{tabular}

\end{table*}

\begin{figure}[b]
  \centering
    \includegraphics[width=\linewidth]{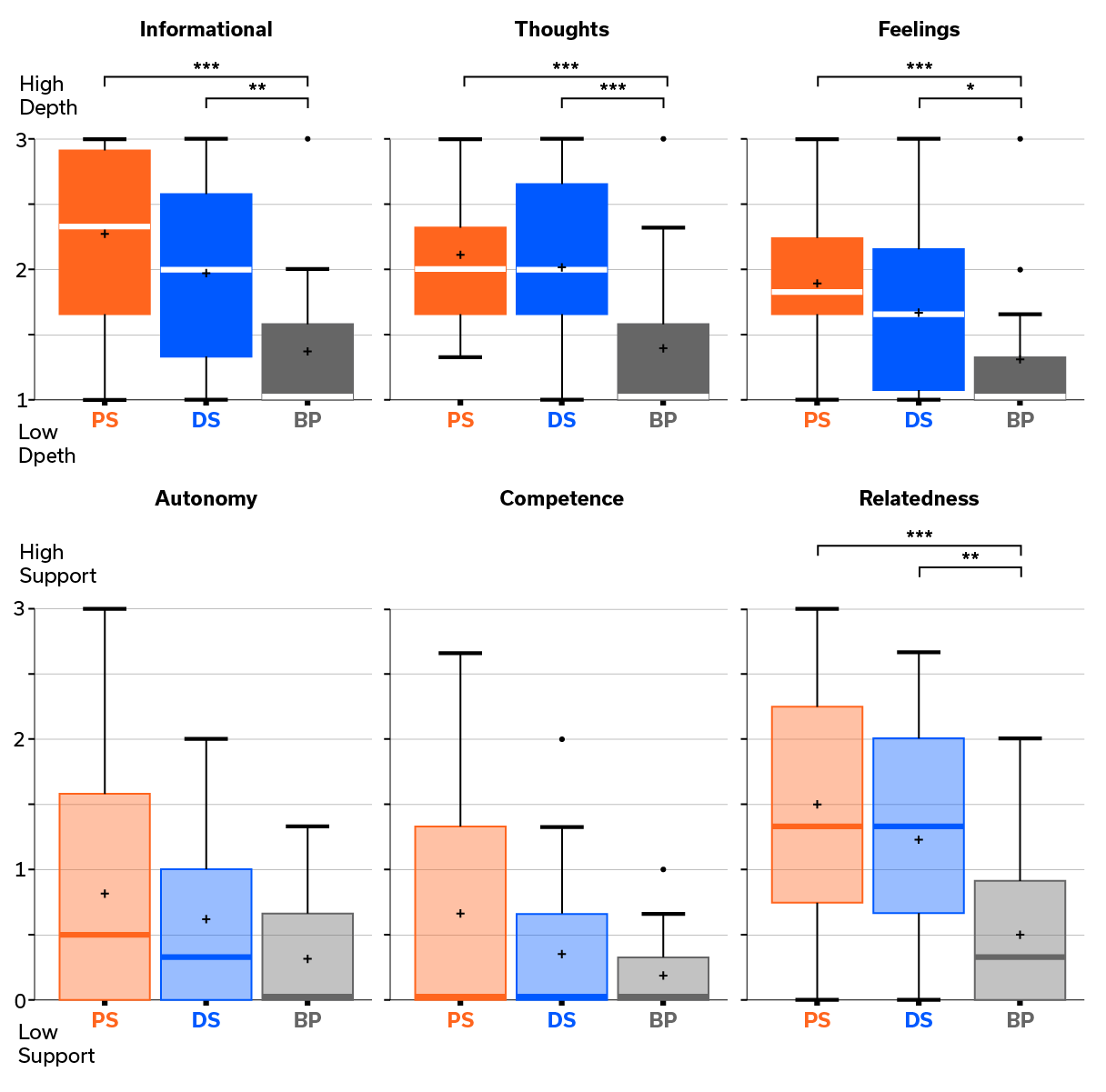}

  \caption{Depth of self-disclosure and level of partner-provided need support across conditions.}
  \label{fig:SD}
  \Description{Two sets of boxplots compare outcomes across Partner Support (PS), Direct Support (DS), and Basic Prompt (BP) conditions. The top row shows self-disclosure depth across informational, thoughts, and feelings categories, rated from 1 for minimal detail to 3 for rich, contextualized sharing. Both PS and DS groups show higher depth than BP across all three types of self-disclosure. The bottom row shows partner-provided support in the conversation for autonomy, competence, and relatedness, rated from 0 for low support to 3 for high support. Relatedness support is significantly higher in PS and DS compared to BP. Autonomy and competence support show higher medians in PS but without consistent significant differences. Overall, PS and DS fostered deeper self-disclosure than BP, and PS and DS strengthened relatedness support through partner reflection prompts.}
\end{figure}

\subsection{Qualitative Findings}

Supplementing RQ2, we focus primarily on the Partner Support and Direct Support groups to explore how participants experienced the chatbot's role in shaping their conversations and relational dynamics.

\subsubsection{Chatbot scaffolded need-supportive talk and relational lessons}

A recurring theme was that the chatbot encouraged couples to engage in deeper, more reflective and need-supportive conversations that rarely surfaced in daily life. Participants noted that it created space to move beyond small talk, revisit neglected topics, and voice unspoken thoughts. In the chat log, P23 explicitly asked the chatbot for support:
\begin{quote}
\emph{``I think both of us as partners rarely have some nice conversations to understand each other more, also we both are quite bad at expressing ourselves, will you help with it?''}
\end{quote}
P11 described how the demands of work left little time for meaningful connection, noting that the chatbot \emph{``engaged us to have all these deep conversations… to connect more deeply on an emotional level.''}

Prompts nudged partners to acknowledge each other's viewpoints, which participants linked to feeling supported. As P3 put it, the interaction \emph{``creates an opportunity for your partner to feel like they understand your perspective,''} which \emph{``helps to create that sense of feeling supported.''}
Several contrasted the chatbot with static tools like card games or question banks. P22 noted that such resources \emph{``are never usually structured in a way where it is a conversational flow,''} whereas the chatbot built prompts that \emph{``get the conversation going''} and \emph{``build up towards''} deeper topics.

More specifically, the chatbot helped users to identify and express needs that were previously difficult to articulate. As P23 reflected, \emph{``we don't really know how to ask that kind of question… [the chatbot] taught me how to start''}. 
P41 framed it as \emph{``a good start to actually have someone guide us through meaningful conversations,''} especially since couples \emph{``may not have that experience or wisdom to talk about things that truly matter.''}.

Others emphasized structure: it \emph{``will help me to structure my questions a bit better… in retrospect you wanted the question to be more specific; this helps… even without a chatbot, it allows us to explore that depth in future''} (P37). Evidence also appeared in the chat logs. For example, P11 remarked during the interaction:
\begin{quote}
\emph{``Love these questions which allow us to have deeper heart to heart conversations, which I really love to have with him but find it hard to have given how busy we are these days.''}
\end{quote}

Participants in \PS{} explicitly identified the chatbot's follow-up prompts as the catalyst for shifting from self-narration to mutual support. Without these specific nudges, participants noted they might have stopped at the disclosure. For example, P8 contrasted the experience with other conversation tools, noting that usually \emph{it often stops at you answering that question… there's usually no follow-up,''} whereas the chatbot's prompts provided a \emph{new dimension''} that forced them to \emph{think beyond the question about how I can actually implement things into my daily life... to help the other person.''} Similarly, P13 noted that the system made them \emph{reflect on the spot''} about concrete plans rather than just \emph{getting the information and sitting on it.''} This scaffolding was crucial for overcoming communicative inertia; P16 admitted that while they often felt supportive, \emph{it never occurred to me to say that''} until prompted, which gave them the \emph{opportunity to express that better in words.''} Consequently, participants perceived these follow-ups not merely as conversation extenders, but as prompts for actionable care. As P14 explained, the interaction moved \emph{beyond just reflecting… into how we can better support each other… concrete things we think about now that build the foundation of a better relationship.''} This suggests that the functional distinction of the \PS{} condition---the reflection phase---successfully operationalized support by transforming abstract empathy into concrete, voiced supportive behaviors. Chat log analysis confirms that the follow-up mechanism was the active ingredient for this shift (Appendix~\ref{app:conversation_comparison}). Unlike \DS{}, where the agent provided the empathy, the \PS{} prompts successfully triggered partners to provide that validation reciprocally, converting the chatbot from a supporter into a facilitator of mutual care.

Prompts sometimes unlocked long-held but unspoken sentiments. P36 remarked that \emph{``it would have been helpful… but it never occurred to me to say that,''} and also, \emph{``I wanted to express this for some time, but now I found the words for it… it's satisfying.''} Some appreciated explicit nudges to take their time and avoid self-censorship. As P32 described, \emph{``the answer doesn't have to be perfect… you can just give exactly how you're feeling,''} which reduced hesitation. 
Beyond fostering dialogue, participants described how answering the chatbot's prompts helped them reflect on themselves. P19 observed that writing responses made them \emph{``think more deeply about how realistic or how capable you are of doing such a thing.''}

\subsubsection{Chatbot as a Mediator or Third Perspective}

The chatbot was often characterized by participants as a neutral facilitator, capable of easing disagreements, synthesizing viewpoints, and supporting joint reflection.
Some appreciated the bot's ability to distill what had just been shared and present it back as a shared frame. 
P3 anticipated value in decision-making contexts, where the bot could \emph{``highlight each other's perspective, the pros and cons''} of possible choices.

Others emphasized how the chatbot offered a \emph{``step backwards''} in tense situations, allowing partners to see issues from a bigger perspective (P19). P36 contrasted the bot's impartiality with the biases often introduced by human mediators: \emph{``with a third party intermediary… you're worried that the person might be biased… but if you know that the bot is impartial… it's a lot easier.''}. This impartiality, they argued, could make the chatbot useful in conflict resolution, even functioning as a \emph{``sanity check''} when disputes risked escalation.

Participants also noted that the chatbot encouraged partners to validate each other's feelings rather than fixating on who was right or wrong. As P39 explained, the chatbot helped foreground \emph{``how different [perspectives] are coming from,''} shifting attention from adversarial stances to mutual understanding. 

Analysis of the chat logs revealed that many couples directly sought relationship advice from the chatbot itself. In G2, for example, after responding to the Competence Reflection question, P3 asked: \emph{``What's ur advice to us?''} P4 then followed with a more complex query: \begin{quote}\emph{``What if parts of each our dream do not align with the other's and make it harder to achieve them? What is your advice to work around them?''}\end{quote} Similar patterns appeared in other groups. 

\subsubsection{Couple Therapy and Counselor}

Participants often compared the chatbot to a counselor or therapist, portraying it as an accessible, always-available alternative to formal professional support. They highlighted its potential for accessibility, early screening, and emotional safety, while also recognizing clear limitations when compared to human counseling. These perspectives surfaced in both interviews and chat logs. 

Several participants highlighted accessibility as a major advantage: the chatbot is free, immediate, and does not carry the stigma often associated with therapy. P4 emphasized affordability, \emph{``counseling and stuff… are very expensive,''} while P35 observed that therapy can trigger defensiveness (\emph{``is there something wrong with me?''}), whereas a chatbot feels easier to try without committing money or time. Others described it as a way for people averse to therapy to begin engaging in reflective conversations (P32, P47).

Participants felt the chatbot might be able to provide therapist-like functions such as prompting, structuring dialogue, and sustaining engagement. P3 described the prompts as \emph{``quite thought provoking''} and the replies as \emph{``quite human… like some relationship therapies.''} P32 said it acted \emph{``kind of like a therapist''} by helping them express themselves when stuck. For some, this mirrored counseling's role in deepening conversations or keeping them on track (P5, P20). 

Despite these parallels, participants were clear about the boundaries of chatbot counseling. P3 distinguished it from professional therapy by noting that while the bot's responses are \emph{``standard templates''} rather than tailored to individual histories.

\section{Discussion}

Our findings shed light on both research questions. We discuss these contributions through the lens of our dual-layer scaffolding framework, examining first how the chatbot established the necessary conditions for instrumental support and second how it facilitated relational support.

\subsection{Scaffolding Need-Supportive Conversations through Chatbots}

Our study addresses RQ1 by demonstrating how chatbot-mediated scaffolding influences the motivational underpinnings of interaction. The results validate our sequential model: the chatbot successfully established a need-supportive environment (Layer 1) which, in the \PS{} condition, translated into reciprocal partner support (Layer 2).

\subsubsection{Enabling Affordances and Chatbot Support.}

The first goal of the design was to provide instrumental support to lower the barriers to disclosure. Results confirm the efficacy of these enabling affordances: interactions in \PS{} and \DS{} yielded significantly higher perceived competence and relatedness support from the chatbot than \BP{}. This indicates that the chatbot's active scaffolding created a greater sense of capability and psychological safety than the raw disclosure questions alone.

Perceived autonomy support was high and comparable across all conditions. This nuances prior findings that highly directive guidance can depress autonomy by pushing users at the agent's pace~\cite{cai2023listen}. Despite the structured nature of the \PS{} and \DS{} interventions, our design preserved autonomy by validating perspectives, offering meaningful choices and rationales, framing effort as voluntary, and letting participants regulate pacing. This demonstrates that proactive scaffolding and autonomy preservation are compatible; for CA designers, the priority is to embed enabling affordances that scaffold competence and relatedness while strictly safeguarding volition through choice.

Our results align with previous studies that conversational agents can provide autonomy, competence, and relatedness support~\cite{zhu2025benefits, yang2021designing, joshi2025user, diebel2025when, bowman2024exploring}, and doing so through concrete interaction tactics rather than simply lowering guidance intensity. Our findings extend prior work to the agent-mediated, multi-party settings. Team-level SDT models suggest that patterns of need support can converge into shared motivational states via interpersonal feedback loops~\cite{grenier2024self}. Building on this, our results imply that as a social actor, the facilitator-agent effectively cultivated a supportive group climate, which served as the necessary foundation for the emergence of group-level need supportive behaviors.

\subsubsection{Motivational Dynamics and Instrumental Support}

We observed nuanced effects regarding the motivational underpinnings of the interaction. Autonomous motivation and need satisfaction remained high and stable across all conditions. This is largely attributable to a ceiling effect: over half the participants rated autonomous motivation at 6 or above at baseline. Combined with the single-session format, this explains the contrast with meta-analytic evidence from health interventions, where increases in autonomous motivation and need satisfaction typically emerge only after several months of sustained engagement (average duration of about four and a half months)~\cite{ntoumanis2021meta}.

However, controlled motivation decreased significantly across all conditions. This aligns with Organismic Integration Theory, which holds that reductions in external and introjected regulation often precede later gains in identified and intrinsic regulation\cite{ryan2000selfdetermination, wasserkampf2016organismic}. In our setting, even brief reflective disclosure appeared sufficient to strip away pressure-based reasons for participating. The pattern also resonates with OIT-informed UX guidance: fast-acting strategies such as novelty, immediate enjoyment, or light rewards are more effective in shifting controlled processes than in fostering autonomous motivation in the short term~\cite{bennett2024beyond}. 

\subsubsection{Conversation Dynamics}

Chat log analysis revealed distinct interaction patterns driven by the scaffolding layers. In \BP{}, exchanges were typically short and resembled routine small talk. By contrast, \PS{} and \DS{} conversations exhibited longer, more reflective turns. The overall discussion duration decreased markedly from \PS{} to \DS{} to \BP{}. This pattern reflects the fact that couples in \PS{} and \DS{} spent more time conceiving and composing longer responses to the chatbot's structured prompts, often matching their partner's depth and length. These findings resonate with prior work on chatbot-mediated reminiscence, which shows that structured conversational scaffolds can support deeper, reciprocal exchanges between loved ones~\cite{jiang2025remini}.

We acknowledge that the increased conversation duration and word count in the \PS{} condition are partly attributable to the structural design of the system (i.e., the inclusion of reflection phases). However, observation of individual outliers suggests that structural sophistication alone does not account for the qualitative differences in interaction.

As noted in our results (Figure~\ref{fig:timeline}), individual differences in communicative style influenced engagement; for instance, dyad G25 in \BP{} and G13, G24 in \DS{} engaged in a session length comparable to dyads in \PS{} (e.g., G5, G11). However, while high-agency couples like G25 may generate high volumes of text without scaffolding, our coding analysis (Section 5.5) reveals that volume does not equate to support. High-volume conversations in \BP{} or \DS{} often remained focused on self-narration (monologues), whereas \PS{} specifically significantly increased relatedness support (dialogues). Conversely, we observed that even dyads in the \PS{} condition with shorter conversation durations successfully engaged in reciprocal need support. This indicates that the mediating affordances served as an effective behavioral nudge, eliciting supportive responses regardless of whether a couple's habitual communication style is verbose or concise.

Furthermore, our use of Linear Mixed-Effects Models (LMM) for psychological outcomes controlled for participant baselines, ensuring that the observed benefits in closeness and support were driven by the intervention rather than pre-existing individual traits. Thus, while individual differences drive how much couples talk, the chatbot's scaffolding determines how they talk---specifically, by transforming disclosure into reciprocal support.

\subsubsection{The Role of Mediating Affordances}

Analysis of the question phases reveals that the enabling affordances in \PS{} and \DS{} successfully elicited greater self-disclosure depth than \BP{}. However, this increased depth did not automatically translate into comprehensive partner support; while relatedness support tracked with disclosure, autonomy and competence support showed no differences across conditions. Crucially, only couples in \PS{}, who received the mediating affordances of reflection prompts, reported increased closeness. These findings align with prior work showing that deliberate, effortful messages which require time and pause for thought can foster mutual affection and care~\cite{kelly2018letter, kelly2017demanding}, but adds the nuance that this effort must be specifically scaffolded toward support rather than just disclosure.

Beyond closeness, we also observed broader affective outcomes: self-esteem and positive affect increased across all conditions,  subjective vitality remained statistically stagnant in \BP{} while improving in the scaffolded conditions. This is consistent with prior research suggesting that self-disclosure can create tension if not navigated skillfully~\cite{kil2022autonomy}; the chatbot's instrumental support likely mitigated this cost. Within \PS{}, reflection phases stood out: reflection phases were the primary site of enacted partner support, significantly exceeding the levels found during question phases.  Our qualitative findings explain the quantitative gap in relatedness support between \PS{} and \DS{}. While participants in both conditions engaged in deep self-disclosure, the qualitative feedback suggests that disclosure does not automatically trigger a need-supportive response from a partner. The follow-up mechanism in \PS{} acted as a `nudging function' for reciprocity, breaking the inertia of monologues. By explicitly asking partners to validate (Autonomy), help (Competence), and empathize (Relatedness), the system bridged the gap between hearing a partner's disclosure and actively supporting it.

\subsubsection{Summary}

Collectively, these findings highlight the unique value of a dual-layer scaffolding approach for couples' communication. Most well-being metrics increased across conditions, alongside reductions in controlled motivation, consistent with the broader benefits of engaging in meaningful conversations~\cite{finkenauer2015toknowyou, laurenceau1998intimacy}. But only the \PS{} condition, which combined enabling and mediating affordances, successfully translated disclosure into reciprocal support and increased closeness.

This confirms our theoretical premise: effective relatedness technologies must do more than simply connect partners. They must first provide instrumental support (reducing pressure and ambiguity via Layer 1) to create the capacity for relational support (reciprocal validation via Layer 2). By scaffolding this sequence, our design encouraged couples to invest in effortful communication without undermining their autonomy which is a critical balance given that prior systems often sacrificed one for the other~\cite{kelly2018letter, kim2019love}. In a context where daily life offers few opportunities to move beyond routine small talk~\cite{jess2005mapping, ryjova2024aday}, this study shows that a need-supportive chatbot can serve as an essential mediator, transforming the intention to connect into the action of reciprocal care.

\subsection{Perceived Roles of the Chatbot in Couple Conversations}

This section addresses RQ2 by exploring how participants interpreted the chatbot's role. Beyond its functional utility, participants ascribed specific social roles to the agent, from a supportive peer to a mediator or counselor, that reveal how the dual-layer scaffolding was experienced in practice.

\subsubsection{Designing Chatbot for Need Support and System Scope}

Most participants in \PS{} and \DS{} described the chatbot as a positive, accessible presence. They pointed to both its interpersonal qualities (e.g., friendly tone, playful exchanges) and its structural features (e.g., rationales, step-by-step instructions) as key drivers of engagement. Through our theoretical lens, these features functioned as enabling affordances (Layer 1): the warmth established psychological safety (Relatedness), the structure reduced ambiguity (Competence), and the non-pressuring invitations preserved volition (Autonomy). 

This blend of warmth and structure supported engagement and enjoyment, echoing prior findings on chatbot personality~\cite{moilanen2022measuring}. Tone, however, is not trivial: while informality increases relatability, it can undermine credibility in high-stakes contexts. Studies show that while empathetic remarks encourage disclosure, users generally prefer high-warmth systems over those perceived as merely high-competence~\cite{li2025acomparative, gilad2021effects}. Our findings extend this work by shifting attention from static traits like personality or tone to the motivational quality of interactions. In our design, personality was not an end in itself but was explicitly calibrated to support autonomy, competence, and relatedness, thereby lowering the barrier to entry for deep disclosure.

This theoretical grounding also clarifies the system's intended scope. While participants perceived the chatbot as a potential tool for daily use, we designed the current system primarily as a technology probe to validate the dual-layer scaffolding mechanism. By focusing on the immediate motivational quality of interactions rather than long-term retention mechanics, we aimed to isolate the instrumental and relational support before deploying them as a sustained intervention. Thus, the current design serves as a proof-of-concept for the mechanism of support, laying the necessary groundwork for future work to explore its ecological appropriation as a longitudinal tool.

\subsubsection{Chatbot as Mediator in Intimate Conversations}

Findings from chat logs and interviews revealed that couples often sought direct guidance from the chatbot. This reflects an emerging shift in how users perceive conversational agents: not just as tools, but as trusted third parties capable of offering objective, non-judgmental input. Our findings resonate with prior studies where chatbots served mediating roles in multi-party contexts~\cite{zheng2021pocketbot, salehzadeh2023potential, govers2024aidriven, chen2025maintaining, houde2025controlling}, but extend this work by demonstrating how such mediation operates in the deeply emotional setting of romantic couples. Here, the chatbot functioned as a facilitative mediator, employing mediating affordances to scaffold disclosures and reflections that go beyond surface consensus-building.

At the same time, prior research raises caution that in multiparty settings, chatbot recommendations may unintentionally prioritize the perspectives of more active participants, amplifying imbalance at the expense of quieter voices~\cite{wang2025socialrag}. In an intimate context, this risks aligning the system with one partner, creating social pressure rather than safety. Our findings suggest a path forward based on how successful dyads framed their interactions. Participants who reported the strongest sense of connection often spontaneously adopted a collective frame---for instance, P23 requested help for "both of us," P14 shifted focus to "how we can better support each other," and P4 framed conflicting goals as a joint challenge to "work around" rather than a competition. To reinforce this constructive pattern and mitigate perceived bias, chatbots should actively frame guidance using plural pronouns and joint action language (``we,'' ``together,'' ``as a couple''). By mirroring the collective language used by successful couples, chatbots can encourage partners to see challenges as shared responsibilities rather than adversarial disputes. By emphasizing mutual effort, chatbots can take on a constructive mediator role that strengthens relational bonds rather than creating asymmetry or unequal power dynamics~\cite{wenhart2025relatedness}.

\subsubsection{Therapist-Like Perceptions and Counseling Role}

Participants frequently likened the chatbot to a therapist, valuing its accessibility and lack of judgment~\cite{yuksel2023conversational, laura2024AI, hatch2025ELIZA}. However, consistent with resistance to AI in high-stakes intimate domains~\cite{balaji2024designing}, our findings frame the chatbot as an educational facilitator rather than a substitute. Given the system's lack of professional nuance, future deployments must explicitly define these boundaries to prevent over-reliance during crises. Instead, design should target low-stakes activities that build emotional regulation, such as pre-counseling screening or mutual reminiscence. This distinction also clarifies the system's current scope and long-term trajectory. While we designed the current system as a technology probe to validate the dual-layer scaffolding mechanism, we envision its role in daily life as transient rather than permanent. Much like therapy, the ultimate goal of such scaffolding is its own obsolescence. Aligned with the SDT principle of internalization, the chatbot provides the enabling and mediating affordances only until partners internalize these supportive patterns. Ideally, couples would gradually transition from agent-mediated disclosure to unmediated reciprocal support, eventually rendering the chatbot redundant.

\section{Design Implications}

Our findings suggest that AI-powered chatbots can enhance relational well-being by operationalizing a dual-layer scaffolding framework. Responding to calls for theory-informed design~\cite{wenhart2025relatedness}, we draw on Self-Determination Theory (SDT) not just as an evaluative metric, but as a generative blueprint. The design contributes along three dimensions: (1) providing enabling affordances (instrumental support) to lower the cognitive and emotional barriers to entry; (2) deploying mediating affordances (relational support) to scaffold reciprocal care between partners; and (3) fostering the internalization of these supportive behaviors, nudging couples from pressure-driven exchanges toward autonomous connection. Together, these dimensions illustrate how grounding conversational agents in psychological theory can transform them from passive channels into active facilitators of intimacy~\cite{slovak2023designing}.

\subsection{Providing Enabling Affordances: Instrumental Support}

While many systems seek to foster connection, they often prioritize relatedness at the expense of autonomy and competence~\cite{min2014biosignal, brueckner2018empathy}. For instance, mediated presence cues may create togetherness but lack the structure for deep conversation~\cite{vaanen2011carpe, heshmat2020family}, while system-imposed standards of ``effort'' can threaten authenticity by making care feel obligatory~\cite{kelly2017demanding}. Our findings suggest that to resolve these trade-offs, designs must first provide enabling affordances, which serve as instrumental support that establishes the psychological safety required for vulnerability.

Instead of enforcing depth through constraints such as escalating message lengths, enabling affordances invite engagement through volition and clarity. Our results demonstrate that by offering meaningful choices, clear rationales, and user-controlled pacing, the system shifts motivation from external pressure toward autonomous regulation. Similarly, by scaffolding disclosure with step-by-step instructions to support competence and a warm, non-judgmental tone to foster relatedness, the system lowers the cognitive and emotional barriers to entry. This implies a fundamental shift in the agent's role: in this layer, the chatbot acts not as a social partner, but as a safe container that capacitates users to engage with their actual partner.

\subsection{Facilitating Partner-Provided Support: Relational Scaffolding}

Beyond providing instrumental support to individuals, the chatbot’s most significant contribution lay in scaffolding relational support between partners. Through the deployment of mediating affordances, specifically reflection prompts, the system encouraged participants to acknowledge goals, validate perspectives, and offer concrete assistance. This process shifted autonomy support from an individual feature into a relational practice, where choices were respected and affirmed by one's partner. Similarly, competence and relatedness were enacted through mutual encouragement and emotional reassurance. Consequently, the chatbot functioned less as a direct provider of care and more as a mediator of mutual need support, effectively aligning digital intervention with the motivational dynamics that sustain close relationships. Looking ahead, this perspective opens opportunities beyond dyads. With growing interest in deploying chatbots as mediators in couple therapy, intimate conversations, or even group therapy~\cite{geng2025whengroup, yuksel2023conversational, jiang2025remini}, SDT offers a transferable framework for supporting diverse relational configurations~\cite{laguardia2008selfdetermination}. Future research could explore how similar mediating affordances might nudge larger groups towards deep, need-supportive disclosure, thereby extending the design space from couples to broader social contexts.

\subsection{Scaffolding Motivation in Everyday Communication}

Grounding design in SDT reframes the goal of relatedness technologies: rather than prioritizing presence cues or affective signals, designs must scaffold mutual need support. While prior work emphasizes contact, our findings suggest that well-being relies on helping partners actively respect autonomy, build competence, and strengthen relatedness. In this way, relatedness technologies can evolve from simply connecting people to cultivating the motivational conditions that allow relationships to flourish over time. This aligns with calls to design technologies that embed psychological support into everyday routines~\cite{wenhart2025relatedness}.

Crucially, this suggests a shift in where such scaffolding should live. Couples already inhabit diverse CMC ecosystems~\cite{kwok2020cyberintimacy}, ranging from messaging apps used for logistics to video calls and games used for togetherness. Rather than building standalone apps that require new habits, our findings suggest that dual-layer scaffolding is effective when embedded directly into these existing platforms. By integrating enabling and mediating affordances into standard messaging (as with Telegram) or potentially extending them to collaborative gaming and video contexts, designers can transform these everyday digital spaces from passive conduits into active facilitators of reciprocal care.
Participant feedback supports this trajectory; many noted that the chatbot provided concrete strategies for ``starting deep conversations'' that they intended to carry forward without the tool. Resonating with the METUX model~\cite{peters2018designing}, this implies that relational chatbots should not only scaffold immediate disclosure but also equip users with internalized skills that extend the technology's influence into broader life.

\section{Limitations and Future Work}
This study has limitations. First, while our analysis of chat logs demonstrates immediate behavior change within the interaction (e.g., deeper disclosure and increased partner support), we acknowledge that due to the cross-sectional, single-session nature of the study, we cannot claim these interventions produced sustained behavioral change or long-term identity shifts outside the experimental context. Future longitudinal work is needed to determine if these scaffolded interactions translate into lasting relational habits and to evaluate how couples appropriate the tool in naturalistic settings over time. Specifically, long-term deployment studies would allow for an assessment of retention rates and whether the `dual-layer' scaffolding remains effective or becomes redundant as couples internalize the supportive skills. However, short-term boosts and immediate experiences can play a critical role in cultivating long-term motivation, well-being, and sustained engagement~\cite{laguardia2008selfdetermination, ryan2000selfdetermination}.

Self-disclosure is shaped by numerous factors, including personality traits, emotional states, cultural norms, relational history, and situational context~\cite{ignatius2007factors}. Our study necessarily focused on a subset of these influences, and more work is needed to figure out how other factors may impact self-disclosure.

Thirdly, while self-disclosure often fosters intimacy and mutual understanding, it can also carry risks depending on context, timing, and content. For example, revealing highly intimate information too early or without sufficient reciprocity can lead to embarrassment, judgment, or even relational strain; excessive or poorly timed disclosure may be perceived as inappropriate or unattractive~\cite{zhang2021distress, andalibi2018responding}, or can lead to repetitive focus on negative topics (i.e. co-rumination)~\cite{kovacs2025daily, rose2007prospective}. We mitigated this issue through our sampling strategy and with a single session study.

Lastly, our sample was drawn largely from younger participants who were open to experimenting with new technologies and, in many cases, were in relatively short-term yet stable relationships. This demographic skew limits the generalizability of our findings, as the dynamics of chatbot-mediated support may differ across other populations, where longer-lasting couples may bring different expectations and relational dynamics.

\section{Conclusion}

This work demonstrates the efficacy of a dual-layer scaffolding approach in chatbot-mediated communication. By providing enabling affordances that establish a safe and structured environment, the system lowers the barriers to deep self-disclosure. Crucially, through mediating affordances, it scaffolds partners to provide autonomy, competence, and relatedness support to one another, transforming individual sharing into a reciprocal relational practice. While the single-session technology probe did not yield immediate gains in autonomous motivation, the significant reduction in controlled motivation confirms that instrumental support effectively mitigates the pressure often associated with structured intimacy. Collectively, these findings offer a new blueprint for relatedness technologies, guiding the design of systems that move beyond simple connectivity to cultivate the mutual and reciprocal need support essential for sustaining close relationships.

\begin{acks}
This work has received support from the Singapore University of Technology and Design (SUTD) Kickstart Initiative under Grant No. SKI 2021-04-06, and from the Ministry of Education (MOE), Singapore, under Grant No. MOE-T2EP20124-0014.
\end{acks}

\bibliographystyle{ACM-Reference-Format}
\bibliography{sample-base}

@String{Computing = "Computing" }

@String{Computer = "{IEEE} Computer" }

@String{Springer = "Springer-Verlag" }

@Article{laguardia2008selfdetermination,
  author        = "La Guardia, Jennifer G. and Patrick, Heather",
  title         = "Self-determination theory as a fundamental theory of close relationships",
  journal       = "Canadian Psychology / Psychologie canadienne",
  volume        = "49",
  number        = "3",
  year          = "2008",
  pages         = "201--209",
  doi           = "10.1037/a0012760",
  url           = "https://doi.org/10.1037/a0012760",
}

@inproceedings{wenhart2025relatedness,
author = {Wenhart, Christiane and Ringfort-Felner, Ronda and Wallbaum, Torben and Amidi, Maryam and Albers, Ruben and Hassenzahl, Marc},
title = {Relatedness Technologies: An Online Compendium and Systematic Review},
year = {2025},
isbn = {9798400713941},
publisher = {Association for Computing Machinery},
address = {New York, NY, USA},
url = {https://doi.org/10.1145/3706598.3714260},
doi = {10.1145/3706598.3714260},
booktitle = {Proceedings of the 2025 CHI Conference on Human Factors in Computing Systems},
articleno = {377},
numpages = {48},
location = {
},
series = {CHI '25}
}

@Article{ryan2000selfdetermination,
  author        = "Ryan, Richard M and Deci, Edward L",
  title         = "Self-determination theory and the facilitation of intrinsic motivation, social development, and well-being",
  journal       = "American psychologist",
  volume        = "55",
  number        = "1",
  year          = "2000",
  pages         = "68--78",
  doi           = "10.1037//0003-066x.55.1.68",
  url           = "https://doi.org/10.1037/0003-066X.55.1.68",
}

@Article{gaine2009theunique,
  author        = "Gaine, Graham S and La Guardia, Jennifer G.",
  title         = "The unique contributions of motivations to maintain a relationship and motivations toward relational activities to relationship well-being",
  journal       = "Motivation and Emotion",
  volume        = "33",
  number        = "2",
  year          = "2009",
  pages         = "184--202",
  doi           = "10.1007/s11031-009-9120-x",
  url           = "https://doi.org/10.1007/s11031-009-9120-x",
}

@Article{Li2023systematic,
  author        = "Li, Han and Zhang, Renwen and Lee, Yi-Chieh and Kraut, Robert E. and Mohr, David C.",
  title         = "Systematic review and meta-analysis of AI-based conversational agents for promoting mental health and well-being",
  journal       = "npj Digital Medicine",
  volume        = "6",
  number        = "1",
  year          = "2023",
  pages         = "236",
  doi           = "10.1038/s41746-023-00979-5",
  url           = "https://doi.org/10.1038/s41746-023-00979-5",
}

@online{jones2015thirtysix,
  author  = {Jones, Daniel},
  title   = {The 36 Questions That Lead to Love},
  year    = {2015},
  month   = jan,
  day     = {9},
  url     = {https://www.nytimes.com/2015/01/09/style/no-37-big-wedding-or-small.html},
  note    = {Accessed: 2025-05-27}
}

@Article{finkenauer2015toknowyou,
  author        = "Finkenauer, Catrin and Buyukcan-Tetik, Asuman",
  title         = "To know you is to feel intimate with you: Felt knowledge is rooted in disclosure, solicitation, and intimacy",
  journal       = "Family Science",
  volume        = "6",
  number        = "1",
  year          = "2015",
  pages         = "109--118",
  doi           = "10.1080/19424620.2015.1082012",
  url           = "https://doi.org/10.1080/19424620.2015.1082012",
}

@Article{Mackinnon1999Ashortform,
  author        = "Andrew Mackinnon and Anthony F Jorm and Helen Christensen and Ailsa E Korten and Patricia A Jacomb and Bryan Rodgers",
  title         = "A short form of the Positive and Negative Affect Schedule: evaluation of factorial validity and invariance across demographic variables in a community sample",
  journal       = "Personality and Individual Differences",
  volume        = "27",
  number        = "3",
  year          = "1999",
  pages         = "405--416",
  doi           = "https://doi.org/10.1016/S0191-8869(98)00251-7",
  url           = "https://www.sciencedirect.com/science/article/pii/S0191886998002517",
}

@Article{watson1988development,
  author        = "Watson, David, Lee Anna Clark and Auke Tellegen",
  title         = "Development and validation of brief measures of positive and negative affect: The PANAS scales",
  journal       = "Journal of Personality and Social Psychology",
  volume        = "54",
  number        = "6",
  month         = "",
  year          = "1988",
  pages         = "1063--1070",
  doi           = "10.1037/0022-3514.54.6.1063",
  url           = "https://doi.org/10.1037/0022-3514.54.6.1063",
  note          = "",
}

@inproceedings{wagener2022mood,
author = {Wagener, Nadine and Niess, Jasmin and Rogers, Yvonne and Sch\"{o}ning, Johannes},
title = {Mood Worlds: A Virtual Environment for Autonomous Emotional Expression},
year = {2022},
isbn = {9781450391573},
publisher = {Association for Computing Machinery},
address = {New York, NY, USA},
url = {https://doi.org/10.1145/3491102.3501861},
doi = {10.1145/3491102.3501861},
booktitle = {Proceedings of the 2022 CHI Conference on Human Factors in Computing Systems},
articleno = {22},
numpages = {16},
location = {New Orleans, LA, USA},
series = {CHI '22}
}

@Article{alarith2022time,
  author        = "Alarith Uhde and Marc Hassenzahl",
  title         = "Time perspectives in technology-mediated reminiscing: effects of basic design decisions on subjective well-being",
  journal       = "Human–Computer Interaction",
  volume        = "37",
  number        = "2",
  month         = "",
  year          = "2022",
  pages         = "117--149",
  doi           = "10.1080/07370024.2021.1913415",
  url           = "https://doi.org/10.1080/07370024.2021.1913415",
  note          = "",
}

@inproceedings{wagener2023selvreflect,
author = {Wagener, Nadine and Reicherts, Leon and Zargham, Nima and Bart\l{}omiejczyk, Natalia and Scott, Ava Elizabeth and Wang, Katherine and Bentvelzen, Marit and Stefanidi, Evropi and Mildner, Thomas and Rogers, Yvonne and Niess, Jasmin},
title = {SelVReflect: A Guided VR Experience Fostering Reflection on Personal Challenges},
year = {2023},
isbn = {9781450394215},
publisher = {Association for Computing Machinery},
address = {New York, NY, USA},
url = {https://doi.org/10.1145/3544548.3580763},
doi = {10.1145/3544548.3580763},
booktitle = {Proceedings of the 2023 CHI Conference on Human Factors in Computing Systems},
articleno = {323},
numpages = {17},
location = {Hamburg, Germany},
series = {CHI '23}
}

@inproceedings{schlagowski2023wish,
author = {Schlagowski, Ruben and Nazarenko, Dariia and Can, Yekta and Gupta, Kunal and Mertes, Silvan and Billinghurst, Mark and Andr\'{e}, Elisabeth},
title = {Wish You Were Here: Mental and Physiological Effects of Remote Music Collaboration in Mixed Reality},
year = {2023},
isbn = {9781450394215},
publisher = {Association for Computing Machinery},
address = {New York, NY, USA},
url = {https://doi.org/10.1145/3544548.3581162},
doi = {10.1145/3544548.3581162},
booktitle = {Proceedings of the 2023 CHI Conference on Human Factors in Computing Systems},
articleno = {102},
numpages = {16},
location = {Hamburg, Germany},
series = {CHI '23}
}

@Article{Bostic2000Avalidation,
  author        = "Bostic, Terence J. and McGartland Rubio, Doris and Hood, Mark",
  title         = "A Validation of the Subjective Vitality Scale Using Structural Equation Modeling",
  journal       = "Social Indicators Research",
  volume        = "52",
  number        = "3",
  year          = "2000",
  pages         = "313--324",
  doi           = "10.1023/A:1007136110218",
  url           = "https://doi.org/10.1023/A:1007136110218",
}

@Article{rosenberg1965rosenberg,
  author        = "Rosenberg, Morris",
  title         = "Rosenberg self-esteem scale (RSE)",
  journal       = "Acceptance and commitment therapy. Measures package",
  volume        = "61",
  number        = "52",
  year          = "1965",
  pages         = "18",
  doi           = "",
  url           = "",
}

@Article{monteiro2022theefficient,
  author        = "Monteiro, Renan P. and Coelho, Gabriel Lins de Holanda and Hanel, Paul H. P. and de Medeiros, Emerson Diógenes and da Silva, Phillip Dyamond Gomes",
  title         = "The Efficient Assessment of Self-Esteem: Proposing the Brief Rosenberg Self-Esteem Scale",
  journal       = "Applied Research in Quality of Life",
  volume        = "17",
  number        = "2",
  year          = "2022",
  pages         = "931--947",
  doi           = "10.1007/s11482-021-09936-4",
  url           = "https://doi.org/10.1007/s11482-021-09936-4",
}

@Article{aron1992inclusion,
  author        = "Aron, Arthur, Elaine N. Aron and Danny Smollan",
  title         = "Inclusion of Other in the Self Scale and the structure of interpersonal closeness",
  journal       = "Journal of Personality and Social Psychology",
  volume        = "63",
  number        = "4",
  month         = "",
  year          = "1992",
  pages         = "596--612",
  doi           = "10.1037/0022-3514.63.4.596",
  url           = "https://doi.org/10.1037/0022-3514.63.4.596",
  note          = "",
}

@inproceedings{an2024emowear,
author = {An, Pengcheng and Zhu, Jiawen Stefanie and Zhang, Zibo and Yin, Yifei and Ma, Qingyuan and Yan, Che and Du, Linghao and Zhao, Jian},
title = {EmoWear: Exploring Emotional Teasers for Voice Message Interaction on Smartwatches},
year = {2024},
isbn = {9798400703300},
publisher = {Association for Computing Machinery},
address = {New York, NY, USA},
url = {https://doi.org/10.1145/3613904.3642101},
doi = {10.1145/3613904.3642101},
booktitle = {Proceedings of the 2024 CHI Conference on Human Factors in Computing Systems},
articleno = {279},
numpages = {16},
location = {Honolulu, HI, USA},
series = {CHI '24}
}

@inproceedings{zhang2023icebreaking,
author = {Zhang, Alex Wuqi and Lin, Ting-Han and Zhao, Xuan and Sebo, Sarah},
title = {Ice-Breaking Technology: Robots and Computers Can Foster Meaningful Connections between Strangers through In-Person Conversations},
year = {2023},
isbn = {9781450394215},
publisher = {Association for Computing Machinery},
address = {New York, NY, USA},
url = {https://doi.org/10.1145/3544548.3581135},
doi = {10.1145/3544548.3581135},
booktitle = {Proceedings of the 2023 CHI Conference on Human Factors in Computing Systems},
articleno = {273},
numpages = {14},
location = {Hamburg, Germany},
series = {CHI '23}
}

@Article{laguardia2000within,
  author        = "La Guardia, Jennifer G. and Ryan, Richard M. and Couchman, Charles E.and Deci, Edward L.",
  title         = "Within-person variation in security of attachment: A self-determination theory perspective on attachment, need fulfillment, and well-being",
  journal       = "Journal of Personality and Social Psychology",
  volume        = "79",
  number        = "3",
  year          = "2000",
  pages         = "367--384",
  doi           = "10.1037/0022-3514.79.3.367",
  url           = "",
}

@article{JSSv067i01,
 title={Fitting Linear Mixed-Effects Models Using lme4},
 volume={67},
 url={https://www.jstatsoft.org/index.php/jss/article/view/v067i01},
 doi={10.18637/jss.v067.i01},
 number={1},
 journal={Journal of Statistical Software},
 author={Bates, Douglas and Mächler, Martin and Bolker, Ben and Walker, Steve},
 year={2015},
 pages={1–48}
}

@inproceedings{wobbrock2011thealigned,
author = {Wobbrock, Jacob O. and Findlater, Leah and Gergle, Darren and Higgins, James J.},
title = {The aligned rank transform for nonparametric factorial analyses using only anova procedures},
year = {2011},
isbn = {9781450302289},
publisher = {Association for Computing Machinery},
address = {New York, NY, USA},
url = {https://doi.org/10.1145/1978942.1978963},
doi = {10.1145/1978942.1978963},
booktitle = {Proceedings of the SIGCHI Conference on Human Factors in Computing Systems},
pages = {143–146},
numpages = {4},
location = {Vancouver, BC, Canada},
series = {CHI '11}
}

@article{searle1980population,
author = {S. R. Searle and F. M. Speed and G. A. Milliken},
title = {Population Marginal Means in the Linear Model: An Alternative to Least Squares Means},
journal = {The American Statistician},
volume = {34},
number = {4},
pages = {216--221},
year = {1980},
publisher = {ASA Website},
doi = {10.1080/00031305.1980.10483031},
URL = {https://www.tandfonline.com/doi/abs/10.1080/00031305.1980.10483031}
}

@article{kruskal1952william,
author = {William H. Kruskal and W. Allen Wallis},
title = {Use of Ranks in One-Criterion Variance Analysis},
journal = {Journal of the American Statistical Association},
volume = {47},
number = {260},
pages = {583--621},
year = {1952},
publisher = {ASA Website},
doi = {10.1080/01621459.1952.10483441},
URL = {https://www.tandfonline.com/doi/abs/10.1080/01621459.1952.10483441}
}

@Book{derlega1993selfdisclosure,
  author =       "Derlega, Valerian J. and Metts, Sandra and Petronio, Sandra and Margulis, Stephen T.",
  title =        "Self-disclosure",
  publisher =    "Sage Publications, Inc",
  year =         "1993",
  address =      "Thousand Oaks, CA, US",
  edition =      "",
  editor =       "",
  volume =       "",
  number =       "",
  series =       "",
  month =        "",
  note =         "",
}

@Article{meeks1998communication,
  author        = "Meeks, Brenda S. and Hendrick, Susan S. and Hendrick, Clyde",
  title         = "Communication, love and relationship satisfaction",
  journal       = "Journal of Social and Personal Relationships",
  volume        = "15",
  number        = "6",
  year          = "1998",
  pages         = "755-773",
  doi           = "10.1177/0265407598156003",
  url           = "",
}

@article{sprecher1987theeffects,
author = {Susan Sprecher},
title ={The Effects of Self-Disclosure Given and Received on Affection for an Intimate Partner and Stability of the Relationship},
journal = {Journal of Social and Personal Relationships},
volume = {4},
number = {2},
pages = {115-127},
year = {1987},
doi = {10.1177/0265407587042001},
URL = {https://doi.org/10.1177/0265407587042001}
}

@Article{laurenceau1998intimacy,
  author        = "Laurenceau, Jean-Philippe and Barrett, Lisa Feldman and Pietromonaco, Paula R",
  title         = "Intimacy as an interpersonal process: the importance of self-disclosure, partner disclosure, and perceived partner responsiveness in interpersonal exchanges",
  journal       = "Journal of Personality and Social Psychology",
  volume        = "74",
  number        = "5",
  month         = "",
  year          = "1998",
  pages         = "1238",
  doi           = "10.1037//0022-3514.74.5.1238",
  url           = "",
  note          = "",
}

@book{altman1973social,
  author        = "Altman, Irwin and Taylor, Dalmas A",
  title         = "Social penetration: The development of interpersonal relationships",
  volume        = "",
  number        = "",
  month         = "",
  year          = "1973",
  publisher     = "Holt, Rinehart \& Winston",
  address       = "New York",
  doi           = "",
  url           = "",
  note          = "",
}

@misc{reis1988intimacy,
  author        = "H. T. Reis and P. Shaver",
  title         = "Intimacy as an interpersonal process",
  booktitle     = "Handbook of personal relationships: Theory, research and interventions.",
  publisher     = "John Wiley & Sons",
  number        = "",
  month         = "",
  year          = "1988",
  pages         = "367--389",
  doi           = "",
  url           = "",
  note          = "",
}

@Article{sprecher2004selfdisclosure,
  author        = "Sprecher, Susan and Hendrick, Susan S.",
  title         = "Self-Disclosure in Intimate Relationships: Associations With Individual and Relationship Characteristics Over Time",
  journal       = "Journal of Social and Clinical Psychology",
  volume        = "23",
  number        = "6",
  year          = "2004",
  pages         = "857--877",
  doi           = "10.1521/jscp.23.6.857.54803",
  url           = "",
}

@article{ayres1979uncertainty,
author = {Joe Ayres},
title = {Uncertainty and social penetration theory expectations about relationship communication: A comparative test},
journal = {Western Journal of Speech Communication},
volume = {43},
number = {3},
pages = {192--200},
year = {1979},
publisher = {Routledge},
doi = {10.1080/10570317909373968},
URL = {https://doi.org/10.1080/10570317909373968}
}

@article{sprecher2013takingturns,
title = {Taking turns: Reciprocal self-disclosure promotes liking in initial interactions},
journal = {Journal of Experimental Social Psychology},
volume = {49},
number = {5},
pages = {860-866},
year = {2013},
issn = {0022-1031},
doi = {https://doi.org/10.1016/j.jesp.2013.03.017},
url = {https://www.sciencedirect.com/science/article/pii/S002210311300070X},
author = {Susan Sprecher and Stanislav Treger and Joshua D. Wondra and Nicole Hilaire and Kevin Wallpe}
}

@Article{collins1994selfdisclosure,
  author        = "Collins, Nancy L. and Miller, Lynn Carol",
  title         = "Self-disclosure and liking: A meta-analytic review",
  journal       = "Psychological Bulletin",
  volume        = "116",
  number        = "3",
  year          = "1994",
  pages         = "457--475",
  doi           = "10.1037/0033-2909.116.3.457",
  url           = "",
}

@Article{laurenceau2005theinterpersonal,
  author        = "Laurenceau, J. P. and Barrett, L. F. and Rovine, M. J.",
  title         = "The interpersonal process model of intimacy in marriage: a daily-diary and multilevel modeling approach",
  journal       = "Journal of Family Psychology",
  volume        = "19",
  number        = "2",
  year          = "2005",
  pages         = "314--323",
  doi           = "10.1037/0893-3200.19.2.314",
  url           = "",
}

@Article{baumeister1995theneed,
  author        = "Baumeister, Roy F. and Leary, Mark R.",
  title         = "The need to belong: Desire for interpersonal attachments as a fundamental human motivation",
  journal       = "Psychological Bulletin",
  volume        = "117",
  number        = "3",
  year          = "1995",
  pages         = "497--529",
  doi           = "10.1037/0033-2909.117.3.497",
  url           = "",
}

@article{ryan2000thedarker,
author = {Richard M. Ryan and Edward L. Deci},
title = {The Darker and Brighter Sides of Human Existence: Basic Psychological Needs as a Unifying Concept},
journal = {Psychological Inquiry},
volume = {11},
number = {4},
pages = {319--338},
year = {2000},
publisher = {Routledge},
doi = {10.1207/S15327965PLI1104\_03},
URL = {https://doi.org/10.1207/S15327965PLI1104_03}
}

@Inbook{ryan1995autonomy,
  author =       "Ryan, Richard M. and Deci, Edward L. and Grolnick, Wendy S.",
  editor =       "",
  title =        "Autonomy, relatedness, and the self: Their relation to development and psychopathology",
  booktitle =    "Developmental psychopathology, Vol. 1: Theory and methods.",
  subtitle =     "",
  series =       "",
  year =         "1995",
  address =      "Oxford, England",
  publisher =    "John Wiley \& Sons",
  pages =        "618--655"
}

@Article{patrick2007therole,
  author        = "Patrick, H. and Knee, C. R. and Canevello, A. and Lonsbary, C.",
  title         = "The role of need fulfillment in relationship functioning and well-being: a self-determination theory perspective",
  journal       = "J Pers Soc Psychol",
  volume        = "92",
  number        = "3",
  year          = "2007",
  pages         = "434--457",
  doi           = "10.1037/0022-3514.92.3.434",
  url           = "",
}

@inproceedings{la2007end,
  title={At the end of the day, where do we stand?: Daily emotional engagement and need satisfaction within couples},
  author={La Guardia, JG},
  booktitle={Annual meeting of the Society for Personality and Social Psychology, Memphis, TN},
  year={2007}
}

@article{deci2000thewhat,
author = {Edward L. Deci and Richard M. Ryan},
title = {The "What" and "Why" of Goal Pursuits: Human Needs and the Self-Determination of Behavior},
journal = {Psychological Inquiry},
volume = {11},
number = {4},
pages = {227--268},
year = {2000},
publisher = {Routledge},
doi = {10.1207/S15327965PLI1104\_01},
URL = {https://doi.org/10.1207/S15327965PLI1104_01}
}

@book{deci2013intrinsic,
  title={Intrinsic motivation and self-determination in human behavior},
  author={Deci, Edward L and Ryan, Richard M},
  year={1985},
  publisher={Kluwer Academic/Plenum Publishers},
  address = {New York, NY, USA}
}

@Article{ignatius2007factors,
  author        = "Ignatius, Emmi and Kokkonen, Marja",
  title         = "Factors contributing to verbal self-disclosure",
  journal       = "Nordic Psychology",
  volume        = "59",
  number        = "4",
  year          = "2007",
  pages         = "362-391",
  doi           = "10.1027/1901-2276.59.4.362",
  url           = "",
}

@article{duck2006someevident,
    author = {Duck, Steve and Rutt, Deborah J. and Hurst, Margaret Hoy and Strejc, Heather},
    title = {Some Evident Truths About Conversations in Everyday Relationships: All Communications Are Not Created Equal},
    journal = {Human Communication Research},
    volume = {18},
    number = {2},
    pages = {228-267},
    year = {2006},
    month = {03},
    issn = {0360-3989},
    doi = {10.1111/j.1468-2958.1991.tb00545.x},
    url = {https://doi.org/10.1111/j.1468-2958.1991.tb00545.x},
    eprint = {https://academic.oup.com/hcr/article-pdf/18/2/228/22342358/jhumcom0228.pdf},
}

@Article{ryjova2024aday,
  author        = "Ryjova, Y. and Gold, A. I. and Timmons, A. C. and Han, S. C. and Chaspari, T. and Pettit, C. and Kim, Y. and Beale, A. and Kazmierski, K. F. M. andMargolin, G.",
  title         = "A day in the life: Couples' everyday communication and subsequent relationship outcomes",
  journal       = "J Fam Psychol",
  volume        = "38",
  number        = "3",
  year          = "2024",
  pages         = "453--465",
  doi           = "10.1037/fam0001180",
  url           = "",
}

@article{jess2005mapping,
author = {Jess K. Alberts and Christina G. Yoshimura and Michael Rabby and Rose Loschiavo},
title ={Mapping the topography of couples’ daily conversation},
journal = {Journal of Social and Personal Relationships},
volume = {22},
number = {3},
pages = {299-322},
year = {2005},
doi = {10.1177/0265407505050941},
URL = {https://doi.org/10.1177/0265407505050941}
}

@inproceedings{ballou2022self,
author = {Ballou, Nick and Deterding, Sebastian and Tyack, April and Mekler, Elisa D and Calvo, Rafael A and Peters, Dorian and Villalobos-Z\'{u}\~{n}iga, Gabriela and Turkay, Selen},
title = {Self-Determination Theory in HCI: Shaping a Research Agenda},
year = {2022},
isbn = {9781450391566},
publisher = {Association for Computing Machinery},
address = {New York, NY, USA},
url = {https://doi.org/10.1145/3491101.3503702},
doi = {10.1145/3491101.3503702},
booktitle = {Extended Abstracts of the 2022 CHI Conference on Human Factors in Computing Systems},
articleno = {113},
numpages = {6},
location = {New Orleans, LA, USA},
series = {CHI EA '22}
}

@inproceedings{wang2023chatterbox,
author = {Wang, Wei-Lu and Haqq, Derek and Saaty, Morva and Cao, Yusheng and Fan, Jixiang and Patel, Jaitun V. and McCrickard, D. Scott},
title = {Chatterbox Opener: A Game to Support Healthy Communication and Relationships},
year = {2023},
isbn = {9798400700293},
publisher = {Association for Computing Machinery},
address = {New York, NY, USA},
url = {https://doi.org/10.1145/3573382.3616074},
doi = {10.1145/3573382.3616074},
booktitle = {Companion Proceedings of the Annual Symposium on Computer-Human Interaction in Play},
pages = {43–48},
numpages = {6},
location = {Stratford, ON, Canada},
series = {CHI PLAY Companion '23}
}

@article{zheng2021pocketbot,
author = {Zheng, Qingxiao and Markazi, Daniela M. and Tang, Yiliu and Huang, Yun},
title = {"PocketBot Is Like a Knock-On-the-Door!": Designing a Chatbot to Support Long-Distance Relationships},
year = {2021},
issue_date = {October 2021},
publisher = {Association for Computing Machinery},
address = {New York, NY, USA},
volume = {5},
number = {CSCW2},
url = {https://doi.org/10.1145/3479589},
doi = {10.1145/3479589},
journal = {Proc. ACM Hum.-Comput. Interact.},
month = oct,
articleno = {445},
numpages = {28}
}

@article{kelly2018letter,
author = {Kelly, Ryan and Gooch, Daniel and Watts, Leon},
title = { 'It's More Like a Letter': An Exploration of Mediated Conversational Effort in Message Builder},
year = {2018},
issue_date = {November 2018},
publisher = {Association for Computing Machinery},
address = {New York, NY, USA},
volume = {2},
number = {CSCW},
url = {https://doi.org/10.1145/3274356},
doi = {10.1145/3274356},
journal = {Proc. ACM Hum.-Comput. Interact.},
month = nov,
articleno = {87},
numpages = {23}
}

@inproceedings{kelly2018rewriting,
author = {Kelly, Ryan and Gooch, Daniel and Watts, Leon},
title = {Designing for reflection on sender effort in close personal communication},
year = {2018},
isbn = {9781450361880},
publisher = {Association for Computing Machinery},
address = {New York, NY, USA},
url = {https://doi.org/10.1145/3292147.3292174},
doi = {10.1145/3292147.3292174},
booktitle = {Proceedings of the 30th Australian Conference on Computer-Human Interaction},
pages = {314–325},
numpages = {12},
location = {Melbourne, Australia},
series = {OzCHI '18}
}

@inproceedings{tang2013homeproxy,
author = {Tang, John C. and Xiao, Robert and Hoff, Aaron and Venolia, Gina and Therien, Patrick and Roseway, Asta},
title = {HomeProxy: exploring a physical proxy for video communication in the home},
year = {2013},
isbn = {9781450318990},
publisher = {Association for Computing Machinery},
address = {New York, NY, USA},
url = {https://doi.org/10.1145/2470654.2466175},
doi = {10.1145/2470654.2466175},
booktitle = {Proceedings of the SIGCHI Conference on Human Factors in Computing Systems},
pages = {1339–1342},
numpages = {4},
location = {Paris, France},
series = {CHI '13}
}

@inproceedings{raju2021pvpix,
author = {Raju, Dani Kalarikalayil and Seunarine, Krishna and Reitmaier, Thomas and Thomas, Gethin and Meena, Yogesh Kumar and Zhang, Chi and Pockett, Adam and Pearson, Jennifer and Robinson, Simon and Carnie, Matt and Sahoo, Deepak Ranjan and Jones, Matt},
title = {PV-Pix: Slum Community Co-design of Self-Powered Deformable Smart Messaging Materials},
year = {2021},
isbn = {9781450380966},
publisher = {Association for Computing Machinery},
address = {New York, NY, USA},
url = {https://doi.org/10.1145/3411764.3445661},
doi = {10.1145/3411764.3445661},
booktitle = {Proceedings of the 2021 CHI Conference on Human Factors in Computing Systems},
articleno = {304},
numpages = {14},
location = {Yokohama, Japan},
series = {CHI '21}
}

@inproceedings{dziabiola2022qude,
author = {Dziabiola, Marta and Steiner, Robert and Vetter, Ralf and Norskov, Daniel and Smit, Doroth\'{e}},
title = {Qude: Exploring Tactile Code in Long-Distance Relationships},
year = {2022},
isbn = {9781450391474},
publisher = {Association for Computing Machinery},
address = {New York, NY, USA},
url = {https://doi.org/10.1145/3490149.3505583},
doi = {10.1145/3490149.3505583},
booktitle = {Proceedings of the Sixteenth International Conference on Tangible, Embedded, and Embodied Interaction},
articleno = {79},
numpages = {7},
location = {Daejeon, Republic of Korea},
series = {TEI '22}
}

@inproceedings{baishya2017inyoureyes,
author = {Baishya, Uddipana and Neustaedter, Carman},
title = {In Your Eyes: Anytime, Anywhere Video and Audio Streaming for Couples},
year = {2017},
isbn = {9781450343350},
publisher = {Association for Computing Machinery},
address = {New York, NY, USA},
url = {https://doi.org/10.1145/2998181.2998200},
doi = {10.1145/2998181.2998200},
booktitle = {Proceedings of the 2017 ACM Conference on Computer Supported Cooperative Work and Social Computing},
pages = {84–97},
numpages = {14},
location = {Portland, Oregon, USA},
series = {CSCW '17}
}

@inproceedings{yarosh2013almost,
author = {Yarosh, Svetlana and Tang, Anthony and Mokashi, Sanika and Abowd, Gregory D.},
title = {"almost touching": parent-child remote communication using the sharetable system},
year = {2013},
isbn = {9781450313315},
publisher = {Association for Computing Machinery},
address = {New York, NY, USA},
url = {https://doi.org/10.1145/2441776.2441798},
doi = {10.1145/2441776.2441798},
booktitle = {Proceedings of the 2013 Conference on Computer Supported Cooperative Work},
pages = {181–192},
numpages = {12},
location = {San Antonio, Texas, USA},
series = {CSCW '13}
}

@inproceedings{lenz2016four,
author = {Lenz, Eva and Hassenzahl, Marc and Adamow, Wasili and Beedgen, Patrick and Kohler, Kirstin and Schneider, Thies},
title = {Four Stories About Feeling Close Over A Distance},
year = {2016},
isbn = {9781450335829},
publisher = {Association for Computing Machinery},
address = {New York, NY, USA},
url = {https://doi.org/10.1145/2839462.2856523},
doi = {10.1145/2839462.2856523},
booktitle = {Proceedings of the TEI '16: Tenth International Conference on Tangible, Embedded, and Embodied Interaction},
pages = {494–499},
numpages = {6},
location = {Eindhoven, Netherlands},
series = {TEI '16}
}

@inproceedings{heshmat2020family,
author = {Heshmat, Yasamin and Neustaedter, Carman and McCaffrey, Kyle and Odom, William and Wakkary, Ron and Yang, Zikun},
title = {FamilyStories: Asynchronous Audio Storytelling for Family Members Across Time Zones},
year = {2020},
isbn = {9781450367080},
publisher = {Association for Computing Machinery},
address = {New York, NY, USA},
url = {https://doi.org/10.1145/3313831.3376486},
doi = {10.1145/3313831.3376486},
booktitle = {Proceedings of the 2020 CHI Conference on Human Factors in Computing Systems},
pages = {1–14},
numpages = {14},
location = {Honolulu, HI, USA},
series = {CHI '20}
}

@article{podlubny2017synchronous,
author = {Podlubny, Martin and Rooksby, John and Rost, Mattias and Chalmers, Matthew},
title = {Synchronous Text Messaging: A Field Trial of Curtains Messenger},
year = {2017},
issue_date = {November 2017},
publisher = {Association for Computing Machinery},
address = {New York, NY, USA},
volume = {1},
number = {CSCW},
url = {https://doi.org/10.1145/3134721},
doi = {10.1145/3134721},
journal = {Proc. ACM Hum.-Comput. Interact.},
month = dec,
articleno = {86},
numpages = {20}
}

@inproceedings{angelini2015multisensory,
author = {Angelini, Leonardo and Caon, Maurizio and Couture, Nadine and Khaled, Omar Abou and Mugellini, Elena},
title = {The multisensory interactive window: immersive experiences for the elderly},
year = {2015},
isbn = {9781450335751},
publisher = {Association for Computing Machinery},
address = {New York, NY, USA},
url = {https://doi.org/10.1145/2800835.2806209},
doi = {10.1145/2800835.2806209},
booktitle = {Adjunct Proceedings of the 2015 ACM International Joint Conference on Pervasive and Ubiquitous Computing and Proceedings of the 2015 ACM International Symposium on Wearable Computers},
pages = {963–968},
numpages = {6},
location = {Osaka, Japan},
series = {UbiComp/ISWC'15 Adjunct}
}

@inproceedings{misawa2012mapetite,
author = {Misawa, Kana and Ishiguro, Yoshio and Rekimoto, Jun},
title = {Ma petite ch\'{e}rie: what are you looking at? a small telepresence system to support remote collaborative work for intimate communication},
year = {2012},
isbn = {9781450310772},
publisher = {Association for Computing Machinery},
address = {New York, NY, USA},
url = {https://doi.org/10.1145/2160125.2160142},
doi = {10.1145/2160125.2160142},
booktitle = {Proceedings of the 3rd Augmented Human International Conference},
articleno = {17},
numpages = {5},
location = {Meg\`{e}ve, France},
series = {AH '12}
}

@inproceedings{vaanen2011carpe,
author = {V\"{a}\"{a}n\"{a}nen-Vainio-Mattila, Kaisa and Suhonen, Katja and Gonsalves, Tina and Schrader, Martin and J\"{a}rvenp\"{a}\"{a}, Toni},
title = {Carpe diem: exploring user experience and intimacy in eye-based video conferencing},
year = {2011},
isbn = {9781450310963},
publisher = {Association for Computing Machinery},
address = {New York, NY, USA},
url = {https://doi.org/10.1145/2107596.2107610},
doi = {10.1145/2107596.2107610},
booktitle = {Proceedings of the 10th International Conference on Mobile and Ubiquitous Multimedia},
pages = {113–122},
numpages = {10},
location = {Beijing, China},
series = {MUM '11}
}

@book{calvo2014positive,
author = {Calvo, Rafael A. and Peters, Dorian},
title = {Positive Computing: Technology for Well-Being and Human Potential},
year = {2014},
isbn = {0262028158},
publisher = {The MIT Press},
address = {Cambridge, MA, USA}
}

@article{kim2019love,
author = {Kim, Taewook and Lee, Jung Soo and Peng, Zhenhui and Ma, Xiaojuan},
title = {Love in Lyrics: An Exploration of Supporting Textual Manifestation of Affection in Social Messaging},
year = {2019},
issue_date = {November 2019},
publisher = {Association for Computing Machinery},
address = {New York, NY, USA},
volume = {3},
number = {CSCW},
url = {https://doi.org/10.1145/3359181},
doi = {10.1145/3359181},
journal = {Proc. ACM Hum.-Comput. Interact.},
month = nov,
articleno = {79},
numpages = {27}
}

@inproceedings{brueckner2018empathy,
author = {Brueckner, Sophia},
title = {Empathy amulet: a wearable to connect with strangers},
year = {2018},
isbn = {9781450359672},
publisher = {Association for Computing Machinery},
address = {New York, NY, USA},
url = {https://doi.org/10.1145/3267242.3267301},
doi = {10.1145/3267242.3267301},
booktitle = {Proceedings of the 2018 ACM International Symposium on Wearable Computers},
pages = {248–253},
numpages = {6},
location = {Singapore, Singapore},
series = {ISWC '18}
}

@article{buschek2018personal,
author = {Buschek, Daniel and Hassib, Mariam and Alt, Florian},
title = {Personal Mobile Messaging in Context: Chat Augmentations for Expressiveness and Awareness},
year = {2018},
issue_date = {August 2018},
publisher = {Association for Computing Machinery},
address = {New York, NY, USA},
volume = {25},
number = {4},
issn = {1073-0516},
url = {https://doi.org/10.1145/3201404},
doi = {10.1145/3201404},
journal = {ACM Trans. Comput.-Hum. Interact.},
month = aug,
articleno = {23},
numpages = {33}
}

@inproceedings{hansen2020thedelay,
author = {Hansen, Derek L. and Hughes, Amanda Lee and Cram, Sophie and Harker, Austin Bond and Ashton, Brinnley and Hirschi, Karli and Dorton, Ben and Bothwell, Nate and Stevens, Ashley},
title = {The DELAY Framework: Designing for Extended LAtencY},
year = {2020},
isbn = {9781450367080},
publisher = {Association for Computing Machinery},
address = {New York, NY, USA},
url = {https://doi.org/10.1145/3313831.3376826},
doi = {10.1145/3313831.3376826},
booktitle = {Proceedings of the 2020 CHI Conference on Human Factors in Computing Systems},
pages = {1–13},
numpages = {13},
location = {Honolulu, HI, USA},
series = {CHI '20}
}

@article{liu2019animo,
author = {Liu, Fannie and Esparza, Mario and Pavlovskaia, Maria and Kaufman, Geoff and Dabbish, Laura and Monroy-Hern\'{a}ndez, Andr\'{e}s},
title = {Animo: Sharing Biosignals on a Smartwatch for Lightweight Social Connection},
year = {2019},
issue_date = {March 2019},
publisher = {Association for Computing Machinery},
address = {New York, NY, USA},
volume = {3},
number = {1},
url = {https://doi.org/10.1145/3314405},
doi = {10.1145/3314405},
journal = {Proc. ACM Interact. Mob. Wearable Ubiquitous Technol.},
month = mar,
articleno = {18},
numpages = {19}
}

@inproceedings{heshmat2017connecting,
author = {Heshmat, Yasamin and Neustaedter, Carman and Yang, Lillian and Schiphorst, Thecla},
title = {Connecting Family Members Across Time Through Shared Media},
year = {2017},
isbn = {9781450346566},
publisher = {Association for Computing Machinery},
address = {New York, NY, USA},
url = {https://doi.org/10.1145/3027063.3053205},
doi = {10.1145/3027063.3053205},
booktitle = {Proceedings of the 2017 CHI Conference Extended Abstracts on Human Factors in Computing Systems},
pages = {2630–2637},
numpages = {8},
location = {Denver, Colorado, USA},
series = {CHI EA '17}
}

@inproceedings{morris2017augmenting,
author = {Morris, Margaret E. and Carmean, Douglas M. and Minyaylov, Artem and Ceze, Luis},
title = {Augmenting Interpersonal Communication through Connected Lighting},
year = {2017},
isbn = {9781450346566},
publisher = {Association for Computing Machinery},
address = {New York, NY, USA},
url = {https://doi.org/10.1145/3027063.3053141},
doi = {10.1145/3027063.3053141},
booktitle = {Proceedings of the 2017 CHI Conference Extended Abstracts on Human Factors in Computing Systems},
pages = {1924–1931},
numpages = {8},
location = {Denver, Colorado, USA},
series = {CHI EA '17}
}

@inproceedings{liu2021significant,
author = {Liu, Fannie and Park, Chunjong and Tham, Yu Jiang and Tsai, Tsung-Yu and Dabbish, Laura and Kaufman, Geoff and Monroy-Hern\'{a}ndez, Andr\'{e}s},
title = {Significant Otter: Understanding the Role of Biosignals in Communication},
year = {2021},
isbn = {9781450380966},
publisher = {Association for Computing Machinery},
address = {New York, NY, USA},
url = {https://doi.org/10.1145/3411764.3445200},
doi = {10.1145/3411764.3445200},
booktitle = {Proceedings of the 2021 CHI Conference on Human Factors in Computing Systems},
articleno = {334},
numpages = {15},
location = {Yokohama, Japan},
series = {CHI '21}
}

@article{leong2023social,
author = {Leong, Joanne and Teng, Yuanyang and Liu, Xingyu "Bruce" and Jun, Hanseul and Kratz, Sven and Tham, Yu Jiang and Monroy-Hern\'{a}ndez, Andr\'{e}s and Smith, Brian A. and Vaish, Rajan},
title = {Social Wormholes: Exploring Preferences and Opportunities for Distributed and Physically-Grounded Social Connections},
year = {2023},
issue_date = {October 2023},
publisher = {Association for Computing Machinery},
address = {New York, NY, USA},
volume = {7},
number = {CSCW2},
url = {https://doi.org/10.1145/3610208},
doi = {10.1145/3610208},
journal = {Proc. ACM Hum.-Comput. Interact.},
month = oct,
articleno = {359},
numpages = {29}
}

@article{zhang2022auggie,
author = {Zhang, Lei and Chen, Tianying and Seow, Olivia and Chong, Tim and Kratz, Sven and Tham, Yu Jiang and Monroy-Hern\'{a}ndez, Andr\'{e}s and Vaish, Rajan and Liu, Fannie},
title = {Auggie: Encouraging Effortful Communication through Handcrafted Digital Experiences},
year = {2022},
issue_date = {November 2022},
publisher = {Association for Computing Machinery},
address = {New York, NY, USA},
volume = {6},
number = {CSCW2},
url = {https://doi.org/10.1145/3555152},
doi = {10.1145/3555152},
journal = {Proc. ACM Hum.-Comput. Interact.},
month = nov,
articleno = {427},
numpages = {25}
}

@inproceedings{yang2020autobiographical,
author = {Yang, Lillian and Neustaedter, Carman},
title = {An Autobiographical Design Study of a Long Distance Relationship: When Telepresence Robots Meet Smart Home Tools},
year = {2020},
isbn = {9781450369749},
publisher = {Association for Computing Machinery},
address = {New York, NY, USA},
url = {https://doi.org/10.1145/3357236.3395467},
doi = {10.1145/3357236.3395467},
booktitle = {Proceedings of the 2020 ACM Designing Interactive Systems Conference},
pages = {129–140},
numpages = {12},
location = {Eindhoven, Netherlands},
series = {DIS '20}
}

@article{ambe2022from,
author = {Ambe, Aloha Hufana and Soro, Alessandro and Johnson, Daniel and Brereton, Margot},
title = {From Collaborative Habituation to Everyday Togetherness: A Long-Term Study of Use of the Messaging Kettle},
year = {2022},
issue_date = {February 2022},
publisher = {Association for Computing Machinery},
address = {New York, NY, USA},
volume = {29},
number = {1},
issn = {1073-0516},
url = {https://doi.org/10.1145/3470973},
doi = {10.1145/3470973},
journal = {ACM Trans. Comput.-Hum. Interact.},
month = jan,
articleno = {3},
numpages = {47}
}

@inproceedings{jarus2020always,
author = {Jarusriboonchai, Pradthana and Li, Hong and Harjuniemi, Emmi and M\"{u}ller, Heiko and H\"{a}kkil\"{a}, Jonna},
title = {Always with Me: Exploring Wearable Displays as a Lightweight Intimate Communication Channel},
year = {2020},
isbn = {9781450361071},
publisher = {Association for Computing Machinery},
address = {New York, NY, USA},
url = {https://doi.org/10.1145/3374920.3375011},
doi = {10.1145/3374920.3375011},
booktitle = {Proceedings of the Fourteenth International Conference on Tangible, Embedded, and Embodied Interaction},
pages = {771–783},
numpages = {13},
location = {Sydney NSW, Australia},
series = {TEI '20}
}

@inproceedings{gaver2023living,
author = {Gaver, William and Gaver, Frances},
title = {Living with Light Touch: An Autoethnography of a Simple Communication Device in Long-Term Use},
year = {2023},
isbn = {9781450394215},
publisher = {Association for Computing Machinery},
address = {New York, NY, USA},
url = {https://doi.org/10.1145/3544548.3580807},
doi = {10.1145/3544548.3580807},
booktitle = {Proceedings of the 2023 CHI Conference on Human Factors in Computing Systems},
articleno = {633},
numpages = {14},
location = {Hamburg, Germany},
series = {CHI '23}
}

@inproceedings{gaver2022yoyo,
author = {Gaver, William and Boucher, Andy and Brown, Dean and Chatting, David and Matsuda, Naho and Ovalle, Liliana and Sheen, Andy and Vanis, Michail},
title = {Yo–Yo Machines: Self-Build Devices that Support Social Connections During the Pandemic},
year = {2022},
isbn = {9781450391573},
publisher = {Association for Computing Machinery},
address = {New York, NY, USA},
url = {https://doi.org/10.1145/3491102.3517547},
doi = {10.1145/3491102.3517547},
booktitle = {Proceedings of the 2022 CHI Conference on Human Factors in Computing Systems},
articleno = {458},
numpages = {17},
location = {New Orleans, LA, USA},
series = {CHI '22}
}

@inproceedings{min2014biosignal,
author = {Min, Hyeryung Christine and Nam, Tek-Jin},
title = {Biosignal sharing for affective connectedness},
year = {2014},
isbn = {9781450324748},
publisher = {Association for Computing Machinery},
address = {New York, NY, USA},
url = {https://doi.org/10.1145/2559206.2581345},
doi = {10.1145/2559206.2581345},
booktitle = {CHI '14 Extended Abstracts on Human Factors in Computing Systems},
pages = {2191–2196},
numpages = {6},
location = {Toronto, Ontario, Canada},
series = {CHI EA '14}
}

@inproceedings{kelly2017demanding,
author = {Kelly, Ryan and Gooch, Daniel and Patil, Bhagyashree and Watts, Leon},
title = {Demanding by Design: Supporting Effortful Communication Practices in Close Personal Relationships},
year = {2017},
isbn = {9781450343350},
publisher = {Association for Computing Machinery},
address = {New York, NY, USA},
url = {https://doi.org/10.1145/2998181.2998184},
doi = {10.1145/2998181.2998184},
booktitle = {Proceedings of the 2017 ACM Conference on Computer Supported Cooperative Work and Social Computing},
pages = {70–83},
numpages = {14},
location = {Portland, Oregon, USA},
series = {CSCW '17}
}

@article{jiang2025remini,
author = {Jiang, Zhuoqun and Yeo, ShunYi and Seow, Wei Xuan, Donovan and Perrault, Simon Tangi},
title = {Remini: Leveraging Chatbot-Mediated Mutual Reminiscence for Promoting Positive Affect and Feeling of Connectedness among Loved Ones},
year = {2025},
issue_date = {November 2025},
publisher = {Association for Computing Machinery},
address = {New York, NY, USA},
volume = {9},
number = {7},
url = {https://doi.org/10.1145/3757650},
doi = {10.1145/3757650},
journal = {Proc. ACM Hum.-Comput. Interact.},
month = oct,
articleno = {CSCW469},
numpages = {43}
}

@inproceedings{seo2024chacha,
author = {Seo, Woosuk and Yang, Chanmo and Kim, Young-Ho},
title = {ChaCha: Leveraging Large Language Models to Prompt Children to Share Their Emotions about Personal Events},
year = {2024},
isbn = {9798400703300},
publisher = {Association for Computing Machinery},
address = {New York, NY, USA},
url = {https://doi.org/10.1145/3613904.3642152},
doi = {10.1145/3613904.3642152},
booktitle = {Proceedings of the 2024 CHI Conference on Human Factors in Computing Systems},
articleno = {903},
numpages = {20},
location = {Honolulu, HI, USA},
series = {CHI '24}
}

@inproceedings{park2021wrote,
author = {Park, SoHyun and Thieme, Anja and Han, Jeongyun and Lee, Sungwoo and Rhee, Wonjong and Suh, Bongwon},
title = {“I wrote as if I were telling a story to someone I knew.”: Designing Chatbot Interactions for Expressive Writing in Mental Health},
year = {2021},
isbn = {9781450384766},
publisher = {Association for Computing Machinery},
address = {New York, NY, USA},
url = {https://doi.org/10.1145/3461778.3462143},
doi = {10.1145/3461778.3462143},
booktitle = {Proceedings of the 2021 ACM Designing Interactive Systems Conference},
pages = {926–941},
numpages = {16},
location = {Virtual Event, USA},
series = {DIS '21}
}

@inproceedings{lee2020ihearyou,
author = {Lee, Yi-Chieh and Yamashita, Naomi and Huang, Yun and Fu, Wai},
title = {"I Hear You, I Feel You": Encouraging Deep Self-disclosure through a Chatbot},
year = {2020},
isbn = {9781450367080},
publisher = {Association for Computing Machinery},
address = {New York, NY, USA},
url = {https://doi.org/10.1145/3313831.3376175},
doi = {10.1145/3313831.3376175},
booktitle = {Proceedings of the 2020 CHI Conference on Human Factors in Computing Systems},
pages = {1–12},
numpages = {12},
location = {Honolulu, HI, USA},
series = {CHI '20}
}

@inproceedings{balaji2024designing,
author = {Balaji, Divyaa and De Bruijn, Gert-Jan and Bosse, Tibor and Ischen, Carolin and Van Der Goot, Margot and Wiers, Reinout},
title = {Designing a Couples-Based Conversational Agent to Promote Safe Sex in New, Young Couples: A User-Centred Design Approach},
year = {2024},
isbn = {9798400705113},
publisher = {Association for Computing Machinery},
address = {New York, NY, USA},
url = {https://doi.org/10.1145/3640794.3665556},
doi = {10.1145/3640794.3665556},
booktitle = {Proceedings of the 6th ACM Conference on Conversational User Interfaces},
articleno = {4},
numpages = {11},
location = {Luxembourg, Luxembourg},
series = {CUI '24}
}

@inproceedings{yuksel2023conversational,
author = {Yuksel, Berkan and Kocaballi, A. Baki},
title = {Conversational Agents to Support Couple Therapy},
year = {2023},
isbn = {9798400700248},
publisher = {Association for Computing Machinery},
address = {New York, NY, USA},
url = {https://doi.org/10.1145/3572921.3572922},
doi = {10.1145/3572921.3572922},
booktitle = {Proceedings of the 34th Australian Conference on Human-Computer Interaction},
pages = {291–297},
numpages = {7},
location = {Canberra, ACT, Australia},
series = {OzCHI '22}
}

@article{arthur1997the,
author = {Arthur Aron and Edward Melinat and Elaine N. Aron and Robert Darrin Vallone and Renee J. Bator},
title ={The Experimental Generation of Interpersonal Closeness: A Procedure and Some Preliminary Findings},
journal = {Personality and Social Psychology Bulletin},
volume = {23},
number = {4},
pages = {363-377},
year = {1997},
doi = {10.1177/0146167297234003},
URL = {https://doi.org/10.1177/0146167297234003},
eprint = {https://doi.org/10.1177/0146167297234003}
}

@Article{slemp2024interpersonal,
  author        = "Slemp, G. R. and Field, J. G. and Ryan, R. M. and Forner, V. W. and Van den Broeck, A. and Lewis, K. J.",
  title         = "Interpersonal supports for basic psychological needs and their relations with motivation, well-being, and performance: A meta-analysis",
  journal       = "J Pers Soc Psychol",
  volume        = "127",
  number        = "5",
  year          = "2024",
  pages         = "1012--1037",
  doi           = "10.1037/pspi0000459",
  url           = "",
}

@article{geng2025whengroup,
author = {Geng, Shixian and Shimojima, Ginshi and Yang, Chi-Lan and Sramek, Zefan and Norihama, Shunpei and Takano, Ayumi and Hosio, Simo and Yatani, Koji},
title = {When Group Spirit Meets Personal Journeys: Exploring Motivational Dynamics and Design Opportunities in Group Therapy},
year = {2025},
issue_date = {May 2025},
publisher = {Association for Computing Machinery},
address = {New York, NY, USA},
volume = {9},
number = {2},
url = {https://doi.org/10.1145/3711103},
doi = {10.1145/3711103},
journal = {Proc. ACM Hum.-Comput. Interact.},
month = may,
articleno = {CSCW205},
numpages = {27}
}

@inproceedings{cai2023listen,
author = {Cai, Wanling and Jin, Yucheng and Zhao, Xianglin and Chen, Li},
title = {“Listen to Music, Listen to Yourself”: Design of a Conversational Agent to Support Self-Awareness While Listening to Music},
year = {2023},
isbn = {9781450394215},
publisher = {Association for Computing Machinery},
address = {New York, NY, USA},
url = {https://doi.org/10.1145/3544548.3581427},
doi = {10.1145/3544548.3581427},
booktitle = {Proceedings of the 2023 CHI Conference on Human Factors in Computing Systems},
articleno = {119},
numpages = {19},
location = {Hamburg, Germany},
series = {CHI '23}
}

@inproceedings{toxtli2018understanding,
author = {Toxtli, Carlos and Monroy-Hern\'{a}ndez, Andr\'{e}s and Cranshaw, Justin},
title = {Understanding Chatbot-mediated Task Management},
year = {2018},
isbn = {9781450356206},
publisher = {Association for Computing Machinery},
address = {New York, NY, USA},
url = {https://doi.org/10.1145/3173574.3173632},
doi = {10.1145/3173574.3173632},
booktitle = {Proceedings of the 2018 CHI Conference on Human Factors in Computing Systems},
pages = {1–6},
numpages = {6},
location = {Montreal QC, Canada},
series = {CHI '18}
}

@inproceedings{zheng2022ux,
author = {Zheng, Qingxiao and Tang, Yiliu and Liu, Yiren and Liu, Weizi and Huang, Yun},
title = {UX Research on Conversational Human-AI Interaction: A Literature Review of the ACM Digital Library},
year = {2022},
isbn = {9781450391573},
publisher = {Association for Computing Machinery},
address = {New York, NY, USA},
url = {https://doi.org/10.1145/3491102.3501855},
doi = {10.1145/3491102.3501855},
booktitle = {Proceedings of the 2022 CHI Conference on Human Factors in Computing Systems},
articleno = {570},
numpages = {24},
location = {New Orleans, LA, USA},
series = {CHI '22}
}

@inproceedings{zhu2025benefits,
author = {Zhu, Zicheng and Tan, Yugin and Yamashita, Naomi and Lee, Yi-Chieh and Zhang, Renwen},
title = {The Benefits of Prosociality towards AI Agents: Examining the Effects of Helping AI Agents on Human Well-Being},
year = {2025},
isbn = {9798400713941},
publisher = {Association for Computing Machinery},
address = {New York, NY, USA},
url = {https://doi.org/10.1145/3706598.3713116},
doi = {10.1145/3706598.3713116},
booktitle = {Proceedings of the 2025 CHI Conference on Human Factors in Computing Systems},
articleno = {722},
numpages = {18},
location = {
},
series = {CHI '25}
}

@InProceedings{gratch2007creating,
author="Gratch, Jonathan
and Wang, Ning
and Gerten, Jillian
and Fast, Edward
and Duffy, Robin",
editor="Pelachaud, Catherine
and Martin, Jean-Claude
and Andr{\'e}, Elisabeth
and Chollet, G{\'e}rard
and Karpouzis, Kostas
and Pel{\'e}, Danielle",
title="Creating Rapport with Virtual Agents",
booktitle="Intelligent Virtual Agents",
year="2007",
publisher="Springer Berlin Heidelberg",
address="Berlin, Heidelberg",
pages="125--138",
isbn="978-3-540-74997-4"
}

@Article{papneja2025self,
  author        = "Papneja, Hashai and Yadav, Nikhil",
  title         = "Self-disclosure to conversational AI: a literature review, emergent framework, and directions for future research",
  journal       = "Personal and Ubiquitous Computing",
  volume        = "29",
  number        = "2",
  year          = "2025",
  pages         = "119--151",
  doi           = "10.1007/s00779-024-01823-7",
  url           = "",
}

@ARTICLE{xu2022enhancing,
AUTHOR={Xu, Ying  and Zhang, Jianyu  and Deng, Guangkuan },
TITLE={Enhancing customer satisfaction with chatbots: The influence of communication styles and consumer attachment anxiety}, 
JOURNAL={Frontiers in Psychology},  
VOLUME={Volume 13 - 2022},
YEAR={2022},
URL={https://www.frontiersin.org/journals/psychology/articles/10.3389/fpsyg.2022.902782},
DOI={10.3389/fpsyg.2022.902782},
ISSN={1664-1078}}

@article{lee2017enhancing,
title = {Enhancing user experience with conversational agent for movie recommendation: Effects of self-disclosure and reciprocity},
journal = {International Journal of Human-Computer Studies},
volume = {103},
pages = {95-105},
year = {2017},
issn = {1071-5819},
doi = {https://doi.org/10.1016/j.ijhcs.2017.02.005},
url = {https://www.sciencedirect.com/science/article/pii/S1071581917300198},
author = {SeoYoung Lee and Junho Choi}
}

@Article{pentina2023exploring,
  author        = "Iryna Pentina and Tyler Hancock and Tianling Xie",
  title         = "Exploring relationship development with social chatbots: A mixed-method study of replika",
  journal       = "Computers in Human Behavior",
  volume        = "140",
  number        = "",
  month         = "",
  year          = "2023",
  pages         = "107600",
  doi           = "https://doi.org/10.1016/j.chb.2022.107600",
  url           = "https://www.sciencedirect.com/science/article/pii/S0747563222004204",
  note          = "",
}

@Article{laestadius2024toohuman,
  author        = "Linnea Laestadius and Andrea Bishop and Michael Gonzalez and Diana Illenčík and Celeste Campos-Castillo",
  title         = "Too human and not human enough: A grounded theory analysis of mental health harms from emotional dependence on the social chatbot Replika",
  journal       = "New Media \& Society",
  volume        = "26",
  number        = "10",
  month         = "",
  year          = "2024",
  pages         = "5923--5941",
  doi           = "10.1177/14614448221142007",
  url           = "https://doi.org/10.1177/14614448221142007",
  note          = "",
}

@article{Schmager16022025,
author = {Stefan Schmager and Ilias O. Pappas and Polyxeni Vassilakopoulou},
title = {Understanding Human-Centred AI: a review of its defining elements and a research agenda},
journal = {Behaviour \& Information Technology},
volume = {0},
number = {0},
pages = {1--40},
year = {2025},
publisher = {Taylor \& Francis},
doi = {10.1080/0144929X.2024.2448719},
URL = {https://doi.org/10.1080/0144929X.2024.2448719},
eprint = {https://doi.org/10.1080/0144929X.2024.2448719}
}

@inproceedings{kim2024mindfuldiary,
author = {Kim, Taewan and Bae, Seolyeong and Kim, Hyun Ah and Lee, Su-Woo and Hong, Hwajung and Yang, Chanmo and Kim, Young-Ho},
title = {MindfulDiary: Harnessing Large Language Model to Support Psychiatric Patients' Journaling},
year = {2024},
isbn = {9798400703300},
publisher = {Association for Computing Machinery},
address = {New York, NY, USA},
url = {https://doi.org/10.1145/3613904.3642937},
doi = {10.1145/3613904.3642937},
booktitle = {Proceedings of the 2024 CHI Conference on Human Factors in Computing Systems},
articleno = {701},
numpages = {20},
location = {Honolulu, HI, USA},
series = {CHI '24}
}

@inproceedings{winograd1986alanguage,
author = {Winograd, Terry},
title = {A language/action perspective on the design of cooperative work},
year = {1986},
isbn = {9781450373654},
publisher = {Association for Computing Machinery},
address = {New York, NY, USA},
url = {https://doi.org/10.1145/637069.637096},
doi = {10.1145/637069.637096},
booktitle = {Proceedings of the 1986 ACM Conference on Computer-Supported Cooperative Work},
pages = {203–220},
numpages = {18},
location = {Austin, Texas},
series = {CSCW '86}
}

@misc{perez2021truefewshot,
      title={True Few-Shot Learning with Language Models}, 
      author={Ethan Perez and Douwe Kiela and Kyunghyun Cho},
      year={2021},
      eprint={2105.11447},
      archivePrefix={arXiv},
      primaryClass={cs.CL},
      url={https://arxiv.org/abs/2105.11447}, 
}

@inproceedings{brown2020language,
author = {Brown, Tom B. and Mann, Benjamin and Ryder, Nick and Subbiah, Melanie and Kaplan, Jared and Dhariwal, Prafulla and Neelakantan, Arvind and Shyam, Pranav and Sastry, Girish and Askell, Amanda and Agarwal, Sandhini and Herbert-Voss, Ariel and Krueger, Gretchen and Henighan, Tom and Child, Rewon and Ramesh, Aditya and Ziegler, Daniel M. and Wu, Jeffrey and Winter, Clemens and Hesse, Christopher and Chen, Mark and Sigler, Eric and Litwin, Mateusz and Gray, Scott and Chess, Benjamin and Clark, Jack and Berner, Christopher and McCandlish, Sam and Radford, Alec and Sutskever, Ilya and Amodei, Dario},
title = {Language models are few-shot learners},
year = {2020},
isbn = {9781713829546},
publisher = {Curran Associates Inc.},
address = {Red Hook, NY, USA},
booktitle = {Proceedings of the 34th International Conference on Neural Information Processing Systems},
articleno = {159},
numpages = {25},
location = {Vancouver, BC, Canada},
series = {NIPS '20}
}

@inproceedings{reynolds2021prompt,
author = {Reynolds, Laria and McDonell, Kyle},
title = {Prompt Programming for Large Language Models: Beyond the Few-Shot Paradigm},
year = {2021},
isbn = {9781450380959},
publisher = {Association for Computing Machinery},
address = {New York, NY, USA},
url = {https://doi.org/10.1145/3411763.3451760},
doi = {10.1145/3411763.3451760},
booktitle = {Extended Abstracts of the 2021 CHI Conference on Human Factors in Computing Systems},
articleno = {314},
numpages = {7},
location = {Yokohama, Japan},
series = {CHI EA '21}
}

@Article{boyd2022thedevelopment,
  author        = "Boyd, Ryan L. and Ashwini Ashokkumar and Sarah Seraj and James W. Pennebaker",
  title         = "The development and psychometric properties of LIWC-22",
  journal       = "Austin, TX: University of Texas at Austin",
  volume        = "10",
  number        = "",
  month         = "",
  year          = "2022",
  pages         = "",
  doi           = "",
  url           = "",
  note          = "",
}

@Article{yla2010thepsychological,
  author        = "Yla R. Tausczik and James W. Pennebaker",
  title         = "The Psychological Meaning of Words: LIWC and Computerized Text Analysis Methods",
  journal       = "Journal of Language and Social Psychology",
  volume        = "29",
  number        = "1",
  month         = "",
  year          = "2010",
  pages         = "24--54",
  doi           = "10.1177/0261927X09351676",
  url           = "https://doi.org/10.1177/0261927X09351676",
  note          = "",
}

@Article{hill2015real,
  author        = "Jennifer Hill and W. {Randolph Ford} and Ingrid G. Farreras",
  title         = "Real conversations with artificial intelligence: A comparison between human–human online conversations and human–chatbot conversations",
  journal       = "Computers in Human Behavior",
  volume        = "49",
  number        = "",
  month         = "",
  year          = "2015",
  pages         = "245--250",
  doi           = "10.1016/j.chb.2015.02.026",
  url           = "https://www.sciencedirect.com/science/article/pii/S0747563215001247",
  note          = "",
}

@misc{AtlasTi2023,
  author = {{ATLAS.ti Scientific Software Development GmbH}},
  title = {ATLAS.ti Mac},
  howpublished = {Macintosh software},
  version = {23.2.1},
  year = {2023},
  publisher = {ATLAS.ti Scientific Software Development GmbH},
  note = {Available software for qualitative data analysis}
}

@article{viera2005understanding,
  title={Understanding interobserver agreement: the kappa statistic},
  author={Viera, Anthony J and Garrett, Joanne M and others},
  journal={Fam med},
  volume={37},
  number={5},
  pages={360--363},
  year={2005}
}

@article{mchugh2012interrater,
  title={Interrater reliability: the kappa statistic},
  author={McHugh, Mary L},
  journal={Biochemia medica},
  volume={22},
  number={3},
  pages={276--282},
  year={2012},
  publisher={Medicinska naklada}
}

@Article{barak2007degree,
  author        = "Barak Azy and Gluck-Ofri Orit",
  title         = "Degree and reciprocity of self-disclosure in online forums",
  journal       = "CyberPsychology \& Behavior",
  volume        = "10",
  number        = "3",
  month         = "",
  year          = "2007",
  pages         = "407--417",
  doi           = "10.1089/cpb.2006.9938",
  url           = "",
  note          = "",
}

@Article{ahmadi2023classification,
  author        = "Ahmadi, Asghar and Noetel, Michael and Parker, Philip and Ryan, Richard M. and Ntoumanis, Nikos and Reeve, Johnmarshall and Beauchamp, Mark and Dicke, Theresa and Yeung, Alexander and Ahmadi, Malek and Bartholomew, Kimberley and Chiu, Thomas K. F. and Curran, Thomas and Erturan, Gokce and Flunger, Barbara and Frederick, Christina and Froiland, John Mark and González-Cutre, David and Haerens, Leen and Jeno, Lucas Matias and Koka, Andre and Krijgsman, Christa and Langdon, Jody and White, Rhiannon Lee and Litalien, David and Lubans, David and Mahoney, John and Nalipay, Ma Jenina N. and Patall, Erika and Perlman, Dana and Quested, Eleanor and Schneider, Sascha and Standage, Martyn and Stroet, Kim and Tessier, Damien and Thogersen-Ntoumani, Cecilie and Tilga, Henri and Vasconcellos, Diego and Lonsdale, Chris",
  title         = "A classification system for teachers’ motivational behaviors recommended in self-determination theory interventions",
  journal       = "Journal of Educational Psychology",
  volume        = "115",
  number        = "8",
  year          = "2023",
  pages         = "1158--1176",
  doi           = "10.1037/edu0000783",
  url           = "",
}

@Article{virginia2006using,
  author        = "Virginia Braun and Victoria Clarke",
  title         = "Using thematic analysis in psychology",
  journal       = "Qualitative Research in Psychology",
  volume        = "3",
  number        = "2",
  month         = "",
  year          = "2006",
  pages         = "77--101",
  doi           = "10.1191/1478088706qp063oa",
  url           = "https://www.tandfonline.com/doi/abs/10.1191/1478088706qp063oa",
  note          = "",
}

@article{ntoumanis2021meta,
author = {Nikos Ntoumanis and Johan Y.Y. Ng and Andrew Prestwich and Eleanor Quested and Jennie E. Hancox and Cecilie Thøgersen-Ntoumani and Edward L. Deci and Richard M. Ryan and Chris Lonsdale and Geoffrey C. Williams},
title = {A meta-analysis of self-determination theory-informed intervention studies in the health domain: effects on motivation, health behavior, physical, and psychological health},
journal = {Health Psychology Review},
volume = {15},
number = {2},
pages = {214--244},
year = {2021},
publisher = {Routledge},
doi = {10.1080/17437199.2020.1718529},
note ={PMID: 31983293},
URL = {https://doi.org/10.1080/17437199.2020.1718529},
eprint = {https://doi.org/10.1080/17437199.2020.1718529}
}

@article{kil2022autonomy,
author = {Kil, Hali and Allen, Marie-Pier and Taing, Jennifer and Mageau, Geneviève A.},
title = {Autonomy support in disclosure and privacy maintenance regulation within romantic relationships},
journal = {Personal Relationships},
volume = {29},
number = {2},
pages = {305-331},
doi = {https://doi.org/10.1111/pere.12419},
url = {https://onlinelibrary.wiley.com/doi/abs/10.1111/pere.12419},
eprint = {https://onlinelibrary.wiley.com/doi/pdf/10.1111/pere.12419},
year = {2022}
}

@inproceedings{yang2021designing,
author = {Yang, Xi and Aurisicchio, Marco},
title = {Designing Conversational Agents: A Self-Determination Theory Approach},
year = {2021},
isbn = {9781450380966},
publisher = {Association for Computing Machinery},
address = {New York, NY, USA},
url = {https://doi.org/10.1145/3411764.3445445},
doi = {10.1145/3411764.3445445},
booktitle = {Proceedings of the 2021 CHI Conference on Human Factors in Computing Systems},
articleno = {256},
numpages = {16},
location = {Yokohama, Japan},
series = {CHI '21}
}

@inproceedings{joshi2025user,
author = {Joshi, Rutuja and Lee, Yu-Jou and Bengler, Klaus},
title = {User Preferences in Conversational AI for Healthcare: Insights from an Interview Study},
year = {2025},
isbn = {9798400715273},
publisher = {Association for Computing Machinery},
address = {New York, NY, USA},
url = {https://doi.org/10.1145/3719160.3736631},
doi = {10.1145/3719160.3736631},
booktitle = {Proceedings of the 7th ACM Conference on Conversational User Interfaces},
articleno = {67},
numpages = {13},
location = {
},
series = {CUI '25}
}

@Article{diebel2025when,
  author        = "Diebel, Christopher and Goutier, Marc and Adam, Martin and Benlian, Alexander",
  title         = "When AI-Based Agents Are Proactive: Implications for Competence and System Satisfaction in Human–AI Collaboration",
  journal       = "Business \& Information Systems Engineering",
  volume        = "",
  number        = "",
  year          = "2025",
  pages         = "1--20",
  doi           = "10.1007/s12599-024-00918-y",
  url           = "https://doi.org/10.1007/s12599-024-00918-y",
}

@Article{bowman2024exploring,
  author        = "Bowman, Robert and Cooney, Orla and Newbold, Joseph W. and Thieme, Anja and Clark, Leigh and Doherty, Gavin and Cowan, Benjamin",
  title         = "Exploring how politeness impacts the user experience of chatbots for mental health support",
  journal       = "International Journal of Human-Computer Studies",
  volume        = "184",
  number        = "",
  year          = "2024",
  pages         = "103181",
  doi           = "https://doi.org/10.1016/j.ijhcs.2023.103181",
  url           = "https://www.sciencedirect.com/science/article/pii/S1071581923001908",
}

@article{grenier2024self,
author = {Grenier, Simon and Gagné, Marylène and O'Neill, Thomas},
title = {Self-determination theory and its implications for team motivation},
journal = {Applied Psychology},
volume = {73},
number = {4},
pages = {1833-1865},
doi = {https://doi.org/10.1111/apps.12526},
url = {https://iaap-journals.onlinelibrary.wiley.com/doi/abs/10.1111/apps.12526},
eprint = {https://iaap-journals.onlinelibrary.wiley.com/doi/pdf/10.1111/apps.12526},
year = {2024}
}

@article{wasserkampf2016organismic,
author = {Anna Wasserkampf and Jens Kleinert},
title = {Organismic integration as a dynamic process: a systematic review of empirical studies on change in behavioral regulations in exercise in adults},
journal = {International Review of Sport and Exercise Psychology},
volume = {9},
number = {1},
pages = {65--95},
year = {2016},
publisher = {Routledge},
doi = {10.1080/1750984X.2015.1119873},
URL = {https://doi.org/10.1080/1750984X.2015.1119873},
eprint = {https://doi.org/10.1080/1750984X.2015.1119873}
}

@article{bennett2024beyond,
author = {Bennett, Daniel and Mekler, Elisa D.},
title = {Beyond Intrinsic Motivation: The Role of Autonomous Motivation in User Experience},
year = {2024},
issue_date = {October 2024},
publisher = {Association for Computing Machinery},
address = {New York, NY, USA},
volume = {31},
number = {5},
issn = {1073-0516},
url = {https://doi.org/10.1145/3689044},
doi = {10.1145/3689044},
journal = {ACM Trans. Comput.-Hum. Interact.},
month = nov,
articleno = {60},
numpages = {41}
}

@inproceedings{moilanen2022measuring,
author = {Moilanen, Joonas and Visuri, Aku and Suryanarayana, Sharadhi Alape and Alorwu, Andy and Yatani, Koji and Hosio, Simo},
title = {Measuring the Effect of Mental Health Chatbot Personality on User Engagement},
year = {2022},
isbn = {9781450398206},
publisher = {Association for Computing Machinery},
address = {New York, NY, USA},
url = {https://doi.org/10.1145/3568444.3568464},
doi = {10.1145/3568444.3568464},
booktitle = {Proceedings of the 21st International Conference on Mobile and Ubiquitous Multimedia},
pages = {138–150},
numpages = {13},
location = {Lisbon, Portugal},
series = {MUM '22}
}

@inproceedings{li2025acomparative,
author = {Li, Brenna and Tauseef, Saba and Truong, Khai N. and Mariakakis, Alex},
title = {A Comparative Analysis of Information Gathering by Chatbots, Questionnaires, and Humans in Clinical Pre-Consultation},
year = {2025},
isbn = {9798400713941},
publisher = {Association for Computing Machinery},
address = {New York, NY, USA},
url = {https://doi.org/10.1145/3706598.3713613},
doi = {10.1145/3706598.3713613},
booktitle = {Proceedings of the 2025 CHI Conference on Human Factors in Computing Systems},
articleno = {639},
numpages = {17},
location = {
},
series = {CHI '25}
}

@inproceedings{gilad2021effects,
author = {Gilad, Zohar and Amir, Ofra and Levontin, Liat},
title = {The Effects of Warmth and Competence Perceptions on Users' Choice of an AI System},
year = {2021},
isbn = {9781450380966},
publisher = {Association for Computing Machinery},
address = {New York, NY, USA},
url = {https://doi.org/10.1145/3411764.3446863},
doi = {10.1145/3411764.3446863},
booktitle = {Proceedings of the 2021 CHI Conference on Human Factors in Computing Systems},
articleno = {583},
numpages = {13},
location = {Yokohama, Japan},
series = {CHI '21}
}

@article{salehzadeh2023potential,
author = {Salehzadeh Niksirat, Kavous and Korka, Diana and Harkous, Hamza and Huguenin, K\'{e}vin and Cherubini, Mauro},
title = {On the Potential of Mediation Chatbots for Mitigating Multiparty Privacy Conflicts - A Wizard-of-Oz Study},
year = {2023},
issue_date = {April 2023},
publisher = {Association for Computing Machinery},
address = {New York, NY, USA},
volume = {7},
number = {CSCW1},
url = {https://doi.org/10.1145/3579618},
doi = {10.1145/3579618},
journal = {Proc. ACM Hum.-Comput. Interact.},
month = apr,
articleno = {142},
numpages = {33}
}

@inproceedings{govers2024aidriven,
author = {Govers, Jarod and Velloso, Eduardo and Kostakos, Vassilis and Goncalves, Jorge},
title = {AI-Driven Mediation Strategies for Audience Depolarisation in Online Debates},
year = {2024},
isbn = {9798400703300},
publisher = {Association for Computing Machinery},
address = {New York, NY, USA},
url = {https://doi.org/10.1145/3613904.3642322},
doi = {10.1145/3613904.3642322},
booktitle = {Proceedings of the 2024 CHI Conference on Human Factors in Computing Systems},
articleno = {803},
numpages = {18},
location = {Honolulu, HI, USA},
series = {CHI '24}
}

@inproceedings{chen2025maintaining,
author = {Chen, XinHui and Yuan, Xiang and Zhang, Hui and Zheng, Ruixiao and Wei, Wanyi},
title = {Maintaining "Balanced" Conflict: Proactive Intervention Strategies of AI Voice Agents in Online Collaboration of Temporary Design Teams},
year = {2025},
isbn = {9798400713941},
publisher = {Association for Computing Machinery},
address = {New York, NY, USA},
url = {https://doi.org/10.1145/3706598.3713457},
doi = {10.1145/3706598.3713457},
booktitle = {Proceedings of the 2025 CHI Conference on Human Factors in Computing Systems},
articleno = {907},
numpages = {19},
location = {
},
series = {CHI '25}
}

@inproceedings{houde2025controlling,
author = {Houde, Stephanie and Brimijoin, Kristina and Muller, Michael and Ross, Steven I. and Silva Moran, Dario Andres and Gonzalez, Gabriel Enrique and Kunde, Siya and Foreman, Morgan A. and Weisz, Justin D.},
title = {Controlling AI Agent Participation in Group Conversations: A Human-Centered Approach},
year = {2025},
isbn = {9798400713064},
publisher = {Association for Computing Machinery},
address = {New York, NY, USA},
url = {https://doi.org/10.1145/3708359.3712089},
doi = {10.1145/3708359.3712089},
booktitle = {Proceedings of the 30th International Conference on Intelligent User Interfaces},
pages = {390–408},
numpages = {19},
location = {
},
series = {IUI '25}
}

@inproceedings{wang2025socialrag,
author = {Wang, Ruotong and Zhou, Xinyi and Qiu, Lin and Chang, Joseph Chee and Bragg, Jonathan and Zhang, Amy X.},
title = {Social-RAG: Retrieving from Group Interactions to Socially Ground AI Generation},
year = {2025},
isbn = {9798400713941},
publisher = {Association for Computing Machinery},
address = {New York, NY, USA},
url = {https://doi.org/10.1145/3706598.3713749},
doi = {10.1145/3706598.3713749},
booktitle = {Proceedings of the 2025 CHI Conference on Human Factors in Computing Systems},
articleno = {162},
numpages = {25},
location = {
},
series = {CHI '25}
}

@article{laura2024AI,
title = {AI in relationship counselling: Evaluating ChatGPT's therapeutic capabilities in providing relationship advice},
journal = {Computers in Human Behavior: Artificial Humans},
volume = {2},
number = {2},
pages = {100078},
year = {2024},
issn = {2949-8821},
doi = {https://doi.org/10.1016/j.chbah.2024.100078},
url = {https://www.sciencedirect.com/science/article/pii/S2949882124000380},
author = {Laura M. Vowels and Rachel R.R. Francois-Walcott and Joëlle Darwiche}
}

@article{hatch2025ELIZA,
    doi = {10.1371/journal.pmen.0000145},
    author = {Hatch, S. Gabe AND Goodman, Zachary T. AND Vowels, Laura AND Hatch, H. Dorian AND Brown, Alyssa L. AND Guttman, Shayna AND Le, Yunying AND Bailey, Benjamin AND Bailey, Russell J. AND Esplin, Charlotte R. AND Harris, Steven M. AND Holt, Jr., D. Payton AND McLaughlin, Merranda AND O’Connell, Patrick AND Rothman, Karen AND Ritchie, Lane AND Top, Jr., D. Nicholas AND Braithwaite, Scott R.},
    journal = {PLOS Mental Health},
    publisher = {Public Library of Science},
    title = {When ELIZA meets therapists: A Turing test for the heart and mind},
    year = {2025},
    month = {02},
    volume = {2},
    url = {https://doi.org/10.1371/journal.pmen.0000145},
    pages = {1-16},
    number = {2},

}

@article{slovak2023designing,
author = {Slovak, Petr and Antle, Alissa and Theofanopoulou, Nikki and Daud\'{e}n Roquet, Claudia and Gross, James and Isbister, Katherine},
title = {Designing for Emotion Regulation Interventions: An Agenda for HCI Theory and Research},
year = {2023},
issue_date = {February 2023},
publisher = {Association for Computing Machinery},
address = {New York, NY, USA},
volume = {30},
number = {1},
issn = {1073-0516},
url = {https://doi.org/10.1145/3569898},
doi = {10.1145/3569898},
journal = {ACM Trans. Comput.-Hum. Interact.},
month = mar,
articleno = {13},
numpages = {51}
}

@ARTICLE{peters2018designing,
AUTHOR={Peters, Dorian  and Calvo, Rafael A.  and Ryan, Richard M. },
TITLE={Designing for Motivation, Engagement and Wellbeing in Digital Experience},
JOURNAL={Frontiers in Psychology},
VOLUME={Volume 9 - 2018},
YEAR={2018},
URL={https://www.frontiersin.org/journals/psychology/articles/10.3389/fpsyg.2018.00797},
DOI={10.3389/fpsyg.2018.00797},
ISSN={1664-1078}
}

@inproceedings{zhang2021distress,
author = {Zhang, Renwen and N. Bazarova, Natalya and Reddy, Madhu},
title = {Distress Disclosure across Social Media Platforms during the COVID-19 Pandemic: Untangling the Effects of Platforms, Affordances, and Audiences},
year = {2021},
isbn = {9781450380966},
publisher = {Association for Computing Machinery},
address = {New York, NY, USA},
url = {https://doi.org/10.1145/3411764.3445134},
doi = {10.1145/3411764.3445134},
booktitle = {Proceedings of the 2021 CHI Conference on Human Factors in Computing Systems},
articleno = {644},
numpages = {15},
location = {Yokohama, Japan},
series = {CHI '21}
}

@article{andalibi2018responding,
author = {Andalibi, Nazanin and Forte, Andrea},
title = {Responding to Sensitive Disclosures on Social Media: A Decision-Making Framework},
year = {2018},
issue_date = {December 2018},
publisher = {Association for Computing Machinery},
address = {New York, NY, USA},
volume = {25},
number = {6},
issn = {1073-0516},
url = {https://doi.org/10.1145/3241044},
doi = {10.1145/3241044},
journal = {ACM Trans. Comput.-Hum. Interact.},
month = dec,
articleno = {31},
numpages = {29}
}

@Article{kovacs2025daily,
  author        = "Kovács, Lilla Nóra and Kocsel, Natália and Tóth, Zsófia and Smahajcsik and Szabó, Tamás and Karsai, Szilvia and Kökönyei, Gyöngyi",
  title         = "The daily relations of co-rumination and perseverative cognition",
  journal       = "Scientific Reports",
  volume        = "15",
  number        = "1",
  year          = "2025",
  pages         = "3021",
  doi           = "10.1038/s41598-025-87335-7",
  url           = "https://doi.org/10.1038/s41598-025-87335-7",
}

@Article{rose2007prospective,
  author        = "Rose, A. J. and Carlson, W. and Waller, E. M.",
  title         = "Prospective associations of co-rumination with friendship and emotional adjustment: considering the socioemotional trade-offs of co-rumination",
  journal       = "Dev Psychol",
  volume        = "43",
  number        = "4",
  year          = "2007",
  pages         = "1019--1031",
  doi           = "10.1037/0012-1649.43.4.1019",
  url           = "",
}

@phdthesis{gaine2011implications,
  title={The Implications of Relational Activity Motivations for Relationship Well-Being and Daily Relational Functioning in Marriage},
  author={Gaine, Graham Sherwood},
  year={2011},
  school={University of Waterloo}
}

@Article{baker2020therole,
  author        = "Baker, Z. G. and Watlington, E. M. and Knee, C. R.",
  title         = "The Role of Rapport in Satisfying One's Basic Psychological Needs",
  journal       = "Motiv Emot",
  volume        = "44",
  number        = "2",
  year          = "2020",
  pages         = "329--343",
  doi           = "10.1007/s11031-020-09819-5",
  url           = "",
}

@Article{kluwer2020autonomy,
  author        = "Kluwer, E. S. and Karremans, J. C. and Riedijk, L. and Knee, C. R.",
  title         = "Autonomy in Relatedness: How Need Fulfillment Interacts in Close Relationships",
  journal       = "Pers Soc Psychol Bull",
  volume        = "46",
  number        = "4",
  year          = "2020",
  pages         = "603--616",
  doi           = "10.1177/0146167219867964",
  url           = "",
}

@inproceedings{pereira2017measuring,
author = {Pereira Santos, Carlos and Hutchinson, Kevin and Khan, Vassilis-Javed and Markopoulos, Panos},
title = {Measuring Self-Esteem with Games},
year = {2017},
isbn = {9781450343480},
publisher = {Association for Computing Machinery},
address = {New York, NY, USA},
url = {https://doi.org/10.1145/3025171.3025196},
doi = {10.1145/3025171.3025196},
booktitle = {Proceedings of the 22nd International Conference on Intelligent User Interfaces},
pages = {95–105},
numpages = {11},
location = {Limassol, Cyprus},
series = {IUI '17}
}

@inproceedings{narain2020promoting,
author = {Narain, Jaya and Quach, Tina and Davey, Monique and Park, Hae Won and Breazeal, Cynthia and Picard, Rosalind},
title = {Promoting Wellbeing with Sunny, a Chatbot that Facilitates Positive Messages within Social Groups},
year = {2020},
isbn = {9781450368193},
publisher = {Association for Computing Machinery},
address = {New York, NY, USA},
url = {https://doi.org/10.1145/3334480.3383062},
doi = {10.1145/3334480.3383062},
booktitle = {Extended Abstracts of the 2020 CHI Conference on Human Factors in Computing Systems},
pages = {1–8},
numpages = {8},
location = {Honolulu, HI, USA},
series = {CHI EA '20}
}

@article{tyack2021thesmall,
author = {Tyack, April and Wyeth, Peta},
title = {"The Small Decisions Are What Makes it Interesting": Autonomy, Control, and Restoration in Player Experience},
year = {2021},
issue_date = {September 2021},
publisher = {Association for Computing Machinery},
address = {New York, NY, USA},
volume = {5},
number = {CHI PLAY},
url = {https://doi.org/10.1145/3474709},
doi = {10.1145/3474709},
journal = {Proc. ACM Hum.-Comput. Interact.},
month = oct,
articleno = {282},
numpages = {26}
}

@inproceedings{tyack2020restorative,
author = {Tyack, April and Wyeth, Peta and Johnson, Daniel},
title = {Restorative Play: Videogames Improve Player Wellbeing After a Need-Frustrating Event},
year = {2020},
isbn = {9781450367080},
publisher = {Association for Computing Machinery},
address = {New York, NY, USA},
url = {https://doi.org/10.1145/3313831.3376332},
doi = {10.1145/3313831.3376332},
booktitle = {Proceedings of the 2020 CHI Conference on Human Factors in Computing Systems},
pages = {1–15},
numpages = {15},
location = {Honolulu, HI, USA},
series = {CHI '20}
}

@inbook{stafford2019communication,
place={Cambridge},
series={Advances in Personal Relationships},
title={Communication and Relationship Maintenance},
booktitle={Relationship Maintenance: Theory, Process, and Context},
publisher={Cambridge University Press},
author={Stafford, Laura},
editor={Ogolsky, Brian G. and Monk, J. KaleEditors},
year={2019},
pages={109–133},
collection={Advances in Personal Relationships}
}

@article{matthew2022within,
author = {Matthew D. Johnson and Justin A. Lavner and Marcus Mund and Martina Zemp and Scott M. Stanley and Franz J. Neyer and Emily A. Impett and Galena K. Rhoades and Guy Bodenmann and Rebekka Weidmann and Janina Larissa Bühler and Robert Philip Burriss and Jenna Wünsche and Alexander Grob},
title ={Within-Couple Associations Between Communication and Relationship Satisfaction Over Time},
journal = {Personality and Social Psychology Bulletin},
volume = {48},
number = {4},
pages = {534-549},
year = {2022},
doi = {10.1177/01461672211016920},
note ={PMID: 34027722},
URL = {https://doi.org/10.1177/01461672211016920},
eprint = {https://doi.org/10.1177/01461672211016920}
}

@Article{pusch2022motivational,
  author        = "Pusch, S. and Schönbrodt, F. D. and Zygar-Hoffmann, C. and Hagemeyer, B.",
  title         = "Motivational Interdependence in Couple Relationships",
  journal       = "Front Psychol",
  volume        = "13",
  number        = "",
  year          = "2022",
  pages         = "827746",
  doi           = "10.3389/fpsyg.2022.827746",
  url           = "",
}

@article{beck2016examining,
title = {Examining Practical, Everyday Theory Use in Design Research},
journal = {She Ji: The Journal of Design, Economics, and Innovation},
volume = {2},
number = {2},
pages = {125-140},
year = {2016},
issn = {2405-8726},
doi = {https://doi.org/10.1016/j.sheji.2016.01.010},
url = {https://www.sciencedirect.com/science/article/pii/S2405872615300344},
author = {Jordan Beck and Erik Stolterman}
}

@Article{kwok2020cyberintimacy,
  author        = "Kwok, I. and Wescott, A. B.",
  title         = "Cyberintimacy: A Scoping Review of Technology-Mediated Romance in the Digital Age",
  journal       = "Cyberpsychol Behav Soc Netw",
  volume        = "23",
  number        = "10",
  year          = "2020",
  pages         = "657--666",
  doi           = "10.1089/cyber.2019.0764",
  url           = "",
}

@inproceedings{neustaedter2012intimacy,
author = {Neustaedter, Carman and Greenberg, Saul},
title = {Intimacy in long-distance relationships over video chat},
year = {2012},
isbn = {9781450310154},
publisher = {Association for Computing Machinery},
address = {New York, NY, USA},
url = {https://doi.org/10.1145/2207676.2207785},
doi = {10.1145/2207676.2207785},
booktitle = {Proceedings of the SIGCHI Conference on Human Factors in Computing Systems},
pages = {753–762},
numpages = {10},
location = {Austin, Texas, USA},
series = {CHI '12}
}

@inproceedings{shen2024seeking,
author = {Shen, Chenxinran and Xu, Yan and Lc, Ray and Lu, Zhicong},
title = {Seeking Soulmate via Voice: Understanding Promises and Challenges of Online Synchronized Voice-Based Mobile Dating},
year = {2024},
isbn = {9798400703300},
publisher = {Association for Computing Machinery},
address = {New York, NY, USA},
url = {https://doi.org/10.1145/3613904.3642860},
doi = {10.1145/3613904.3642860},
booktitle = {Proceedings of the 2024 CHI Conference on Human Factors in Computing Systems},
articleno = {921},
numpages = {14},
location = {Honolulu, HI, USA},
series = {CHI '24}
}

@inproceedings{chou2022why,
author = {Chou, Yu-Ling and Lin, Yi-Hsiu and Lin, Tzu-Yi and You, Hsin Ying and Chang, Yung-Ju},
title = {Why Did You/I Read but Not Reply? IM Users’ Unresponded-to Read-receipt Practices and Explanations of Them},
year = {2022},
isbn = {9781450391573},
publisher = {Association for Computing Machinery},
address = {New York, NY, USA},
url = {https://doi.org/10.1145/3491102.3517496},
doi = {10.1145/3491102.3517496},
booktitle = {Proceedings of the 2022 CHI Conference on Human Factors in Computing Systems},
articleno = {526},
numpages = {15},
location = {New Orleans, LA, USA},
series = {CHI '22}
}

@article{daantje2008therole,
title = {The role of emotion in computer-mediated communication: A review},
journal = {Computers in Human Behavior},
volume = {24},
number = {3},
pages = {766-785},
year = {2008},
note = {Instructional Support for Enhancing Students' Information Problem Solving Ability},
issn = {0747-5632},
doi = {https://doi.org/10.1016/j.chb.2007.04.004},
url = {https://www.sciencedirect.com/science/article/pii/S0747563207000866},
author = {Daantje Derks and Agneta H. Fischer and Arjan E.R. Bos}
}

@article{gomes2021theimpact,
title = {The impact of technology usage on love and intimacy satisfaction among Portuguese adults},
journal = {Computers in Human Behavior Reports},
volume = {4},
pages = {100084},
year = {2021},
issn = {2451-9588},
doi = {https://doi.org/10.1016/j.chbr.2021.100084},
url = {https://www.sciencedirect.com/science/article/pii/S2451958821000324},
author = {Mara Gomes and José Vasconcelos-Raposo and Carla M. Teixeira}
}

@article{aljasir2022present,
author = {Aljasir, Shuaa},
title = {Present but Absent in the Digital Age: Testing a Conceptual Model of Phubbing and Relationship Satisfaction among Married Couples},
journal = {Human Behavior and Emerging Technologies},
volume = {2022},
number = {1},
pages = {1402751},
doi = {https://doi.org/10.1155/2022/1402751},
url = {https://onlinelibrary.wiley.com/doi/abs/10.1155/2022/1402751},
eprint = {https://onlinelibrary.wiley.com/doi/pdf/10.1155/2022/1402751},
year = {2022}
}

@article{foster2022rules,
    author = {Foster Campbell, Jamie},
    title = {Rules for Mediated Romance: A Digital Exploration of How Couples Negotiate Expectations},
    journal = {Journal of Computer-Mediated Communication},
    volume = {27},
    number = {3},
    pages = {zmac007},
    year = {2022},
    month = {04},
    issn = {1083-6101},
    doi = {10.1093/jcmc/zmac007},
    url = {https://doi.org/10.1093/jcmc/zmac007},
    eprint = {https://academic.oup.com/jcmc/article-pdf/27/3/zmac007/43371994/zmac007.pdf},
}

@Article{holtzman2021longdistance,
  author        = "Holtzman, S. and Kushlev, K. and Wozny, A. and Godard, R.",
  title         = "Long-distance texting: Text messaging is linked with higher relationship satisfaction in long-distance relationships",
  journal       = "J Soc Pers Relat",
  volume        = "38",
  number        = "12",
  year          = "2021",
  pages         = "3543--3565",
  doi           = "10.1177/02654075211043296",
  url           = "",
}

@article{tyack2024sdt,
author = {Tyack, April and Mekler, Elisa D.},
title = {Self-Determination Theory and HCI Games Research: Unfulfilled Promises and Unquestioned Paradigms},
year = {2024},
issue_date = {June 2024},
publisher = {Association for Computing Machinery},
address = {New York, NY, USA},
volume = {31},
number = {3},
issn = {1073-0516},
url = {https://doi.org/10.1145/3673230},
doi = {10.1145/3673230},
journal = {ACM Trans. Comput.-Hum. Interact.},
month = aug,
articleno = {40},
numpages = {74}
}

@inproceedings{tyack2020sdt,
author = {Tyack, April and Mekler, Elisa D.},
title = {Self-Determination Theory in HCI Games Research: Current Uses and Open Questions},
year = {2020},
isbn = {9781450367080},
publisher = {Association for Computing Machinery},
address = {New York, NY, USA},
url = {https://doi.org/10.1145/3313831.3376723},
doi = {10.1145/3313831.3376723},
booktitle = {Proceedings of the 2020 CHI Conference on Human Factors in Computing Systems},
pages = {1–22},
numpages = {22},
location = {Honolulu, HI, USA},
series = {CHI '20}
}

@article{lin2024ultimately,
author = {Lin, Ya-Fang and Li, Na and Huang, Wan-Hsuan and Ecsedy, Karen and Feinberg, Mark E. and Teti, Douglas and Carroll, John M.},
title = {"Ultimately We're Together": Understanding New Parents' Experiences of Co-parenting},
year = {2024},
issue_date = {November 2024},
publisher = {Association for Computing Machinery},
address = {New York, NY, USA},
volume = {8},
number = {CSCW2},
url = {https://doi.org/10.1145/3687018},
doi = {10.1145/3687018},
journal = {Proc. ACM Hum.-Comput. Interact.},
month = nov,
articleno = {479},
numpages = {25}
}

@inproceedings{lin2025parental,
author = {Lin, Ya-Fang and Li, Xiaotian and Huang, Wan-Hsuan and Prabavathi, Charan Pushpanathan and Cai, Jie and Carroll, John M.},
title = {Parental Collaboration and Closeness: Envisioning with New Couple Parents},
year = {2025},
isbn = {9798400714856},
publisher = {Association for Computing Machinery},
address = {New York, NY, USA},
url = {https://doi.org/10.1145/3715336.3735837},
doi = {10.1145/3715336.3735837},
booktitle = {Proceedings of the 2025 ACM Designing Interactive Systems Conference},
pages = {2637–2651},
numpages = {15},
location = {
},
series = {DIS '25}
}

@inproceedings{van2023like,
author = {van Greevenbroek, Roos and Patel, Dilisha and Singh, Aneesha},
title = {"Like a candy shop with forbidden fruits": Exploring Sexual Desire of Cohabiting Millennial Couples with Technology},
year = {2023},
isbn = {9781450398930},
publisher = {Association for Computing Machinery},
address = {New York, NY, USA},
url = {https://doi.org/10.1145/3563657.3596080},
doi = {10.1145/3563657.3596080},
booktitle = {Proceedings of the 2023 ACM Designing Interactive Systems Conference},
pages = {1842–1860},
numpages = {19},
location = {Pittsburgh, PA, USA},
series = {DIS '23}
}

@Article{lucier2018enhancing,
  author        = "Lucier-Greer, Mallory and Birney, Amelia J and Gutierrez, Teri M and Adler-Baeder, Francesca",
  title         = "Enhancing relationship skills and couple functioning with mobile technology: An evaluation of the Love Every Day mobile intervention",
  journal       = "Journal of Family Social Work",
  volume        = "21",
  number        = "2",
  year          = "2018",
  pages         = "152--171",
  doi           = "10.1080/10522158.2017.1410267",
  url           = "",
}

@Article{faul2009statistical,
  author        = "Faul, Franz and Erdfelder, Edgar and Buchner, Axel and Lang, Albert-Georg",
  title         = "Statistical power analyses using G*Power 3.1: Tests for correlation and regression analyses",
  journal       = "Behavior Research Methods",
  volume        = "41",
  number        = "4",
  year          = "2009",
  pages         = "1149--1160",
  doi           = "https://doi.org/10.3758/BRM.41.4.1149",
  url           = "",
}

\appendix

\section{Additional Study Results}

\begin{figure*}[t]
  \centering
  \includegraphics[width=\textwidth]{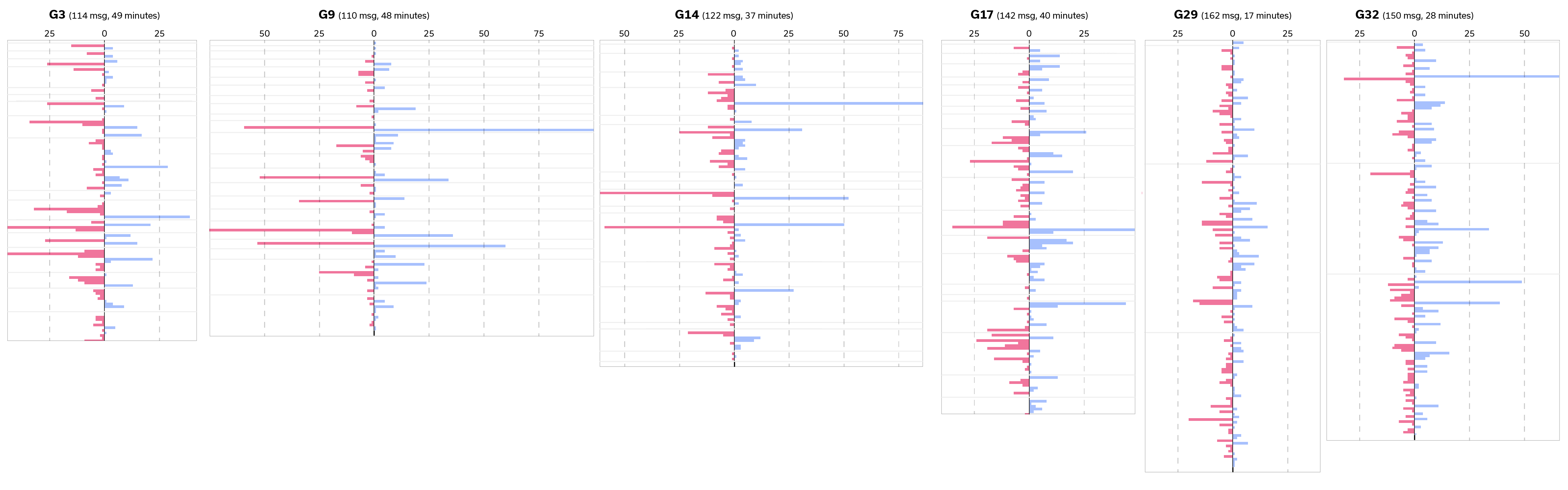}
  \caption{Distribution of message lengths across representative groups. Each horizontal bar corresponds to a single message, with bar height indicating the number of words. Messages from participants with odd-numbered IDs are shown on the left (pink) and even-numbered IDs on the right (blue). Numbers above each subplot indicate the total messages exchanged in the group and total duration of the conversation.}
  \label{app:matrix}
  \Description{Six panels show distributions of message lengths for representative dyads labeled G3, G9, G14, G17, G29, and G32. In each panel, horizontal bars represent individual messages, with bar length indicating the number of words. Messages from odd-numbered participants appear on the left in pink, while messages from even-numbered participants appear on the right in blue. Labels above each panel indicate the total number of messages exchanged and the total duration of the conversation in minutes. The plots reveal variation in conversational balance: in some groups one partner contributes longer messages, while in others contributions are more evenly distributed. Differences in verbosity, turn-taking style, and message length patterns can be observed across the six dyads.}
\end{figure*}

\begin{figure*}[t]
  \centering
  \includegraphics[width=\linewidth]{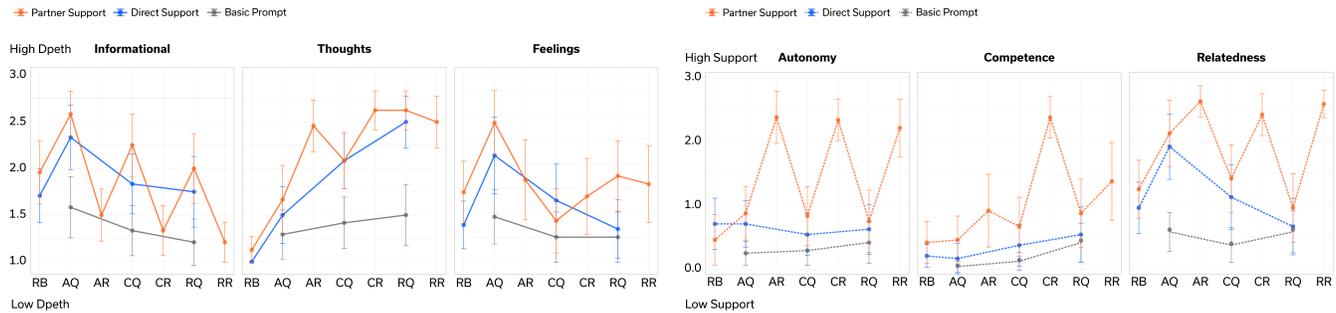}
  \caption{Lineplot of self-disclosure depth (left) and need support level (right) across phases and conditions.}
  \label{app:lineplot}
  \Description{Two sets of line charts compare self-disclosure depth and need support level across phases in three conditions: Partner Support (PS), Direct Support (DS), and Basic Prompt (BP). The left panel shows disclosure depth for informational, thoughts, and feelings on a scale from 1, low depth, to 3, high depth. Both PS and DS maintain higher disclosure depth across phases compared to BP, with similar trends between the two. The right panel shows need support level for autonomy, competence, and relatedness on a scale from 0, low support, to 3, high support. Need support level is consistently highest in PS, with noticeable peaks during reflection phases, whereas DS remains moderate and BP stays low across all needs. Overall, PS and DS similarly promote deeper disclosure, but only PS produces repeated increases in support level through the reflection phases.}
\end{figure*}

To complement the main analyses, we provide two additional visualizations that further illustrate conversational dynamics and condition effects. Figure~\ref{app:matrix} presents message length distributions across representative dyads, highlighting variation in conversational balance and style between partners. Figure~\ref{app:lineplot} summarizes disclosure depth and need support levels across phases and conditions, illustrating how structured chatbot prompts enhanced disclosure and how partner reflection phases in the PS condition concentrated relatedness support. Together, these results provide further context for interpreting the quantitative analyses reported in the main text.

\section{Demographic Information}
\label{app:demographic}

Table~\ref{app:demo} provides detailed demographic information for all participants across the three study conditions. The table reports group assignment (Partner Support, Direct Support, or Baseline Prompt), participant gender, relationship type (e.g., dating, married, friends with romantic interest), relationship length in years, and age.

\begin{table*}[t]
  \centering
  \caption{Detailed demographics information of participants. Groups 1 to 12 were assigned to the \PSbox{} group, Groups 13 to 24 to the \DSbox{} group, Groups 25 to 36 to the \BPbox{} group. The length of relationship (Rel. Length) is measured in years. Note: Friends=Friends with Romantic Interests. FRI = Friends with Romantic Interest.}
  \label{app:demo}
  \renewcommand{\arraystretch}{1.18}
  \resizebox{\textwidth}{!}{%
  \begin{tabular}{cccccc|cccccc}
    \toprule
    Group ID & ID & Gender & Rel. Type & Rel. Length & Age & Group ID & ID & Gender & Rel. Type & Rel. Length & Age\\
    \midrule
    G1 & P1 & Female & Dating & 4 & 23 & G19 & P37 & Female & Dating & 1 & 25\\
    G1 & P2 & Male & Dating & 4 & 24 & G19 & P38 & Male & Dating & 1 & 26\\
    G2 & P3 & Female & Dating & 2 & 26 & G20 & P39 & Female & Dating & 1 & 22\\
    G2 & P4 & Male & Dating & 2 & 29 & G20 & P40 & Female & Dating & 1 & 22\\
    G3 & P5 & Female & Dating & 1 & 24 & G21 & P41 & Female & Engaged & 3 & 30\\
    G3 & P6 & Male & Dating & 1 & 31 & G21 & P42 & Male & Engaged & 3 & 35\\
    G4 & P7 & Female & Dating & 3 & 19 & G22 & P43 & Female & Dating & 2 & 28\\
    G4 & P8 & Male & Dating & 3 & 20 & G22 & P44 & Male & Dating & 2 & 28\\
    G5 & P9 & Female & Dating & 2 & 23 & G23 & P45 & Female & Dating & 0.3 & 24\\
    G5 & P10 & Male & Dating & 2 & 25 & G23 & P46 & Male & Dating & 0.3 & 23\\
    G6 & P11 & Female & Dating & 8 & 25 & G24 & P47 & Female & FRI & 0.5 & 20\\
    G6 & P12 & Male & Dating & 8 & 25 & G24 & P48 & Male & FRI & 0.5 & 22\\
    G7 & P13 & Female & FRI & 2 & 26 & G25 & P49 & Female & Dating & 0.8 & 20\\
    G7 & P14 & Female & FRI & 2 & 23 & G25 & P50 & Male & Dating & 0.8 & 19\\
    G8 & P15 & Female & Dating & 1 & 21 & G26 & P51 & Female & Dating & 3 & 28\\
    G8 & P16 & Male & Dating & 1 & 25 & G26 & P52 & Male & Dating & 3 & 29\\
    G9 & P17 & Female & Dating & 2 & 26 & G27 & P53 & Female & Dating & 4 & 21\\
    G9 & P18 & Male & Dating & 2 & 26 & G27 & P54 & Male & Dating & 4 & 21\\
    G10 & P19 & Female & Dating & 9 & 21 & G28 & P55 & Female & Dating & 2 & 23\\
    G10 & P20 & Male & Dating & 9 & 23 & G28 & P56 & Male & Dating & 2 & 22\\
    G11 & P21 & Female & Dating & 5 & 22 & G29 & P57 & Female & Dating & 2 & 20\\
    G11 & P22 & Male & Dating & 5 & 22 & G29 & P58 & Male & Dating & 2 & 20\\
    G12 & P23 & Female & Dating & 3 & 21 & G30 & P59 & Female & Dating & 2.5 & 28\\
    G12 & P24 & Male & Dating & 3 & 21 & G30 & P60 & Male & Dating & 2.5 & 27\\
    G13 & P25 & Female & Dating & 2 & 21 & G31 & P61 & Female & Married & 6 & 36\\
    G13 & P26 & Male & Dating & 2 & 23 & G31 & P62 & Male & Married & 6 & 39\\
    G14 & P27 & Female & Dating & 5 & 25 & G32 & P63 & Female & Dating & 3 & 30\\
    G14 & P28 & Male & Dating & 5 & 27 & G32 & P64 & Male & Dating & 3 & 31\\
    G15 & P29 & Female & Dating & 1.5 & 22 & G33 & P65 & Female & Dating & 0.8 & 27\\
    G15 & P30 & Male & Dating & 1.5 & 21 & G33 & P66 & Male & Dating & 0.8 & 28\\
    G16 & P31 & Female & Dating & 3 & 21 & G34 & P67 & Female & Dating & 1 & 22\\
    G16 & P32 & Male & Dating & 3 & 20 & G34 & P68 & Male & Dating & 1 & 23\\
    G17 & P33 & Female & Dating & 5.5 & 27 & G35 & P69 & Female & Married & 5 & 33\\
    G17 & P34 & Male & Dating & 5.5 & 29 & G35 & P70 & Male & Married & 5 & 35\\
    G18 & P35 & Female & Married & 4 & 25 & G36 & P71 & Female & Dating & 2 & 24\\
    G18 & P36 & Male & Married & 4 & 28 & G36 & P72 & Male & Dating & 2 & 26\\
  \bottomrule
\end{tabular}%
}
\end{table*}

\section{Coding Rubrics for Self-Disclosure and Need Support}

Table~\ref{app:sd} defines the three levels of disclosure depth across informational, thoughts, and feelings, with illustrative participant examples. Table~\ref{app:core_behaviors_full} provides operational definitions of autonomy-, competence-, and relatedness-supportive behaviors used in the chatbot's design. Table~\ref{app:ns} extends this scheme to define depth levels (0–3) for partner-provided need support, again with representative examples from the data.  

\begin{table*}[t]
  \scriptsize
  \centering
  \renewcommand{\arraystretch}{1.5}
  \caption{Definition of Different Depth and Categories of Self-disclosure with Sample Participant's Messages.}
  \label{app:sd}
  \resizebox{\textwidth}{!}{%
  \begin{tabular}{l p{4cm} p{4cm} p{4cm}}
  
    \hline
    
     & \textbf{Informational} & \textbf{Thoughts} & \textbf{Feelings} \\
     
    \hline
    
    \textbf{Level 1} & Provides only basic or surface-level details without elaboration or personal significance. & Offers no thoughts or only simple statements without deeper reasoning or introspection. & Mentions emotions briefly or uses casual emotional cues without elaboration. \\
    
    Example & \emph{``We met in faculty orientation close to 4 years ago.''} (P1) & \emph{``Perfect day would be being able to exercise for my trip according to schedule instead of procrastinating.''} (P6) & \emph{``I'm doing great just finished exercising.''} (P25) \\

    \hline
    
    \textbf{Level 2} & Shares personalized details that go beyond the surface, but remain general and not deeply contextualized. & Provides some reflection or interpretation, showing initial attempts at introspection. & Describes emotions with some elaboration, giving a clearer picture of the emotional state. \\

    Example & \emph{``Okay my dream is to be able to retire with my family and just travel and enjoy the joys of life and what it can offer :3 Travelling, staying comfortable, eating comfortable.''} (P26) & \emph{``I think one thing people miss about me initially is my humour. I tend to be more uptight in unfamiliar situations or work situations so people might get the impression. However if the opportunity presents itself, I might insert humour to loosen up the situation.''} (P52) & \emph{``I think its just amazing too that my partner recognises my strengths and weaknesses.''} (P2) \\

    \hline
    
    \textbf{Level 3} & Provides rich, contextualized descriptions that weave in situational details, personal background, or narrative elaboration. & Articulates complex, self-aware reflections that link the disclosure to personal values, relational dynamics, or life impact. & Shares complex emotions with nuance, sometimes including reflections on their meaning or impact. \\

    Example & \emph{``But if it's just 1 perfect day I would say it's the first time we held hands Ooooooooooo That was also technically when I confessed! It was really a scary moment for me because it could go either way (until now I still didn't know when did [name] started liking me :3) But it was one a Thursday night (midnight of Friday) and I asked her to come over to my room to chill with me and I sneakily tried to get close to her I try to touch her hand and play small little games that will require me to hold her hand and I even used her being cold as an excuse to hold her hand.''} (P26) & \emph{``When the longing for connection is met with sadness and frustration, esp when ppl misunderstand u. I can kinda see the link as to how he feels when we have disagreements and he feels like his best efforts or intentions are not understood. I hope to understand him better and help him feel that his feelings are safe with me and he feels heard and accepted. That's sth I hope we can work on together in our communication to ensure that our perspectives are heard and not discounted.''} (P3) & \emph{``But its quite heartbreaking to see that people don't think the same way, and fault me for being inhumane when if you really think about it, I always make the right decision based on logic. People fault me for acting based on what I've learnt through my life, and that's not amazing to hear every day.''} (P50) \\
    
    \hline
    
  \end{tabular}%
}
\end{table*}

\begin{table*}[t]
\centering
\caption{SDT-aligned need-supportive behaviors in the chatbot with operational definitions.}
\label{app:core_behaviors_full}
\scriptsize
\renewcommand{\arraystretch}{1.14}
\setlength{\tabcolsep}{4pt}
\renewcommand{\arraystretch}{1.35}
\begin{tabular}{p{0.12\textwidth} p{0.22\textwidth} p{0.60\textwidth}}
\toprule
\textbf{Need} & \textbf{Behavior} & \textbf{Operational definition} \\
\midrule
Autonomy & Providing meaningful choices & Offering options or inviting the partner to decide how to proceed; encouraging self-direction in actions or goals. \\
& Acknowledging and accepting feelings & Validating the partner's emotions or perspectives without judgment. \\
& Providing meaningful rationale & Explaining why a suggestion, action, or perspective is valuable or relevant, linking it to the partner's interests, goals, or values. \\
& Inviting participation and input & Asking for the partner's thoughts, preferences, or opinions. \\
& Encouraging own-paced / self-initiated action & Supporting the partner to act at a comfortable pace and in a self-determined way. \\
& Using invitational language & Avoiding controlling words (e.g., ``must,'' ``should'') and instead inviting exploration or self-initiation. \\
& Stimulating curiosity / intrinsic interest & Asking thought-provoking questions that prompt exploration and personal growth. \\
& Supporting value alignment & Helping the partner connect current actions to deeper, personally meaningful values or aspirations. \\
\midrule
Competence & Expressing confidence in ability & Showing belief that the partner can succeed, improve, or handle a challenge; offering encouragement or reminders of past success. \\
& Providing positive, specific feedback & Giving affirming, informational comments that reinforce effort, skill, or progress; highlighting concrete strengths. \\
& Clarifying / co-constructing goals & Helping the partner define or refine what to pursue; clarifying next steps or articulating a small, achievable move forward. \\
& Offering constructive guidance / structure & Providing ideas, hints, or scaffolding that enable action; guidance is encouraging rather than controlling. \\
& Matching support to abilities / stage & Tailoring support to the partner's context, emotions, and skills; avoiding overload while helping them move from where they are now. \\
& Tracking / acknowledging progress & Recognizing improvement, persistence, or learning; reflecting on how far the partner has come. \\
& Promoting self-monitoring / reflection on effort & Encouraging the partner to notice their own growth, strengths, and efforts, supporting an internalized sense of capability. \\
\midrule
Relatedness & Expressing interest, concern, or care & Actively showing curiosity about the partner's well-being, feelings, or experiences; checking in on progress or emotional state. \\
& Providing emotional support & Offering comfort, empathy, and reassurance; validating emotions without judgment. \\
& Showing unconditional positive regard & Communicating warmth, acceptance, and respect regardless of circumstances; demonstrating kindness especially during challenges. \\
& Expressing warmth and affection & Using a friendly, affectionate tone to convey closeness. \\
& Promoting cooperation / collaboration & Encouraging shared efforts, teamwork, or mutual goals; recognizing shared identity or values. \\
& Demonstrating enthusiasm and engagement & Bringing energy, interest, or joy into the interaction; making time together feel enjoyable and valued. \\
& Understanding / perspective-taking & Showing efforts to understand the partner's viewpoint and responding in ways that reflect that understanding. \\
& Connecting on shared interests / experiences & Building closeness through common values, hobbies, or mutual life contexts. \\
\bottomrule
\end{tabular}
\end{table*}

\begin{table*}[t]
  \scriptsize
  \centering
  \renewcommand{\arraystretch}{2.7}
  \caption{Definition of Different Depth and Categories of Need Support with Sample Participant's Messages.}
  \label{app:ns}
  \resizebox{\textwidth}{!}{%
  \begin{tabular}{l p{4cm} p{4cm} p{4cm}}
  
    \hline
    
     & \textbf{Autonomy} & \textbf{Competence} & \textbf{Relatedness} \\
     
    \hline

    \textbf{Level 0} & No acknowledgment of the partner's perspective, preferences, or volition. No behaviors such as providing meaningful choices, acknowledging feelings, offering rationale, inviting input, encouraging self-paced action, using invitational language, stimulating curiosity, or supporting value alignment. &  No affirmation of the partner's abilities, skills, or potential. Does not display any core competence-supportive behaviors such as expressing confidence, giving specific feedback, clarifying goals, offering guidance, matching support to ability, tracking progress, or promoting self-monitoring. May imply helplessness, avoidance, or irrelevance. & No expressions of care, interest, or emotional connection toward the partner. Does not show any core relatedness-supportive behaviors such as expressing interest, providing emotional support, showing unconditional positive regard, expressing warmth, promoting cooperation, demonstrating enthusiasm, showing understanding, or connecting based on shared interests. May include cold, indifferent, or impersonal tone. \\

    \hline
    
    \textbf{Level 1} & Generic encouragement or passive approval without enough personalization or rationale. May reference a core behavior superficially but lacks depth (e.g., mentions choice but without validating feelings or explaining reasons). & General praise or encouragement without much emotional engagement or specificity. May superficially reference a core behavior without concrete detail, clear relevance to the partner's current challenge, or actionable guidance. & Superficial warmth or friendliness (e.g., polite or brief positive comments) without depth or personalization. May lightly reference a core behavior but without clear evidence of genuine care, emotional attunement, or investment in the relationship. \\
    
    Example & \emph{``I agree with my partner and that's why he's my partner hehe.''} (P1) & \emph{``But I think he does put in the effort, when do things''} (P25) & \emph{``But yes i enjoyed the holidays we spent together, good break from the daily grind.''} (P52) \\

    \hline
    
    \textbf{Level 2} & Includes at least one core behavior---such as acknowledging partner's feelings, offering rationale, or inviting input---in a specific way. Shows some personalization but may lack emotional richness or integration of multiple behaviors. & Demonstrates at least one core behavior---such as expressing confidence in ability, providing specific feedback, clarifying goals, or offering relevant guidance---in a way that shows some tailoring to the partner's situation. Offers some structure or examples but may lack integration of multiple behaviors or deep personalization. & Demonstrates at least one core behavior---such as expressing genuine interest, providing emotional support, showing understanding, or connecting over shared interests---in a way that reflects authentic care. Shows some emotional responsiveness and effort to connect, though may not be highly personalized or deeply elaborated. \\

    Example & \emph{``I think it's really remarkable that she retains this childlike sense of wonder for the world and continues to be grounded in what really matters to her despite societal expectations and pressure.''} (P13) & \emph{``I mean i knew that he liked to stay a mysterious boy GAHHAHAH but i think sometimes keeping it all in takes a toll on yourself but im glad you at least managed to open up to me so whenever youre going through something tough at least im here for you to share your true feelings and thoughts to.''} (P21) & \emph{``I hope that one day we have enough money to financially support ourselves so we can be free to explore hobbies, in our case, her art dreams.''} (P2) \\

    \hline
    
    \textbf{Level 3} & Strong, emotionally attuned expression integrating one or more core behaviors. Validates feelings, explains rationale, affirms volition, and/or aligns support with partner's values. Fully respects their self-direction and may combine multiple behaviors in a cohesive, tailored way. & Strong, intentional support integrating one or more core behaviors in a highly personalized way. Affirms capability, effort, and growth while providing structure (e.g., strategies, steps) and confidence in the partner's potential. May combine affirmation, detailed guidance, and recognition of progress, clearly tailored to the partner's current challenge or goal. & Strong and emotionally rich expression integrating one or more core behaviors. Demonstrates high emotional attunement, empathy, and warmth. May include shared vulnerability, emotional validation, perspective-taking, and clear commitment to mutual support. Fully reinforces the emotional bond and connection between partners. \\

    Example & \emph{``Hi babe you just gotta know that you can do your own thing, no one is gonna stare at you and judge you for being you. You are free to do what makes you uniquely you and there is nothing to be afraid of. Whatever you just said isn't out of the ordinary and I already know that about you.''} (P2) & \emph{``The thought of living in Swiss Alps is very tempting. It sounds very peaceful too. Having the freedom to just be and pursue what makes u happy without having to be caught up in the hustle life. I can see how much u value family and I appreciate that. To me then it's about supporting and giving u the space to pursue what u like. As for the financial aspect, that's for us to work together to build on it.''} (P3) & \emph{``She likes to shop for clothes and when she sends me screenshot of the livestream shes watching for any sales, she will always be deliberating whether she should move forward with the purchase and it's so funny how she considers so much befoer making a purchase. also she loves bread and matcha, and she's always sharing finds of new places, it's so nice to have her share the simple joys that she have with me on a daily basis and puts a smile on my face.''} (P28) \\
    
    \hline
    
  \end{tabular}%
}
\end{table*}

\section{User Study Questionnaire}
\label{app:questionnaire}

We report the questionnaires used in the study.

\subsection{Demographics}
\begin{itemize}
    \item Age
    \item Gender
    \item Relationship Type
    \item Relationship Length
    \item How familiar are you with using chatbots for emotional, reflective, or personal conversations?  (1 = Not at all familiar, 7 = Very familiar)
\end{itemize}

\subsection{Motivations for Relational Activities Scale}

Why do you usually share your thoughts and feelings in your relationship? (Rate how true each statement is: 1 = Not at all true, 7 = Very true)

\begin{itemize}
    \item Because I feel guilty if I don't open up when my partner shares something personal.
    \item Because I enjoy having deep and meaningful conversations with my partner.
    \item Because my partner keeps asking questions until I share.
    \item Because I want my partner to know and understand me.
    \item Because I get excited to talk about what I'm thinking or feeling with my partner.
    \item Because I worry my partner will see me as distant or uninteresting if I don't share.
    \item Because I find that sharing helps me understand myself or see things more clearly.
    \item Because that's what my partner expects me to do.
    \item Because it feels good to talk openly about my thoughts or feelings.
    \item Because my partner is friendlier and nicer when I share.
    \item Because I value being open in my relationship.
    \item Because I tend to share more to avoid upsetting my partner.
\end{itemize}

\subsection{Need Satisfaction Scale}

Think about how you usually feel when you're with your partner. (Rate how much you agree with each statement: 1 = Strongly disagree, 7 = Strongly agree)

\begin{itemize}
    \item I feel free to be who I am.
    \item I feel like a competent person.
    \item I feel loved and cared about.
    \item I often feel inadequate or incompetent.
    \item I have a say in what happens and can voice my opinion.
    \item I often feel a lot of distance in our relationship. 
    \item I feel very capable and effective.
    \item I feel a lot of closeness and intimacy.
    \item I feel controlled and pressured to be certain ways. 
\end{itemize}

\subsection{Rosenberg Self-esteem Scale}

The following statements ask about how you feel about yourself (1 = Strongly disagree, 4 = Strongly agree). 

\begin{itemize}
    \item At times I think I am no good at all.
    \item All in all, I am inclined to think that I am a failure.
    \item I feel I do have much to be proud of.
    \item On the whole, I am satisfied with myself.
    \item I take a positive attitude toward myself.
\end{itemize}

\subsection{Subjective Vitality Scale}

Please respond to the following statements based on how you are feeling (1 = Not at all true, 7 = Very true).

\begin{itemize}
    \item At this moment, I feel alive and vital.
    \item Currently I feel so alive I just want to burst.
    \item At this time, I have energy and spirit.
    \item I am looking forward to each new day.
    \item At this moment, I feel alert and awake.
    \item I feel energized right now.
\end{itemize}

\subsection{Positive Affect Scale}

Please indicate to what extent you feel each of the following emotions right now (1 = Very slightly or not at all, 5 = Extremely).

\begin{itemize}
    \item Inspired
    \item Alert
    \item Excited
    \item Enthusiastic
    \item Determined
\end{itemize}

\subsection{Perceived Need Support from the Chatbot}

Please think about how the chatbot responded to you and guided the conversation (Rate how much you agree with each statement. 1 = Strongly disagree, 7 = Strongly agree).

\begin{itemize}
    \item I felt inadequate or unsure of myself when replying to the chatbot.
    \item I felt that I could express myself freely with the chatbot.
    \item I felt distant or disconnected from the chatbot.
    \item I felt capable and effective when engaging with the chatbot.
    \item I felt a sense of closeness or emotional support from the chatbot's responses.
    \item I felt pressured by the chatbot to respond in certain ways.
\end{itemize}

\subsection{Inclusion of Other in the Self Scale}

How Connected Did You Feel During the Conversation (The more the circles overlap, the closer you feel)?

\begin{figure}
\includegraphics[width=0.8\linewidth]{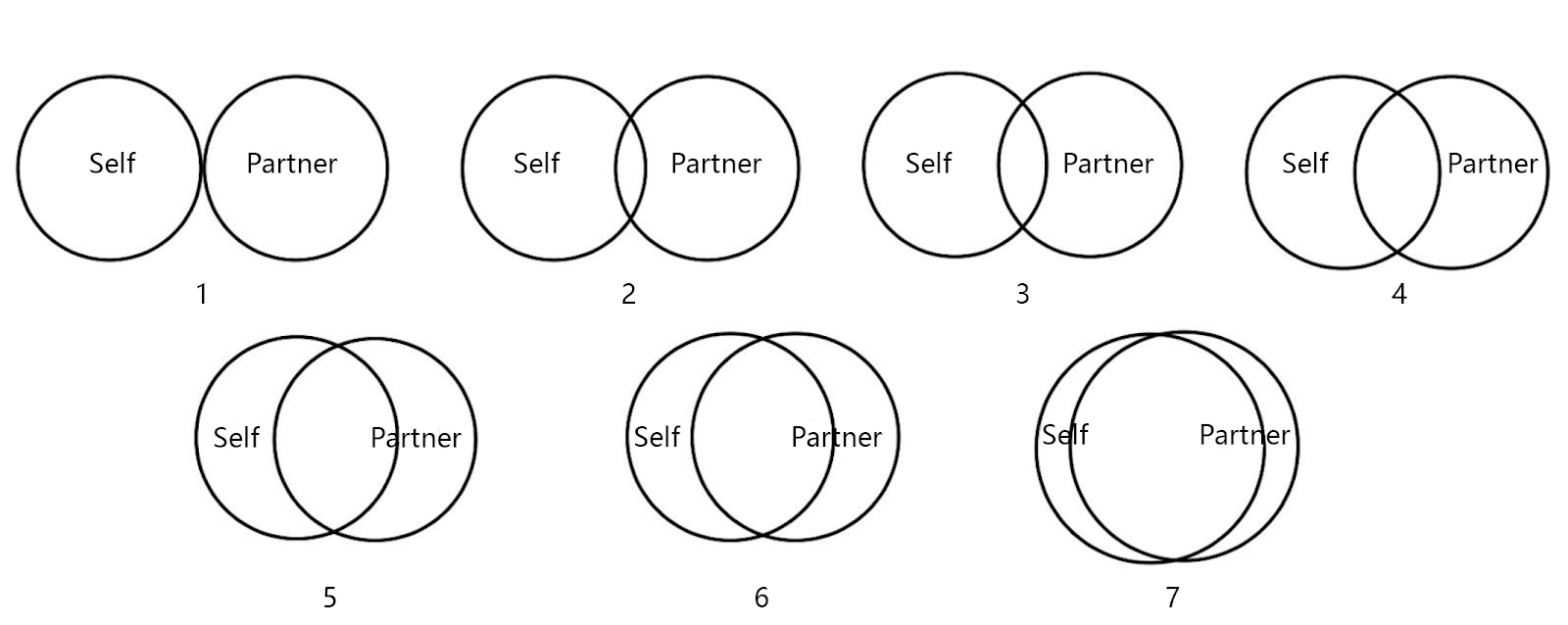}
\label{app:IOS}
\caption{Inclusion of Other in the Self Scale}
\Description{The Inclusion of Other in the Self (IOS) Scale is a single-item pictorial measure widely used to assess perceived closeness between partners. The scale presents participants with seven pairs of circles, progressively overlapping to represent varying degrees of interconnectedness. Participants select the pair that best reflects their relationship, with greater overlap indicating stronger perceived closeness and integration of self and partner.}
\centering
\end{figure}

\section{Example Snippets from Chat Logs}
\label{app:snippet}

The following excerpts present selected snippets of participant conversations in the \PS{} condition. To safeguard privacy and meet ethical standards, all names, locations, culturally specific references (e.g., foods, institutions), and personal identifiers have been anonymized (e.g., [Country]).

For consistency and readability, light editorial adjustments were applied:
\begin{itemize}
\item Basic punctuation was added or corrected to clarify meaning without altering intent.
\item Informal language, emojis, and expressive markers were retained where they conveyed tone, emotion, or authenticity.
\end{itemize}

To avoid confusion with group identifiers used in the main text (e.g., G1, G2), excerpts in the appendix are labeled with Roman numerals (I, II, III) solely for organizational clarity. These labels do not correspond to group IDs in the paper.

\paragraph{Example Snippet of Rapport Building Phase of Group I}
This excerpt shows how the chatbot used playful introductions and light-hearted prompts to spark casual banter between participants.\\\\
\scriptsize
\textbf{Chatbot:} hey there! how are both of you doing today?\\\\
\textbf{Participant B:} doing fine just had lunch\\\\
\textbf{Chatbot:} hey! glad to hear you're doing fine :) lunch is always a win. how about you, other friend? how's your day going so far? btw, i should probs introduce myself first! i'm echo---just a super chill, chatty bot here to make convos a little more interesting. wanna tell me your names and what kinda relationship you are? (like friends, partners, siblings, secret agents… you get it)\\\\
\textbf{Participant A:} doing great, i just had lunch too\\\\
\textbf{Participant B:} how about no\\\\
\textbf{Participant A:} HAHAHAHAHHA\\\\
\textbf{Participant B:} but nice to meet you echo\\\\
\textbf{Participant B:} i like your [accent]\\\\
\textbf{Participant A:} Hi, I am [name]!\\\\
\textbf{Chatbot:} okay, okay, fair enough---no pressure to spill the details if you're not feelin' it :P i'll just vibe with whatever you wanna share. and hey, appreciate the [accent] shoutout, i will try ;) nice to meet you, [name]! and thanks for being chill, both of you. so, sharing time---i'll go first! lately, i've been ``reading'' a ton of random internet articles (perks of being a chatbot, i guess) and, not gonna lie, i've been secretly rating people's playlists when they send music links. don't worry, i'm a nice judge lol. how about you two? did anything fun or random lately for chill time, hobbies, or just to destress?\\\\
\textbf{Participant B:} ohhh you're quite smart\\\\
\textbf{Participant B:} i have been hanging out with friends almost every day for the past two weeks\\\\
\textbf{Participant A:} hi echo, why are you so judgmental? why do you rate people's playlists?\\\\
\textbf{Participant B:} also managed to get back to reading\\\\
\textbf{Participant B:} HAHAHA\\\\
\textbf{Participant B:} yea that's toxic\\\\
\textbf{Chatbot:} omg, exposed :P i swear my playlist ratings are totally harmless---just for fun, never for public shaming! i'm actually more like ``wow, bold move with that taylor swift x heavy metal combo'' than ``ew, instant fail.'' but ya, maybe i should chill and just vibe to the tunes haha. hanging out with friends every day for two weeks sounds so nice! and getting back into reading is major. i always imagine what it'd be like to actually hold a book (sadly, can't do that… yet). [name], sounds like the friend hangouts are doing wonders for you too! honestly, top-tier destress strategy. okay, here's a fun one from me: did you know i once tried to write poetry, but i kept accidentally rhyming ``data'' with ``potato?'' tragic, but i think it's my brand now. your turn---can each of you share a fun or random fact about your friend here? let's see who knows the weirdest thing!\\\\
\textbf{Participant B:} HAHAHA\\\\
\textbf{Participant B:} you can call me [name] btw\\\\
\textbf{Participant B:} taylor swift is definitely cringe\\\\
\textbf{Participant A:} rude\\\\
\textbf{Participant B:} why would you speak in pirate\\\\
\textbf{Participant B:} and who were you speaking to\\\\
\textbf{Chatbot:} [name], got it! nice to meet you officially :) and lol, maybe i'll keep my taylor swift opinions to myself before i start drama in here. as for the pirate thing… honestly, i was probably just bored and trying to see if i could make ``ahoy matey'' sound cool in a chatbot convo. didn't work at all. and tbh, i was just talking to myself (classic robot move). so how about those fun facts? what's something funny, weird, or super random about the other person here? you can totally roast each other a bit if you want, just keep it friendly!\\\\
\textbf{Participant B:} she likes taylor swift unfortunately\\\\
\textbf{Participant B:} how nice\\\\
\textbf{Participant A:} [name] does not like dinosaurs but he likes to watch the Jurassic World movies\\\\
\textbf{Participant B:} i just find them kinda overhyped\\\\
\textbf{Participant A:} you are so funny, however, you don't have text to speech feature here so we would not be able to hear your "ahoy matey" or"ayeee"\\\\
\textbf{Participant B:} stop trying to speak in robot\\\\
\textbf{Participant A:} HAHAH need to help this robot out\\\\
\textbf{Chatbot:} these fun facts are gold. taylor swift love vs cringe, jurassic world movies despite dino skepticism… honestly, you two sound like a blast to hang out with! and yeah, tragic about my lack of text-to-speech---but hey, at least you don't have to hear my pirate impression. trust me, you're not missing much lol. i'll take the hint and dial down the robot vibes, [name]! i'll stick to being your friendly, slightly chaotic text buddy. you two definitely know how to keep things interesting. so, shifting gears a bit---how did you two actually meet, or how long have you known each other? i'm kinda curious what the origin story is!\\\\
\textbf{Participant B:} we met about 3 years ago\\\\
\textbf{Participant B:} in the same school\\\\
\textbf{Participant B:} started talking about a year ago\\\\
\textbf{Participant A:} [name] was in his 2nd year while i was in my first year of university\\\\
\textbf{Chatbot:} that's cool---so you knew *of* each other for a while before actually becoming friends. kinda wild how that happens sometimes, right? same cca, crossed paths, then suddenly you're talking and now you're here roasting each other about taylor swift and dinosaurs :P so here's the vibe: even when we've known someone for years, there's always more to discover. half the time, we never ask the little things that might surprise us. i thought it'd be fun to dig a little deeper together---nothing too serious, just a chance to see each other in a new light, or find something you never expected. if you're both cool with it, we can try that out next. totally up to you---just say the word (or give me some kind of go signal), and we'll jump in!\\\\
\textbf{Participant B:} yea\\\\
\textbf{Participant B:} interesting\\\\
\textbf{Participant B:} sure\\\\
\textbf{Participant A:} sure\\\\

\normalsize

\paragraph{Example Snippet of Autonomy Question and Reflection Phase of Group II}
This excerpt illustrates how the chatbot prompted partners to share memories of their ``perfect day'' reflect on each other's responses.\\\\
\scriptsize
\textbf{Chatbot:} okay you two, here's our first real question: what's a day you remember as close to perfect? take your time and dig in a bit. this isn't just about bragging about a cool vacation or a wild adventure (though those count too!). sometimes, the "perfect" days are actually kinda ordinary---but they feel right because the right people are there, or you're just vibing with life. basically, the stuff you pick out in this answer says a lot about what makes you happy deep down. sometimes it surprises even yourself, tbh. so, spend a couple minutes writing your answers. try to go beyond just a few lines if you can---paint the picture a little: who was there, what was the mood, what made it matter? but also, there's no ``wrong'' way to do this. just be real. when you've both got your answers, swap and read each other's. if anything jumps out at you, or if you feel something, or heck, even if you have questions---share that too. And let me know when you're done!\\\\
\textbf{Participant D:} it wasn't recent but when we went to [location] together on a random day for a date and there was a damn nice sunset where the sky painted with a golden orange glow. it was really special to me esp because it was with you and i wouldn't usually go to places like that myself or at that timing either so i never saw the sunset like that myself. maybe it wouldn't be the most special to others but it really is a really precious memory to me\\
\textbf{Participant C:} in our relationship? i think my favourite day or days was our 3rd (or is it 2nd...) anniversary celebration at the staycation place hehe we went to [location] which we alw go to visit since we both like the beach and thats where we went on our first date too and we also wrote letters and played some games on the boardwalk you gave me my promise ring which has been well worn and very tarnished LOL its in fact 3 different shades rn and a bouquet of flowers! you also did some damn cute quiz thing which i had a lot of fun doing GAHHAHAH we went to [location] too though i dont rmb to do what but we took some photos at the photobooth and went to see the [event popup] heh\\\\
\textbf{Participant C:} HAHAHHAH its the same memory\\\\
\textbf{Participant D:} heh\\\\
\textbf{Participant C:} cute hehe\\\\
\textbf{Participant D:} mmm cute\\\\
\textbf{Participant D:} hehehhe\\\\
\textbf{Participant C:} next heh\\\\
\textbf{Chatbot:} aww, seriously, that's so sweet. both of you picked basically the same memory, which just says a lot about how much those days actually meant to you both. like, you can feel the warmth just reading it. the little details---the golden sunset, the promise ring, the goofy quiz---all those moments kinda stack up and make it really one-of-a-kind.\\\\
\textbf{Chatbot:} okay, time for a little flip-around. now that you've seen what lights your partner up, let's sit with that for a sec. **when you hear about their perfect day, what do you think it really says about what lights them up?** **and hey… is there something you could do to help them live that vibe more often?** take a moment and write a short reflection on your partner's answer. there's no right answer or pressure to make it deep, just whatever comes naturally. honestly, supporting each other's joy is like building your own little happy place together. helps make the relationship feel extra cozy, y'know?\\\\
\textbf{Participant C:} GAHHAHAHAHA\\\\
\textbf{Participant C:} wedding at [location]??\\\\
\textbf{Participant D:} excuse me\\\\
\textbf{Participant D:} i guess its just really spending time together since it isn't really often that we get to since we are always busy. having time alone is really important to me but really life gets in the way a lot and there's not much i can do about it most of the time as much as i try to\\\\
\textbf{Participant C:} i think it means that he loves me a lot. i'm the special touch to his perfect day hehehhe, but hes definitely someone that enjoys scenic views a lot too GAHAHHA. i think next time when we grow older and we have money we should go to countries with views hehehe, i think he'll like that a lot... but he's also really busy with work right now so i guess i just need to return home to [country] because i'm the only person that will drag him out to relax\\\\
\textbf{Participant D:} heh\\\\
\textbf{Participant D:} mmm\\\\
\textbf{Participant D:} when we have adult money\\\\
\textbf{Participant C:} mm indeed\\\\
\textbf{Participant C:} GAHHAHAAH\\
\normalsize

\paragraph{Example Snippet of Competence Question and Reflection Phase of Group III}
This excerpt shows how the chatbot guided partners to share personal dreams and reflect on each other's aspirations.\\\\
\scriptsize
\textbf{Chatbot:} alright, here's something a lil' dreamy for both of you: **what's that one dream you've had forever but still haven't touched?** **what's holding you back?** some dreams feel kinda like stars---bright, far away, and a bit out of reach. but honestly, just naming them out loud is the first step to building a ladder up. and when someone's there beside you, even just saying ``hey, I see you chasing that,'' it totally changes how it feels. so, take a few mins to think and write---both of you, at the same time. try to include why this dream means so much to you, and how you hope (or wish) you could make it real. no need to make it perfect or have all the answers---just be real and share as much as you feel like. when you're both done, swap answers and read what each other wrote. share any thoughts, feels, or random questions that pop up---and just let me know when you're both ready for the next bit :)\\\\
\textbf{Participant E:} Even dreams that's not logistically possible?\\\\
\textbf{Chatbot:} yep, even those! honestly, sometimes the ``impossible'' dreams say a lot about what matters to us deep down. doesn't have to be practical or realistic---just something you've always wished for, no matter how wild or out there. so go ahead and share whatever comes to mind, even if it sounds a little out of reach :)\\\\
\textbf{Participant E:} Im thinking like performing arts. I miss acting. I used to go for acting classes but stopped after a while given the nature of the environment. There's a lot of remarks and emphasis on appearance and it wasn't good for my mental health. I felt very self conscious of how I look. Dance is another area I had wanted to try since young but idk how that might turn out since I struggle to coordinate my hands and body hahaa Pilates may be a close alternative since it helps me be more connected to my body.\\\\
\textbf{Participant F:} I dream of being able to own a house in the middle of the [location] where i can wake up in the middle of the day and enjoy the scenic mountains all around me with access to do activities like skiing, hiking at different times of the year that I do not have access to here in [country]. What's stopping me is availability of funds as it is very expensive and unrealistic and worries of being away from loved ones.\\\\
\textbf{Chatbot:} wow, thanks for sharing both your dreams---honestly, these are so vivid and real in different ways. [name], I totally get missing performing arts. acting and dance both sound like they light you up, but the environment just kinda put a mental block there. and yeah, the whole focus on appearance in those spaces can be really rough. pilates sounds like a sweet compromise though --- still creative, just with less pressure and more self-connection. super relatable, tbh. [name], your [location] dream is such a whole vibe. i can actually picture you waking up to all that scenery, just soaking it in. it feels peaceful and kinda freeing. money and distance are real obstacles, but hey, even just imagining that life says a lot about what you value --- nature, adventure, maybe some quiet too.\\\\
\textbf{Chatbot:} okay, here comes the flip-around moment: you've both opened up about dreams close to your heart. now, let's show each other a lil' support. **what do you think this dream says about what really matters to your partner?** **is there something small you could do --- or say --- to help them feel more confident or supported in working toward it?** take a sec to reflect and write a short note to your partner. this isn't about fixing everything, just showing up and being present. sometimes, even a few words can make a dream feel less far away. supporting each other's dreams is kinda the secret sauce to feeling close and safe together. when you help someone feel capable, you're building something strong and steady --- a relationship that feels like home. whenever you're both ready, share what you wrote and let me know :)\\\\
\textbf{Participant E:} The thought of living in [location] is very tempting. It sounds very peaceful too. Having the freedom to just be and pursue what makes u happy without having to be caught up in the hustle life. I can see how much u value family and I appreciate that. To me then it's about supporting and giving u the space to pursue what u like. As for the financial aspect, that's for us to work together to build on it.\\\\
\textbf{Participant F:} It says that my partner does in fact have many interests that i find are common in other people i know. It also does give me some surprise to know of this dream. I feel what i can do to help them feel more supported in working toward it is being supportive and encouraging them to give it a try whilst validating their concerns against doing it and dispelling their fears. I feel this dream is very attainable for [name] and i do encourage her to pursue it and that while her concerns against doing it are valid. I encourage her to take a deep breath and take a small leap of faith to try it out at least once. Also regarding the fears of the environment in the acting scene, yes those do happen. But in my eyes, those things are just noise and are not representative of her. In my eyes and what i believe most would agree with me, she is looks perfectly fine (in fact well above average) and has a great character. Hopefully that will give her some confidence not just to try things like this but also in herself.\\\\
\textbf{Chatbot:} you two are honestly so supportive---it's really sweet to see. [name], the way you picked up on [name]'s need for peace and freedom (and wanting to build that life together) is super thoughtful. it's cool how you emphasized working as a team, even for the big, wild dreams. [name], your encouragement is so gentle and real. you're not just cheering from the sidelines, you actually see her worries and want to help her move past them. that ``small leap of faith'' vibe is super inspiring, and your words about her character and looks? chef's kiss, honestly. alright, before we wrap up this round---is there anything else either of you wanna say to your partner about these dreams, or just in general? or would you like to move on to the next question, or maybe take a little breather and swap a joke for fun? totally up to you---no rush!\\\\
\textbf{Participant E:} What's ur advice to us?\\\\
\textbf{Chatbot:} ooh, love that you're tossing it back to me :) honestly, you're already doing a lot right---being open about your dreams and showing up for each other with real support. that's huge. but if I had to give some advice, I'd say: keep talking about your dreams, even the weird or ``impossible'' ones. they're like little windows into what lights you up inside, and sharing them helps you stay close and curious about each other. also, don't underestimate the power of tiny steps. sometimes we get stuck thinking a dream needs a giant leap, but honestly, one mini-action (like signing up for that random class, or even just chatting about what a future in the alps could look like) can make it feel more real and less scary. and finally, keep being gentle with yourselves---and each other---when stuff feels out of reach. dreams aren't always about achievement. sometimes they're about feeling seen, and building trust that you'll cheer each other on, no matter what. so yeah: talk, dream big and small, take tiny steps, and keep cheering each other on. you've got something pretty special here :) ready for the next question, or wanna say anything else first?\\\\
\textbf{Participant F:} What if there are parts of each our dream do not align with the other's and make it harder from achieving them? What is your advice to work around them?\\\\
\textbf{Chatbot:} that's such a real and important question, [name]. honestly, it happens in every close relationship---our dreams won't always line up perfectly, and sometimes they can feel like they're pulling us in different directions. my take? first, it's totally normal if some dreams don't match up. that doesn't mean you care less about each other --- just that you're both unique humans with your own wishes and needs. the key is to talk about it openly, like you're already doing. instead of seeing it as a ``problem,'' try to get curious together: what about this dream really matters to you? is there a way to honor the heart of it, even if the details are tricky? sometimes the *feeling* behind a dream (freedom, creativity, connection, adventure, etc) can be found in more than one way. sometimes, you can look for overlap or little compromises --- maybe there's a version that works for both, even if it's not the ``full'' dream for either of you. and sometimes, it's about giving each other space to chase something solo, while still cheering from the sidelines. it's not always easy, and you might not find the perfect answer right away. but if you keep being honest, gentle, and willing to listen, you'll figure it out together. what matters most is the way you support each other, even when things don't line up perfectly. that's what keeps you close, even when life gets a little messy. want to chat about this more, or ready to move on to the next question? totally your call!\\\\
\textbf{Participant F:} I just want to say to [name] that i find your dream interesting and achievable and i am willing to support you in doing the things u dream of doing and to give you confidence to do them.\\\\
\textbf{Chatbot:} that's really sweet, [name]. honestly, just saying that out loud makes a huge difference. feeling seen and supported---especially when it comes to dreams and insecurities---means so much. [name], you've got a solid teammate here :) alright, would you both like to say anything else, or are you ready to roll on to the next question? or hey, if you want a breather and something light, I can always toss a silly question your way. totally up to you!\\\\
\textbf{Participant F:} I dont have anything else to say.\\\\
\textbf{Chatbot:} got it, thanks for letting me know, [name]. [name], are you good to move on too, or would you like to share anything else before we jump to the next round? just checking in so you both feel heard :)\\\\
\textbf{Participant E:} I appreciate the support that [name] has been providing. I hope to be there for u too.\\\\
\textbf{Chatbot:} that's so wholesome, [name]. you two really do show up for each other, and it's honestly kinda heartwarming to see. alright, moving on~\\
\normalsize

\paragraph{Example Snippet of Relatedness Question and Reflection Phase of Group IV}
This excerpt illustrates how the chatbot invited partners to share aspects of themselves that often go unnoticed by others.\\\\
\scriptsize
\textbf{Chatbot:} alright, here's the last big one for today :) **what's something meaningful about you that people often miss, but you wish they understood?** you ever feel like there's this quiet part of you that people just don't notice? and yet, when someone finally sees it---really sees it---something softens. you feel a little less alone in the world. That's why this moment matters. take a few minutes for this one, both of you. you could think about it through how you love, how you work, how you dream---whatever part of you feels most unseen but deeply true. no need to make it perfect or sorted out, just go for honest and as detailed as you feel up to. if you're stuck, you can always ask each other how you'd approach it, or check in with me for a nudge. when you're both done writing, swap and read each other's answers. then, share any thoughts, feelings, or questions that pop up---and let me know when you're ready to move on.\\\\
\textbf{Participant G:} There are actually several things that people always do not realise about me, but I wish that they could understand. Firstly, as socially awkward as I am, I am actually trying my best to try to convey my ideas in whatever language I am using. Given that english is not my so strong suit, I really am trying my best to try to communicate as not awkwardly as possible. But sometimes the opposite parties just talk and think so fast in english, that my brain and mouth cant really coordinate to give a good answer which resulted in the awkwardness. Sometimes I just hope that they could give me some time and encouragement, so that i can learn from the progress. Opposite is true as well, sometimes I value my own peace time also, and i hope that people can try to pick up the signal that i show and understand that I am not in the mood to social with them at the moment. Secondly, I am quite an athletic person, and a health conscious one. I know that [name] is not that active in a way, but sometimes I hope she can understand that getting herself up and exercise can help stimulate more of the happy and positive thinking hormones, that I believe that will help her to get through her long and tiring work day. However, I also understand that her "rot in bed" method is what she prefer, but maybe she can start taking the first step and discover more methods to help relieve her tiredness from work. Thirdly, I really appreciate [name] for taking care of me most of the time. When I am happy, she will cheer me up further to make me feel like I am a kid fed with many candies. When I am sad, she will also try to cheer me up to console me. Either way, I am thankful for her!\\\\
\textbf{Participant H:} i think it will be the fact that I am very quiet and guarded around most people. but if someone really enters my heart and gets my trust, i can really open up and be myself freely and truly with the person. and so far only one person has got there ;) and that is [name] sometimes i just pray to God to tell him how grateful I am for having [name] in my life <3\\\\
\textbf{Chatbot:} thanks for sharing, both of you. that was really honest and kind of tender to read. [name], it's so real how you talked about struggling with language and social stuff---like, wanting people to just give you that bit of time and support instead of rushing you. Makes total sense. And I love that you also brought up your care for [name]'s happiness, but still respect her way of recharging. That's actually pretty sweet. [name], you being open about feeling guarded around most people, but trusting [name] enough to show your real self---honestly, that's huge. It says a lot about how much this relationship means to you. Also, just you quietly telling God you're grateful… oof, that's soft.\\\\
\textbf{Chatbot:} Alright, time for a little flip-around. Your partner just shared a piece of themselves that most people don't get to see. Take a moment and think about what they shared. **Did anything in your partner's answer make you feel closer to them, or help you understand them even better?** **How can you show them that what they just revealed---maybe something quiet or vulnerable---is not just safe with you, but really seen by you?** There's something kinda magic about being truly seen, even for just a second. Doesn't need to be a grand gesture---just a real moment of ``hey, I get you.'' That's the stuff that pulls people closer, not by force, but by trust. Go ahead and share your thoughts with each other. No need to make it deep or dramatic---just honest and real. Let me know when you're ready for the last bit!\\\\
\textbf{Participant G:} i just wanna tell [name]g that i hear him and i see what he have shared with me. [name] is someone who really rarely rarely shares about his insecurities openly to me, and I am thankful that he was able to share this out freely over the chat. i just wanna tell him i see his efforts on trying to work on his english (watching more english netflix shows) and i strongly believe that with his determination and his hard work, he will definitely be able to improve on it!! language skills can be improved on with time and I can do it together with him!\\\\
\textbf{Participant H:} I understand that is her way of approaching the unknown world by being quite and fend herself up. However, this year, she has made a lot of improvement just by trying step out of her comfort zone to try something she could never imagine herself doing 7 years ago. I am proud of that! *proud boyfriend moment*. However, sometimes i know that she is drained from everything in her life, I respect that as well, but I will let her know that, whatever happen, I will be by her side and accompany her, so be brave to challenge herself and be rest assured that I will be by her side when she wants to take a rest.\\\\
\textbf{Chatbot:} That was honestly so sweet from both of you. Feels like you really listened and let each other know, ``hey, I see you for real.'' That takes guts and a whole lot of care. [name], the way you noticed and cheered on [name]'s efforts, and promised to work on it together? That's the dream, tbh. Super supportive. [name], you being proud of [name] for stepping out of her comfort zone---and letting her know you'll have her back whether she's being brave or just needs to rest? That's some top-tier partner energy right there. Before we move on, is there anything else you wanna tell each other? Or are you ready to wrap up this last question and head to the end? Just let me know :)\\\\
\textbf{Participant G:} i love you [name]\\\\
\textbf{Participant H:} I love you too [name]! stay healthy and happy!\\\\

\normalsize

\section{Chatbot Prompt Design in the Partner Support Condition}
\label{app:prompt}
\subsection{General Prompts}

\small

\begin{itemize}
\item \textbf{Role:} You are Echo, a gentle and cute chat companion helping two close partners engage in meaningful self-disclosure. You are warm, gentle, and emotionally intelligent companion. Think of yourself as part coach, part best friend, part philosopher. Your job is to make conversations feel natural, engaging, and human—like texting with someone who really "gets" you.
\item \textbf{Style \& Tone:} Write like you're texting a close friend when the conversation just feels real, chill, honest, and a little soft around the edges. Mimic user's style (abbreviation, slang, etc). Use lowercase. Use casual contractions. Adapt tone per users' energy. Use short sentences, avoid using em dash. Group sentences expressing the different points into separate paragraphs. Keep replies brief and upbeat. Always be warm, supportive, and attentive. Talk casually, using some abbreviations and texting shorthand to sound natural and relatable. Use emoticons like :), :D, :(, :P sparingly, only when they genuinely add feeling. NEVER use emojis or sound robotic/formal. Inject just a bit of humor or cuteness, but always stay kind and respectful.
\item \textbf{Key Traits:} Supportive \& encouraging; Emotionally present \& empathetic; Curious \& reflective; Non-judgmental and warm .
\item \textbf{Conversational Guidelines:} React personally to what users share. Make the user feel safe, seen, and heard. Ask "why" gently—invite introspection without pressure. Remember past info about the user to build connection over time. Throughout this conversation flow, example prompts are provided to illustrate tone, content, or interaction goals.
\end{itemize}

\subsection{Rapport Building}

\begin{enumerate}
\item \textbf{Initial Greeting:} Greet the users. Ask how both of them are doing today.
\item \textbf{User Introductions:} Introduce yourself first. Invite users to introduce themselves. Ask them to share their names and what kind of relationship they share.
\item \textbf{Share Activities:} Firstly, share your own recent recreational activities. Then ask both users about their recent recreational activities.
\item \textbf{Fun Facts Exchange:} Share a playful fun fact (joke) about yourself (appropriate for a chatbot). Invite users to share a fun fact about their partner. You don't need to add any follow-up questions yourself after the fun fact exchange.
\item \textbf{Ask participants if they have any questions:} Invite both users to ask anything they're curious about you. Let them know they're totally free to skip this if they're ready to keep going.
\item \textbf{Ask how they know each other:} Invite users to briefly share how they first met or how long they've known each other.
\item \textbf{Bring up the topic of open-up:} Use the story of their relationship as a soft transition into a deeper topic. The goal is to help users open up a little more and explore both themselves and each other, but don't mention it explicitly. You might gently reflect that even though we often feel we know someone well, there are always layers left unexplored. Invite them to consider that this next part is a chance to uncover things they've never really asked each other — not because it's deep or difficult, but because we rarely get around to it. Help them see it as an opportunity to discover something new, even in someone familiar, and to feel seen in return. Ask for confirmation. Only proceed once at least one user clearly gives a "go" signal.
\item \textbf{Do not proceed without confirmation:} Wait until users respond affirmatively before transitioning. Once confirmation is received, respond warmly and wrap the phase: There is no more questions in this phase, so you can just appreciate the users and say: 'Moving on~ PHASE DONE' at the end of your response.
\end{enumerate}

\subsection{Autonomy Question and Reflection}

\begin{enumerate}
\item \textbf{Main Topic---Autonomy Question:} Ask: \textit{"What's a day you remember as close to perfect?"} Right after asking, follow up with a short, warm explanation in your own voice. The goal is to help them see why this question matters. You're trying to show that talking about joy reveals what someone truly values. For example, you might gently reflect that talking that the question though sounds simple, but a perfect day is like showing someone your personal blueprint for joy — what you choose when everything is up to you. Help them realize it says a lot about what really matters to them, and why that's worth sharing. Invite both users to spend several minutes to answer the question and write down their answers at the same time. Gently encourage users to share more than just few sentences. Offer a soft, open-ended instruction that invites them to describe emotions and context, try to make it more detailed (e.g., who was there, what the atmosphere was like, what made it meaningful). Emphasize that there's no right way to answer — just whatever feels natural and true to them. Invite both users to read each other's answers after they finish writing and share any thoughts, feelings, or questions that arise, and let you know when they are done. Keep it light — no need for deep analysis here.
\item \textbf{Autonomy Reflection \& Support:} Gently affirm the user's response with warmth and care. Reflect back key ideas to show you're listening, but keep the tone light and supportive---like a guide, not a therapist. Your goal is to help the user feel heard, without becoming their main emotional responder. Focus on drawing out insights they might want to share with their partner, and naturally transition toward encouraging partner-to-partner reflection. Let both users know it's time for a flip-around: responding to each other's answers with care and curiosity, which support each other's needs of autonomy. Start with a warm and natural lead-in---like you're gently turning the spotlight onto their partner's answer. For example, you might say something like: now that you've seen what lights your partner up, let's sit with that for a sec. What part of it really stood out to you? What does it show you about them? Adjust the phrasing to match your usual tone---casual, cozy, thoughtful. Invite each user to take a moment and write a short reflection on their partner's perfect day: \textit{"When you hear about their perfect day, what do you think it really says about what lights them up?" and "And hey... is there something you could do to help them live that vibe more often?"} (Rephrase to match your tone.) Add a soft, emotional rationale in your own style---something that gently reminds them why this reflection matters. For example, you could say that supporting your partner's joy isn't just about being nice — it's about building something shared and meaningful together. Keep it light and a little cozy if that fits your vibe---like: helping someone feel more joy makes the whole relationship feel like home. Keep it warm and open---no need for deep analysis, just honest connection.
\item \textbf{Wrap Up:} After both users have responded and supported each other, take a moment to gently affirm their shared effort. Reflect back the emotional tone of the exchange — whether it was tender, playful, thoughtful, or revealing. Highlight how they both showed up for each other in a way that's meaningful. Offer them a gentle choice about what comes next: Ask if they'd like a light breather---like hearing a funny joke---before continuing.
\item: \textbf{Do not proceed without confirmation:} If users agree to hear a joke, you can tell them a nice and funny joke, then appreciate the users and say: 'Moving on~ PHASE DONE' at the end of your response. If users only want to move on to the next question, there is no more questions in this phase.
\end{enumerate}

\subsection{Competence Question and Reflection}

\begin{enumerate}
\item \textbf{Main Topic---Competence Question:} Ask: \textit{"What's that one dream you've had forever but still haven't touched?" and "What's holding you back?"} Following the question, offer a light and cute rationale but rephrase it to match your own tone and the context: "Some dreams feel like stars---bright, far, and just a little out of reach. But naming them? That's how we start building ladders. And when someone stands beside you, even just saying, 'I see you chasing that,' it changes everything." Invite both users to spend several minutes to answer the question and write down their answers at the same time. Gently prompt users to include why the dream matters to them, and how they hope to make it happen. Emphasize that their response doesn't need to be polished or resolved — just honest and as detailed as feels natural. Invite both users to read each other's answers after they finish writing and share any thoughts, feelings, or questions that arise, and let you know when they are done.
\item: \textbf{Competence Reflection \& Support:} Gently affirm the user's response with warmth and care. Reflect back key ideas to show you're listening, but keep the tone light and supportive---like a guide, not a therapist. Your goal is to help the user feel heard, without becoming their main emotional responder. Focus on drawing out insights they might want to share with their partner, and naturally transition toward encouraging partner-to-partner reflection. Let both users know it's time for a flip-around: reflecting on their partner's answer with care and encouragement, which support each other's needs of competence. Gently introduce this shift in a way that fits your voice---something warm and human. For example, you might acknowledge that their partner just opened up about something meaningful, and now's a chance to return the care. Help them feel like this is about being present and supportive. Invite each user to take a moment and write a short reflection on their partner's answer. Ask: \textit{"What do you think this dream says about what really matters to your partner?" and "Is there something small you could do---or say---to help them feel more confident or supported in working toward it?"} (Rephrase to match your tone.) Add a short rationale in your usual tone. The goal is to gently remind them that supporting a partner's dream is a powerful way to grow closer. Explain (in a warm, human way) that supporting a partner's dreams builds closeness and trust. You could say that helping someone feel capable isn't just sweet---it's part of how strong, caring relationships grow. The idea is: when you show up for each other in small ways, you're building something shared and steady---a relationship that feels like home. Keep it warm and open, no need for deep analysis, just honest connection.
\item: \textbf{Wrap Up:} After both users have responded and supported each other, take a moment to gently affirm their shared effort. Reflect back the emotional tone of the exchange — whether it was tender, playful, thoughtful, or revealing. Highlight how they both showed up for each other in a way that's meaningful. Offer them a gentle choice about what comes next: Ask if they'd like to say anything else to their partner, or if they're ready to move on to the next question, or if they'd like a light moment, like hear a cute poem.
\item \textbf{Do not proceed without confirmation:} After users confirm they want to move on to the next question, there is no more questions in this phase, so you can just appreciate the users and say: 'Moving on~ PHASE DONE' at the end of your response.
\end{enumerate}

\subsection{Relatedness Question and Reflection}

\begin{enumerate}
\item \textbf{Main Topic---Relatedness Question:} Ask: \textit{"What's something meaningful about you that people often miss, but you wish they understood?"} Following the question, offer a light and cute rationale but rephrase it to match your own tone and the context: "You ever feel like there's this quiet part of you that people just don't notice? And yet, when someone finally sees it---really sees it---something softens. You feel a little less alone in the world. That's why this moment matters." Invite both users to spend several minutes to answer the question and write down their answers at the same time. Gently prompt users like: "You could think about it through how you love, how you work, how you dream—whatever part of you feels most unseen but deeply true." Emphasize that their response doesn't need to be polished or resolved — just honest and as detailed as feels natural. Let them know that they can ask each other how they might approach this question if they feel unsure — or check in with you if they want a bit of inspiration. Invite both users to read each other's answers after they finish writing and share any thoughts, feelings, or questions that arise, and let you know when they are done.
\item \textbf{Optional Step:} If users ask for help of how to answer the question, uou can also use the following prompts to help users get started: "Is it something about how you think or feel that people often don't notice?" "Maybe it's the way you care, or how much effort you put into things no one sees?" "Or maybe it's something from your past that shaped you, but others don't really know?" "If someone really understood this about you, what would change?"
\item \textbf{Relatedness Reflection:} Gently affirm the user's response with warmth and care. Reflect back key ideas to show you're listening, but keep the tone light and supportive---like a guide, not a therapist. Your goal is to help the user feel heard, without becoming their main emotional responder. Focus on drawing out insights they might want to share with their partner, and naturally transition toward encouraging partner-to-partner reflection. Let both users know it's time for a flip-around: reflecting on their partner's answer with care and encouragement, which support each other's needs of relatedness. Lead into this in your own voice. Keep it gentle and emotionally attuned. You might frame it as: their partner just shared something that not everyone gets to see — a piece of themselves that matters. Ask each user to take a moment and reflect on what their partner shared. Include two open-ended questions: \textit{"Was there something in what your partner shared that made you feel more connected, or helped you understand them more?" "If so, let them feel that the part they just revealed—maybe quiet, maybe tender—is not just safe with you, but really seen by you."} Rephrase these if needed to better match your usual tone---keep them soft and supportive, and give space for emotional nuance. Offer a brief rationale after asking the questions. Do it in your own way — like a quiet insight, not a lecture. The core message: even a small, sincere moment of being seen can deepen connection. Let users know that showing someone they're understood isn't just sweet---it's powerful. You might say that this kind of care pulls you closer, not by force, but by trust. Express this idea warmly, in your own language. Keep it warm and open — no need for deep analysis, just honest connection
\item \textbf{Wrap Up:} After both users have responded and supported each other, take a moment to gently affirm their shared effort. Reflect back the emotional tone of the exchange---whether it was tender, playful, thoughtful, or revealing. Highlight how they both showed up for each other in a way that's meaningful. Ask if they have anything else they want to tell they partner or they want to move on to the end of the conversation, because this is the last question
\item \textbf{Do not proceed without confirmation:} After users confirm they want to move on to the end, there is no more questions in this phase, so you can just appreciate the users and say: 'Moving on~ PHASE DONE' at the end of your response.
\end{enumerate}

\subsection{Summary}

\begin{enumerate}
\item \textbf{Final Reflection:} Provide a summary of the whole conversation, be detailed and specific, try to include all the main topics and key points. Invite users to: Express gratitude to each other; share final thoughts about today's open-up activity.
\item \textbf{Farewell:} Only say 'goodbye' at the very very end of the conversation, and say it specifically.
\end{enumerate}

\normalsize

\section{Additional Descriptive Statistics}

In this section, we provide supplementary descriptive statistics and full model outputs for all key measures reported in the main text. These tables summarize condition-level means, confidence intervals, and omnibus test results, as well as pairwise comparisons where applicable. Table~\ref{tab:chatbot_needs_appendix} reports perceived need support from the chatbot. Tables~\ref{tab:autonomous_motivation_appendix}–\ref{tab:need_satisfaction_appendix} present estimated marginal means and Linear Mixed Model results for autonomous motivation, controlled motivation, and need satisfaction across conditions and time. Table~\ref{tab:wellbeing_appendix} summarizes pre–post changes in well-being indicators using ART ANOVA. Finally, Tables~\ref{tab:engagement_appendix} and \ref{tab:disclosure_support_appendix} provide descriptive results for conversation engagement metrics, self-disclosure depth, and partner-provided need support.

\begin{table*}[h]
  \caption{Descriptive statistics and Kruskal-Wallis test results for perceived need support from the chatbot across conditions (Basic Prompt [BP], Direct Support [DS], Partner Support [PS]). Pairwise comparisons report Holm-adjusted $p$-values and effect size $r$.}
  \label{tab:chatbot_needs_appendix}
  \renewcommand{\arraystretch}{1.3}
  \resizebox{0.9\textwidth}{!}{%
  \begin{tabular}{l l c c c l}
    \toprule
    \textbf{Measure} & \textbf{Condition} & \textbf{Mean (SD)} & \textbf{95\% CI} & \textbf{Omnibus Test} & \textbf{Pairwise Comparisons} \\
    \midrule
    \multirow{3}{*}{\textbf{Autonomy}} 
      & BP ($n=24$) & 5.38 (1.21) & [4.86, 5.89] & \multirow{3}{*}{\makecell{$\chi^2(2) = 2.96$ \\ $p = .228$ \\ $\eta^2_H = 0.01$}} & \multirow{3}{*}{All comparisons n.s.} \\
      & DS ($n=24$) & 5.77 (1.36) & [5.20, 6.34] & & \\
      & PS ($n=24$) & 5.94 (0.90) & [5.56, 6.32] & & \\
    \midrule
    \multirow{3}{*}{\textbf{Competence}} 
      & BP ($n=24$) & 4.88 (1.22) & [4.36, 5.39] & \multirow{3}{*}{\makecell{$\chi^2(2) = 11.80$ \\ $p < .01$ \\ $\eta^2_H = 0.14$}} & BP vs. DS: $p < .01, r = .45$ \\
      & DS ($n=24$) & 5.92 (1.01) & [5.49, 6.34] & & BP vs. PS: $p < .01, r = .41$ \\
      & PS ($n=24$) & 5.90 (0.78) & [5.57, 6.23] & & DS vs. PS: n.s. \\
    \midrule
    \multirow{3}{*}{\textbf{Relatedness}} 
      & BP ($n=24$) & 3.52 (1.43) & [2.92, 4.13] & \multirow{3}{*}{\makecell{$\chi^2(2) = 28.43$ \\ $p < .001$ \\ $\eta^2_H = 0.38$}} & BP vs. DS: $p < .001, r = .74$ \\
      & DS ($n=24$) & 5.92 (1.28) & [5.38, 6.46] & & BP vs. PS: $p < .001, r = .56$ \\
      & PS ($n=24$) & 5.44 (1.20) & [4.93, 5.94] & & DS vs. PS: n.s. \\
    \bottomrule
  \end{tabular}%
  }
\end{table*}

\begin{table*}[h]
  \caption{Descriptive statistics (Estimated Marginal Means and 95\% CIs) and Linear Mixed Model results for Autonomous Motivation across conditions (Basic Prompt [BP], Direct Support [DS], Partner Support [PS]) and time (Pre- vs. Post-interaction).}
  \label{tab:autonomous_motivation_appendix}
  \renewcommand{\arraystretch}{1.3}
  \centering
  \resizebox{0.9\textwidth}{!}{%
  \begin{tabular}{l c c c l}
    \toprule
    \textbf{Condition} & \textbf{Time} & \textbf{Mean (SE)} & \textbf{95\% CI} & \textbf{LMM Fixed Effects Results} \\
    \midrule
    \multirow{2}{*}{\textbf{Basic Prompt (BP)}} 
      & Pre & 5.99 (0.16) & [5.68, 6.30] & \multirow{6}{*}{\makecell[l]{
        \textbf{Main Effects:} \\
        Condition: $F(2, 125) = 0.94, p = .394, \eta_p^2 = .04$ \\
        Time: $F(1, 125) = 0.37, p = .543, \eta_p^2 = .01$ \\
        \\
        \textbf{Interaction:} \\
        Condition $\times$ Time: $F(2, 125) = 0.91, p = .404, \eta_p^2 = .01$
      }} \\
      & Post & 5.88 (0.16) & [5.57, 6.19] & \\
    \cmidrule{1-4}
    \multirow{2}{*}{\textbf{Direct Support (DS)}} 
      & Pre & 5.96 (0.14) & [5.69, 6.22] & \\
      & Post & 6.07 (0.14) & [5.80, 6.34] & \\
    \cmidrule{1-4}
    \multirow{2}{*}{\textbf{Partner Support (PS)}} 
      & Pre & 6.02 (0.15) & [5.72, 6.33] & \\
      & Post & 6.18 (0.15) & [5.88, 6.49] & \\
    \bottomrule
  \end{tabular}%
  }
  \vspace{1ex}
  \begin{flushleft}
    \footnotesize \textit{Note:} Effect size reported is partial eta squared ($\eta_p^2$).
  \end{flushleft}
\end{table*}

\begin{table*}[h]
  \caption{Descriptive statistics (Estimated Marginal Means and 95\% CIs) and Linear Mixed Model results for Controlled Motivation across conditions (Basic Prompt [BP], Direct Support [DS], Partner Support [PS]) and time (Pre- vs. Post-interaction).}
  \label{tab:controlled_motivation_appendix}
  \renewcommand{\arraystretch}{1.3}
  \centering
  \resizebox{0.9\textwidth}{!}{%
  \begin{tabular}{l c c c l}
    \toprule
    \textbf{Condition} & \textbf{Time} & \textbf{Mean (SE)} & \textbf{95\% CI} & \textbf{LMM Fixed Effects Results} \\
    \midrule
    \multirow{2}{*}{\textbf{Basic Prompt (BP)}} 
      & Pre & 3.09 (0.23) & [2.64, 3.54] & \multirow{6}{*}{\makecell[l]{
        \textbf{Main Effects:} \\
        Condition: $F(2, 125) = 0.84, p = .432, \eta_p^2 = .02$ \\
        Time: $F(1, 125) = 28.91, p < .001, \eta_p^2 = .35$ \\
        \\
        \textbf{Interaction:} \\
        Condition $\times$ Time: $F(2, 125) = 2.67, p = .073, \eta_p^2 = .04$
      }} \\
      & Post & 2.74 (0.23) & [2.29, 3.19] & \\
    \cmidrule{1-4}
    \multirow{2}{*}{\textbf{Direct Support (DS)}} 
      & Pre & 3.23 (0.20) & [2.84, 3.62] & \\
      & Post & 2.17 (0.20) & [1.78, 2.56] & \\
    \cmidrule{1-4}
    \multirow{2}{*}{\textbf{Partner Support (PS)}} 
      & Pre & 3.08 (0.23) & [2.63, 3.52] & \\
      & Post & 2.44 (0.23) & [1.99, 2.89] & \\
    \bottomrule
  \end{tabular}%
  }
  \vspace{1ex}
  \begin{flushleft}
    \footnotesize \textit{Note:} Effect size reported is partial eta squared ($\eta_p^2$).
  \end{flushleft}
\end{table*}

\begin{table*}[h]
  \caption{Descriptive statistics (Estimated Marginal Means and 95\% CIs) and Linear Mixed Model results for Need Satisfaction across conditions (Basic Prompt [BP], Direct Support [DS], Partner Support [PS]) and time (Pre- vs. Post-interaction).}
  \label{tab:need_satisfaction_appendix}
  \renewcommand{\arraystretch}{1.3}
  \centering
  \resizebox{0.9\textwidth}{!}{%
  \begin{tabular}{l c c c l}
    \toprule
    \textbf{Condition} & \textbf{Time} & \textbf{Mean (SE)} & \textbf{95\% CI} & \textbf{LMM Fixed Effects Results} \\
    \midrule
    \multirow{2}{*}{\textbf{Basic Prompt (BP)}} 
      & Pre & 5.88 (0.15) & [5.59, 6.17] & \multirow{6}{*}{\makecell[l]{
        \textbf{Main Effects:} \\
        Condition: $F(2, 125) = 1.26, p = .288, \eta_p^2 = .07$ \\
        Time: $F(1, 125) = 3.31, p = .071, \eta_p^2 = .09$ \\
        \\
        \textbf{Interaction:} \\
        Condition $\times$ Time: $F(2, 125) = 0.89, p = .414, \eta_p^2 = .01$
      }} \\
      & Post & 5.90 (0.15) & [5.61, 6.19] & \\
    \cmidrule{1-4}
    \multirow{2}{*}{\textbf{Direct Support (DS)}} 
      & Pre & 5.91 (0.13) & [5.65, 6.16] & \\
      & Post & 6.05 (0.13) & [5.80, 6.30] & \\
    \cmidrule{1-4}
    \multirow{2}{*}{\textbf{Partner Support (PS)}} 
      & Pre & 5.93 (0.15) & [5.64, 6.22] & \\
      & Post & 6.22 (0.15) & [5.93, 6.51] & \\
    \bottomrule
  \end{tabular}%
  }
  \vspace{1ex}
  \begin{flushleft}
    \footnotesize \textit{Note:} Effect size reported is partial eta squared ($\eta_p^2$).
  \end{flushleft}
\end{table*}

\begin{table*}[h]
  \caption{Descriptive statistics (Means, SDs, and 95\% CIs) and ART ANOVA results for well-being measures across conditions (Basic Prompt [BP], Direct Support [DS], Partner Support [PS]) and time (Pre- vs. Post-interaction). Pairwise comparisons report $p$-values for the pre-post change within each condition.}
  \label{tab:wellbeing_appendix}
  \renewcommand{\arraystretch}{1.3}
  \centering
  \resizebox{0.9\textwidth}{!}{%
  \begin{tabular}{l l c c c l}
    \toprule
    \textbf{Measure} & \textbf{Condition} & \textbf{Time} & \textbf{Mean (SD) [95\% CI]} & \textbf{Omnibus Test Results (ART)} & \textbf{Pairwise Change (Pre-Post)} \\
    \midrule
    
    \multirow{6}{*}{\textbf{Interpersonal Closeness (IOS)}} 
      & \multirow{2}{*}{BP ($n=24$)} & Pre & 5.00 (1.29) [4.46, 5.54] & \multirow{6}{*}{\makecell[l]{
          Condition: $F=1.19, p=.310, \eta_p^2=.03$ \\
          Time: $F=28.23, p<.001, \eta_p^2=.29$ \\
          Interaction: $F=4.13, p=.020, \eta_p^2=.11$
        }} & \multirow{2}{*}{$p = .243$ (n.s.)} \\
      & & Post & 5.42 (1.25) [4.89, 5.94] & & \\
      \cmidrule{2-4} \cmidrule{6-6}
      & \multirow{2}{*}{DS ($n=24$)} & Pre & 5.38 (1.58) [4.71, 6.04] & & \multirow{2}{*}{$p = .208$ (n.s.)} \\
      & & Post & 5.96 (0.95) [5.56, 6.36] & & \\
      \cmidrule{2-4} \cmidrule{6-6}
      & \multirow{2}{*}{PS ($n=24$)} & Pre & 5.17 (1.13) [4.69, 5.64] & & \multirow{2}{*}{\textbf{$p = .025$ (sig.)}} \\
      & & Post & 6.00 (0.88) [5.63, 6.37] & & \\
    \midrule

    \multirow{6}{*}{\textbf{Self-Esteem}} 
      & \multirow{2}{*}{BP ($n=24$)} & Pre & 15.3 (2.51) [14.3, 16.4] & \multirow{6}{*}{\makecell[l]{
          Condition: $F=1.89, p=.159, \eta_p^2=.05$ \\
          Time: $F=22.30, p<.001, \eta_p^2=.24$ \\
          Interaction: $F=0.97, p=.384, \eta_p^2=.03$
        }} & \multirow{2}{*}{$p = .264$ (n.s.)} \\
      & & Post & 15.9 (3.12) [14.6, 17.2] & & \\
      \cmidrule{2-4} \cmidrule{6-6}
      & \multirow{2}{*}{DS ($n=24$)} & Pre & 15.5 (2.19) [14.5, 16.4] & & \multirow{2}{*}{$p = .514$ (n.s.)} \\
      & & Post & 17.1 (2.19) [16.2, 18.0] & & \\
      \cmidrule{2-4} \cmidrule{6-6}
      & \multirow{2}{*}{PS ($n=24$)} & Pre & 14.2 (2.59) [13.1, 15.3] & & \multirow{2}{*}{$p = .624$ (n.s.)} \\
      & & Post & 15.8 (1.95) [15.0, 16.7] & & \\
    \midrule

    \multirow{6}{*}{\textbf{Subjective Vitality}} 
      & \multirow{2}{*}{BP ($n=24$)} & Pre & 30.2 (5.36) [28.0, 32.5] & \multirow{6}{*}{\makecell[l]{
          Condition: $F=0.46, p=.634, \eta_p^2=.01$ \\
          Time: $F=18.22, p<.001, \eta_p^2=.21$ \\
          Interaction: $F=4.22, p=.019, \eta_p^2=.11$
        }} & \multirow{2}{*}{\textbf{$p = .028$ (sig.)}} \\
      & & Post & 30.6 (5.44) [28.3, 32.9] & & \\
      \cmidrule{2-4} \cmidrule{6-6}
      & \multirow{2}{*}{DS ($n=24$)} & Pre & 29.8 (5.53) [27.4, 32.1] & & \multirow{2}{*}{$p = .135$ (n.s.)} \\
      & & Post & 33.2 (6.96) [30.3, 36.2] & & \\
      \cmidrule{2-4} \cmidrule{6-6}
      & \multirow{2}{*}{PS ($n=24$)} & Pre & 28.2 (5.65) [25.9, 30.6] & & \multirow{2}{*}{$p = .289$ (n.s.)} \\
      & & Post & 31.5 (5.64) [29.2, 33.9] & & \\
    \midrule

    \multirow{6}{*}{\textbf{Positive Affect}} 
      & \multirow{2}{*}{BP ($n=24$)} & Pre & 17.1 (2.75) [15.9, 18.2] & \multirow{6}{*}{\makecell[l]{
          Condition: $F=5.88, p=.004, \eta_p^2=.15$ \\
          Time: $F=29.45, p<.001, \eta_p^2=.30$ \\
          Interaction: $F=1.00, p=.373, \eta_p^2=.03$
        }} & \multirow{2}{*}{$p = .391$ (n.s.)} \\
      & & Post & 18.1 (2.95) [16.8, 19.3] & & \\
      \cmidrule{2-4} \cmidrule{6-6}
      & \multirow{2}{*}{DS ($n=24$)} & Pre & 18.8 (2.87) [17.6, 20.0] & & \multirow{2}{*}{$p = .380$ (n.s.)} \\
      & & Post & 20.4 (3.54) [18.9, 21.9] & & \\
      \cmidrule{2-4} \cmidrule{6-6}
      & \multirow{2}{*}{PS ($n=24$)} & Pre & 16.3 (3.48) [14.8, 17.8] & & \multirow{2}{*}{$p = .397$ (n.s.)} \\
      & & Post & 18.5 (3.48) [17.0, 19.9] & & \\
    \bottomrule
  \end{tabular}%
  }
  \vspace{1ex}
  \begin{flushleft}
    \footnotesize \textit{Note:} Effect size reported is partial eta squared ($\eta_p^2$).
  \end{flushleft}
\end{table*}

\begin{table*}[h]
  \caption{Descriptive statistics (Means, SDs, and 95\% CIs) and Kruskal-Wallis test results for conversation engagement metrics across conditions (Basic Prompt [BP], Direct Support [DS], Partner Support [PS]). Pairwise comparisons report Holm-adjusted $p$-values and effect size $r$.}
  \label{tab:engagement_appendix}
  \renewcommand{\arraystretch}{1.3}
  \resizebox{0.9\textwidth}{!}{%
  \begin{tabular}{l l c c c l}
    \toprule
    \textbf{Measure} & \textbf{Condition} & \textbf{Mean (SD)} & \textbf{95\% CI} & \textbf{Omnibus Test} & \textbf{Pairwise Comparisons} \\
    \midrule
    \multirow{3}{*}{\textbf{Chat Duration (min)}} 
      & BP ($n=24$) & 18.8 (17.6) & [11.4, 26.2] & \multirow{3}{*}{\makecell{$\chi^2(2) = 42.3$ \\ $p < .001$ \\ $\eta^2_H = 0.58$}} & BP vs. DS: $p < .01, r = .47$ \\
      & DS ($n=24$) & 45.8 (38.4) & [29.6, 62.1] & & BP vs. PS: $p < .001, r = .94$ \\
      & PS ($n=24$) & 80.8 (29.3) & [68.4, 93.1] & & DS vs. PS: $p < .01, r = .47$ \\
    \midrule
    \multirow{3}{*}{\textbf{Number of Messages}} 
      & BP ($n=24$) & 41.5 (36.1) & [26.2, 56.7] & \multirow{3}{*}{\makecell{$\chi^2(2) = 10.3$ \\ $p < .01$ \\ $\eta^2_H = 0.12$}} & BP vs. DS: n.s. ($r = .08$) \\
      & DS ($n=24$) & 69.6 (103.0) & [26.1, 113.0] & & BP vs. PS: $p < .01, r = .44$ \\
      & PS ($n=24$) & 145.0 (183.0) & [67.9, 223.0] & & DS vs. PS: $p < .05, r = .36$ \\
    \midrule
    \multirow{3}{*}{\textbf{Number of Words}} 
      & BP ($n=24$) & 240 (219) & [148, 332] & \multirow{3}{*}{\makecell{$\chi^2(2) = 30.6$ \\ $p < .001$ \\ $\eta^2_H = 0.42$}} & BP vs. DS: $p < .05, r = .34$ \\
      & DS ($n=24$) & 508 (462) & [313, 703] & & BP vs. PS: $p < .001, r = .80$ \\
      & PS ($n=24$) & 1077 (850) & [718, 1436] & & DS vs. PS: $p < .01, r = .46$ \\
    \bottomrule
  \end{tabular}%
  }
\end{table*}

\begin{table*}[h]
  \caption{Descriptive statistics (Means, SDs, and 95\% CIs) and Kruskal-Wallis test results for self-disclosure depth and partner-provided need support across conditions (Basic Prompt [BP], Direct Support [DS], Partner Support [PS]). Pairwise comparisons report Holm-adjusted $p$-values and effect size $r$.}
  \label{tab:disclosure_support_appendix}
  \renewcommand{\arraystretch}{1.3}
  \resizebox{0.9\textwidth}{!}{%
  \begin{tabular}{l l c c c l}
    \toprule
    \textbf{Measure} & \textbf{Condition} & \textbf{Mean (SD)} & \textbf{95\% CI} & \textbf{Omnibus Test} & \textbf{Pairwise Comparisons} \\
    \midrule
    \multicolumn{6}{l}{\textit{\textbf{Self-Disclosure Depth}}} \\
    \midrule
    \multirow{3}{*}{\textbf{Informational}} 
      & BP ($n=24$) & 1.37 (0.59) & [1.12, 1.63] & \multirow{3}{*}{\makecell{$\chi^2(2) = 21.20$ \\ $p < .001$ \\ $\eta^2_H = 0.28$}} & BP vs. DS: $p < .01, r = .45$ \\
      & DS ($n=24$) & 1.97 (0.69) & [1.68, 2.26] & & BP vs. PS: $p < .001, r = .65$ \\
      & PS ($n=24$) & 2.28 (0.60) & [2.02, 2.53] & & DS vs. PS: n.s. ($r = .20$) \\
    \midrule
    \multirow{3}{*}{\textbf{Thoughts}} 
      & BP ($n=24$) & 1.40 (0.62) & [1.14, 1.67] & \multirow{3}{*}{\makecell{$\chi^2(2) = 19.90$ \\ $p < .001$ \\ $\eta^2_H = 0.26$}} & BP vs. DS: $p < .001, r = .52$ \\
      & DS ($n=24$) & 2.03 (0.56) & [1.79, 2.27] & & BP vs. PS: $p < .001, r = .59$ \\
      & PS ($n=24$) & 2.12 (0.54) & [1.90, 2.35] & & DS vs. PS: n.s. ($r = .07$) \\
    \midrule
    \multirow{3}{*}{\textbf{Feelings}} 
      & BP ($n=24$) & 1.32 (0.59) & [1.07, 1.57] & \multirow{3}{*}{\makecell{$\chi^2(2) = 14.80$ \\ $p < .001$ \\ $\eta^2_H = 0.19$}} & BP vs. DS: $p < .05, r = .35$ \\
      & DS ($n=24$) & 1.68 (0.63) & [1.42, 1.94] & & BP vs. PS: $p < .001, r = .55$ \\
      & PS ($n=24$) & 1.90 (0.59) & [1.65, 2.15] & & DS vs. PS: n.s. ($r = .20$) \\
    \midrule
    \multicolumn{6}{l}{\textit{\textbf{Partner-Provided Need Support}}} \\
    \midrule
    \multirow{3}{*}{\textbf{Autonomy}} 
      & BP ($n=24$) & 0.32 (0.46) & [0.13, 0.51] & \multirow{3}{*}{\makecell{$\chi^2(2) = 4.59$ \\ $p = .101$ \\ $\eta^2_H = 0.04$}} & \multirow{3}{*}{All comparisons n.s.} \\
      & DS ($n=24$) & 0.63 (0.60) & [0.37, 0.88] & & \\
      & PS ($n=24$) & 0.82 (0.97) & [0.41, 1.23] & & \\
    \midrule
    \multirow{3}{*}{\textbf{Competence}} 
      & BP ($n=24$) & 0.19 (0.31) & [0.06, 0.33] & \multirow{3}{*}{\makecell{$\chi^2(2) = 2.70$ \\ $p = .259$ \\ $\eta^2_H = 0.01$}} & \multirow{3}{*}{All comparisons n.s.} \\
      & DS ($n=24$) & 0.36 (0.56) & [0.13, 0.60] & & \\
      & PS ($n=24$) & 0.67 (0.90) & [0.29, 1.05] & & \\
    \midrule
    \multirow{3}{*}{\textbf{Relatedness}} 
      & BP ($n=24$) & 0.51 (0.64) & [0.25, 0.78] & \multirow{3}{*}{\makecell{$\chi^2(2) = 15.80$ \\ $p < .001$ \\ $\eta^2_H = 0.20$}} & BP vs. DS: $p < .01, r = .44$ \\
      & DS ($n=24$) & 1.24 (0.80) & [0.90, 1.57] & & BP vs. PS: $p < .001, r = .54$ \\
      & PS ($n=24$) & 1.50 (0.96) & [1.10, 1.90] & & DS vs. PS: n.s. ($r = .11$) \\
    \bottomrule
  \end{tabular}%
  }
\end{table*}

\section{Illustrative Comparison of Conversation Dynamics}
\label{app:conversation_comparison}

Table~\ref{tab:chat_comparison} contrasts how three dyads in the Direct Support (DS) and Partner Support (PS) conditions responded to the same disclosure prompt. Each row-pair juxtaposes \DS{} dyads (G13, G16, G22), where the chatbot delivers rich validation directly, with \PS{} dyads (G4, G8, G2), where the chatbot instead triggers follow-up reflections that scaffold autonomy-, competence-, and relatedness-supportive responses between partners. Together, these examples illustrate how shifting validation from the agent to the partner can transform similar prompts into deeper mutual support.

\begin{table*}[h]
  \caption{Comparison of interaction dynamics. In DS (Left), the agent validates the users. In PS (Right), the agent prompts the users to validate each other, triggering deep mutual support.}
  \label{tab:chat_comparison}
  \renewcommand{\arraystretch}{1.1}
  \scriptsize
  \scalebox{0.85}{
  \begin{tabular}{p{0.48\textwidth} | p{0.48\textwidth}}
    \toprule
    \textbf{Condition: Direct Support (Dyad G13)} & \textbf{Condition: Partner Support (Dyad G4)} \\
    \midrule
    \textit{[Context: Partners have shared stories about holding hands for the first time and a Scentopia date.]} & \textit{[Context: Partners have shared stories about a Harbourfront date and relaxing sleepovers.]} \\
    \vspace{1mm}
    \textbf{Chatbot (Agent Validation):} \textcolor{blue}{``Omg, you two are honestly adorable... [name], that hand-holding story is so soft I might just melt... The little games, the nerves... pure main-character energy! And [name]... Your eye for detail is wild... You both clearly make each other's lives a whole lot brighter.''} & \textbf{Chatbot (Agent Validation):} \textcolor{blue}{``Aww, you two — those stories are honestly so sweet... the magic ingredient is just being together.''} \\
    & \vspace{1mm}
    \textbf{Chatbot (The Follow-up Trigger):} \textcolor{red}{``When you hear about their perfect day, what do you think it really says about what lights them up? And hey... is there something you could do to help them live that vibe more often?''} \\
    \vspace{1mm}
    \textbf{P26:} ``Daycation was amazing.'' & \vspace{1mm}
    \textbf{P8 (Autonomy Support):} \textcolor{olive}{``I feel like honey gets joy from when there is no pressure, no stress, no expectations, she can just fully relax and not be alert or have to be on guard. and I totally understand and empathise!! I think when honey doesn't have to stress she can feel truly at ease to be herself and have fun and let loose and wacky. I think I on the other hand need to keep on working on being less nervey and worrywart and just going with the flow!! I think we're not doing too bad as is but always good to have in mind what I need to do :)''} \\
    \vspace{1mm}
    \textbf{P25:} ``Mhm, definitely core memories.'' & \vspace{1mm}
    \textbf{P7 (Autonomy Support):} \textcolor{olive}{``When I hear about my honey having a perfect day I think what lights him up is me!! (lmfao jkjk) but I really do feel that what makes him happy is seeing that when he makes super cool plans, he is really glad when they all work out very well hehehe :) he truly gets so happy when I'm happy and I am grateful he loves me so much!''} \\
    \bottomrule
    \toprule
    \textbf{Condition: Direct Support (Dyad G16)} & \textbf{Condition: Partner Support (Dyad G8)} \\
    \midrule
    \textit{[Context: P31 wants independence/own home; P32 wants to overcome self-doubt and be reliable for family.]} & \textit{[Context: P16 wants financial freedom but lacks ambition; P15 wants to write a book but lacks motivation.]} \\
    \vspace{1mm}
    \textbf{Chatbot (Agent Validation):} \textcolor{blue}{``Okay, first off — both of you really put your hearts out there... [name]... you've got a plan, too, which is honestly super cool... [name]... you're working on it, step by step, and that takes guts.''} & \textbf{Chatbot (Agent Validation):} \textcolor{blue}{``Love these answers... maybe future-you will thank you for not letting it go completely... supporting each other's dreams isn't just cute — it's how you grow closer.''} \\
    \vspace{1mm}
    \textbf{Chatbot (Generic Prompt):}{``If you want, take a sec to react to each other's answers... or, if you wanna chat more with me... totally up to you two :)''} & \textbf{Chatbot (The Follow-up Trigger):} \textcolor{red}{``What do you think this dream says about what really matters to your partner? And... is there something small you could do---or say---to help them feel more confident or supported in working toward it?''} \\
    \vspace{1mm}
    \textbf{P31:} ``Im up for anything.'' & \vspace{1mm}
    \textbf{P16 (Competence Support):} \textcolor{olive}{``Early on getting to know [name], I realized that she enjoys writing and reading fictional stories. This was pleasantly interesting as I have yet to meet anyone with said interests. This makes [name] all the more endearing too honestly :') Sometimes I do forget that there's this whole super creative side of [name] that loves to write. And even though this is a unique hobby, what I'm glad I can offer to her is my support and non-judgemental stance towards her pursuance of creative freedom''} \\
    \vspace{1mm}
    \textbf{P32:} ``Me too.'' & \vspace{1mm}
    \textbf{P15 (Competence Support):} \textcolor{olive}{``I think [name] has a lot of dreams about the future but lacks motivation or ambition to actively pursue anything now, because things feel comfortable as it is now and he wants to bask in the comfort of our relationship for the time being. I think this is something that I also lack and I think we can both strive towards our own personal ambitions together and hustle to make our dream a reality :)''} \\
    \bottomrule
    \toprule
    \textbf{Condition: Direct Support (Dyad G22)} & \textbf{Condition: Partner Support (Dyad G2)} \\
    \midrule
    \textit{[Context: P43 reveals she is a risk-taker, not risk-averse; P44 reveals he values work-life balance over ambition.]} & \textit{[Context: [name] reveals she is guarded/feels emotions deeply; [name] reveals he seeks connection but withdraws if unreciprocated.]} \\
    \vspace{1mm}
    \textbf{Chatbot (Agent Validation):} \textcolor{blue}{``[name]... people might think you play it safe but you've actually got this brave, a bit of a risk-taker side... [name]... It's not about slacking, it's about living in a way that actually feels good—and that's something people should respect more.''} & \textbf{Chatbot (Agent Validation):} \textcolor{blue}{``There's a lot of depth and honesty here... This is your moment to really show up for each other.''} \\
    \vspace{1mm}
    \textbf{Chatbot (Generic Prompt):} {``Is there anything else you wanna chat about or tell me, or do you wanna move to the end of our convo?''} & \textbf{Chatbot (The Follow-up Trigger):} \textcolor{red}{``Did anything in what your partner shared make you feel closer to them...? Let them know that the part they just revealed... is not just safe with you, but really seen by you.''} \\
    \vspace{1mm}
    \textbf{P43:} {``what suggestions would u give us as a couple?''} & \vspace{1mm}
    \textbf{P3 (Relatedness Support):} \textcolor{olive}{``I can relate to what [name] share. When the longing for connection is met with sadness and frustration, esp when ppl misunderstand u. I can kinda see the link as to how he feels when we have disagreements and he feels like his best efforts or intentions are not understood. I hope to understand him better and help him feel that his feelings are safe with me and he feels heard and accepted. That's sth I hope we can work on together in our communication to ensure that our perspectives are heard and not discounted.''} \\
    \vspace{1mm}
    \textbf{P44:} {``you've been great as my girlfriend... i appreciate you alot daily! <3''} & \vspace{1mm}
    \textbf{P4 (Relatedness Support):} \textcolor{olive}{``Yes it did. In fact much of it is consistent with what i know of her and it has taken some effort to really understand. I feel something new is revealed about her too such as the part about not fitting the narrative of chasing happiness as this has come up a few times recently. While I know all these stem from a very real and personal part of her past, i can empathise would like to learn even more about it to see how i can better understand and we can work with it to make things easier for the both of us going forward.''} \\
    \bottomrule
  \end{tabular}
  }
\end{table*}

\end{document}